\documentclass{aa}  
\usepackage[para,online,flushleft]{threeparttable}
\usepackage{subcaption}
\usepackage{hyperref}
\DeclareUnicodeCharacter{2212}{-}
\usepackage{graphicx}
\usepackage{txfonts}
\usepackage{xcolor}
\usepackage{float}
\usepackage{soul}
\usepackage{placeins}
\usepackage{flafter} 
\usepackage{dblfloatfix}
\usepackage{lineno}

\begin{document}

   \title{Probing the faint end of simulated galaxy counts at $z\gtrsim3$}

  \author{Flaminia Fortuni \inst{1}
        \and Emiliano Merlin \inst{1}
        \and Marco Castellano \inst{1}
        \and Adriano Fontana \inst{1}
        \and Paola Santini \inst{1}
        }

   \institute{INAF - Osservatorio Astronomico di Roma, via Frascati 33, 00078 Monte Porzio Catone (Roma), Italy.
        \email{flaminia.fortuni@inaf.it}
             }

   \date{Received ; accepted }

\titlerunning{Galaxy counts with FORECAST}
\authorrunning{Fortuni et al.}

  \abstract
   {Simulations and observations now probe comparable redshift regimes with unprecedented accuracy. This overlap allows us to test their consistency under the same observational conditions using the forward-modeling technique.}
   {In our previous work, we identified a faint-end discrepancy between the observed and simulated near-infrared galaxy counts in CANDELS GOODS-South. Here, we assess whether this mismatch arises from the forward-modeling process or from intrinsic limitations of the underlying simulations, and we characterize the galaxy populations that give rise to the difference.}
   {Using the FORECAST forward-modeling code, we generated ten independent realizations of light cones and mock images emulating the CANDELS fields from the TNG100 and EAGLE simulations. We compared the light-cone catalogs, that is, the input universe and the detections on mock images with observations, testing dependences on fields and redshift. We performed validation tests to assess the robustness of the forward-modeling pipeline, including checks on stellar mass, comparisons with independent mock datasets, and a multiband analysis of the spectral energy distribution modeling incorporating a CEERS band.}
   {The faint-end deficit was found in all CANDELS fields; it emerges at $z\gtrsim$3 in both simulations. After we corrected GOODS-South counts for completeness, they exceeded input universe counts already where the survey is 50\% complete, indicating that the missing population is not simply hidden below the detection threshold. An artificial deepening of the mock images recovered the observed counts near the peak, but overpredicted the faintest counts, showing that depth alone cannot resolve the difference. The analysis of the structural parameters revealed that the faint compact galaxies with bright central cores in GOODS-South are underproduced in simulations, which are biased toward diffuse systems with a low surface brightness.}
   {The shortfall of faint sources partly originates from detection losses of diffuse galaxies with a low surface brightness, but more fundamentally, from the inability of current hydrodynamical simulations to produce enough faint compact galaxies with bright cores at $z>3$. This difference highlights the need to refine the modeling of early galaxy formation, in particular, the treatment of star formation and feedback, and to improve the treatment of dust in the forward-modeling process.}
   

   \maketitle
   \nolinenumbers

\section{Introduction}

The formation and evolution of galaxies involve a complex interplay of nonlinear physical processes acting across a wide range of scales. Cosmological hydrodynamical simulations (CHSs) provide a powerful framework for investigating these processes within a cosmological context. They give insights into the mechanisms driving galaxy evolution by combining large-scale structure formation with explicit hydrodynamical modeling of baryons via grid-based or particle-based algorithms, and they include subgrid prescriptions for star formation, stellar feedback, and Active Galactic Nucleus (AGN) feedback \citep{evrard88, hernquistkatz89, springelhernquist02, springel10, dubois14, schaye15, pillepich18, dave19, dubois21, kannan22, bird22, schaye23,pakmor23}. Because of this, their predictive power is considered more rigorous than that of semi-analytical models \citep{laceysilk91, whitefrenk91,cole94,somerville99,cole00,lacey01,somerville15, lacey16}. The latter are computationally more efficient and thus suitable for exploring a large parameter space, but they are limited by the less accurate prescriptions that are adopted to model baryonic processes in addition to dark matter N-body simulations \citep[e.g.,][]{efstathiou85}.\\
\indent As highly sophisticated tools for predicting the statistical properties of galaxies in the Universe, CHSs are in principle expected to reproduce the very basic observational features of the galactic populations at all cosmic epochs. However, differences between theoretical predictions and observations have been known for a long time \citep[see][]{crain23}. 
A persistent challenge for CHSs is to accurately reproduce pivotal diagnostics such as the galaxy stellar mass function (GSMF or MF) across cosmic time. Historically, simulations tended to overproduce stars at the low- and high-mass end, particularly at high redshifts, when cosmic inflows were high \citep{navarro91}. The introduction of subgrid models for feedback, such as star-driven winds and AGN feedback, has significantly improved low-redshift predictions, but to this day, CHSs typically show severe inconsistencies with observations and also between different models \citep{suresh26}, especially at $z>$1 \citep{weaver23}. A related issue is the underprediction of passive galaxy populations at $z>3$, which highlights difficulties in properly modeling early quenching mechanisms \citep{merlin19,santini21,merlin25}. Galaxy sizes are another point of difference: simulated galaxies have been found to be larger than their observed counterparts, with discrepances up to a factor of two in some models \citep{snyder15}. CHSs also struggle to match the observed distribution of gas in galaxies. They overestimate the abundance of cold gas in massive halos and underestimate that of molecular gas in low-mass systems \citep{dave20}. 
Traditionally, these issues have been addressed through the a posteriori fine-tuning of free parameters in the subgrid modeling of the physical processes of star formation and feedback, which are not always physically justified in any case and may affect local (i.e., $z=0$) related predictions such as the local stellar mass function and star formation rates.\\ 
\indent In this context, deep-imaging surveys such as the Cosmic Assembly Near-infrared Deep Extragalactic Legacy Survey \citep[CANDELS]{grogin11, koekemoer11}, which provide well-characterized galaxy samples over a broad redshift range, offer a crucial benchmark for assessing the predictive power of the most advanced simulations.\\
\indent Beyond physical modeling, a relevant source of difference lies in the method that is adopted to compare simulations and observations. While the first yield intrinsic physical properties of the galaxies, in the context of imaging surveys, the latter provide multiband photometric fluxes from which physical quantities are inferred through spectral energy distribution (SED) fitting techniques. These inferences depend on several modeling assumptions, including stellar population synthesis models, dust attenuation laws, star formation histories, and the adopted initial mass function, and they are further affected by observational uncertainties and selection effects. As a result, comparisons performed directly in the physical parameter space may introduce systematic biases that are difficult to quantify \citep[see][]{pacifici23}.\\
\indent Forward modeling provides a physically consistent way to bridge this gap. It ensures that theoretical predictions are evaluated under observational conditions that closely mimic those of real data, while keeping track of the underlying modeling assumptions. In recent years, this approach has been increasingly adopted  \citep[e.g.,][]{laigle19,drakos22,marshall22, snyder23,lovell25}. This method consists of generating synthetic observations from theoretical predictions by transforming the physical output of simulations (e.g., stellar mass, ages, metallicities, gas, and dust properties) into observational quantities such as integrated fluxes or even spectra \citep{wu21}, and by incorporating realistic observational effects including noise and the instrumental point spread function (PSF). Forward modeling is therefore a powerful tool for quantifying and constraining observational biases and uncertainties \citep[see][]{nanni24}. For this reason, it has been applied to a wide range of astrophysical problems: \cite{lovell21} studied submillimeter galaxy number counts to test the realism of dust models and star formation rates in simulations; \cite{lachance25} used mock observations to analyze the galaxy morphologies and their evolution with redshift and revealed key differences between theory and data; \cite{cochrane23} investigated AGN-driven winds, highlighting the role of feedback in shaping galaxies; \cite{cochrane24} explored Hubble Space Telescope (HST)-dark but James Webb Space Telescope (JWST)-bright star-forming galaxies and examined how dust and orientation angle affect their detectability; and finally, \cite{ishikawa24} explored cosmological applications by building mock luminous red galaxy catalogs for the Subaru Hyper Suprime-Cam Strategic Survey Program (HSC-SSP) \citep{aihara18}, analyzing galaxy clustering and predicting the baryonic acoustic oscillation detectability.\\
\indent Within this framework, we developed FORECAST \citep[][F23 hereafter]{fortuni23}, a dedicated forward-modeling tool that converts the CHS output into realistic mock astronomical images that can be used for direct comparison with observational data (see Sect. \ref{methods}). In F23, we used this tool to reproduce the CANDELS GOODS-South (GS) field \citep{grogin11,koekemoer11} from the IllustrisTNG100 simulation \citep[TNG100 for short,][]{pillepich18,weinberger18}. We performed a direct comparison between simulated and observed galaxy number counts in the F160W ($H$) band using the input universe (IU), that is, the catalog of simulated sources in the mock light cone, including their intrinsic physical properties and the corresponding noiseless observer-frame photometric fluxes derived from them, and fully forward-modeled catalogs obtained by applying PSF convolution, noise, and standard detection techniques to the mock images.
We found that for magnitudes fainter than $H \sim 26$, the mock detections yielded lower counts than the CANDELS GS data, falling short of $\sim$50\% with respect to the real counts at those magnitudes. This result suggested a potential shortfall of faint sources in the underlying hydrodynamical simulation, but the analysis was limited to only one field and focused on the detection band. It also lacked a systematic exploration of possible other sources of discrepancy.

This work presents an in-depth analysis of these findings. We extend the study to cover all five CANDELS fields \citep{guo13,nayyeri17,stefanon17,galametz13,barro19}, and we examine how the discrepancy evolves with redshift using two independent large-volume cosmological simulations, TNG100 and EAGLE \citep{schaye15}. We assess the effect of completeness and simulated noise, and we identify the structural properties of the galaxies that are missing from the simulations and those that are lost at the detection stage.

The paper is organized as follows. In Sect. \ref{methods} we describe the observational and theoretical datasets and the methods adopted to create and analyze mock datasets. In Sect. \ref{results} we present the results. In Sect. \ref{discussion} we discuss their implications, and finally, in Sect. \ref{conclusions}, we summarize our main findings. In Appendix \ref{validation_tests} we provide validation tests for the forward-modeling process.\\
We assumed a flat $\Lambda$CDM cosmology with $H_0$, $\Omega_{\Lambda}$, and $\Omega_{m}$, and we adopted standard AB magnitudes \citep{okegunn83}.

\section{Data and methods}\label{methods}

We used observational data from the CANDELS survey, exploiting publicly available photometric catalogs. Theoretical data were obtained from the output snapshots of the TNG100 and EAGLE cosmological hydrodynamical simulations. In this section, we describe the datasets and the methods we used for the comparisons in detail.

\subsection{Observational data: The CANDELS survey}
The CANDELS program, the largest Multi-Cycle Treasury initiative on the Hubble Space Telescope (HST), provides imaging data across five key extragalactic fields: Ultra Deep Survey (UDS), Cosmological Evolution Survey (COSMOS), Extended Groth Strip (EGS), Great Observatories Origins Deep Survey South (GOODS-South, GS), and Great Observatories Origins Deep Survey North (GOODS-North, GN), covering nearly 10000 sq. arcmin. These fields were observed with HST's Wide Field Camera 3 (WFC3) and Advanced Camera for Surveys (ACS) in $\sim$20 photometric bands, spanning ultraviolet to near-infrared wavelengths. The official CANDELS catalogs adopt the F160W (or $H$) band as the detection band, using a combination of hot and cold mode for the detection and deblending of extended and bright sources, as well as for the identification of faint sources with \textsc{SExtractor} \citep[see][]{galametz13}. These catalogs complement HST photometry with multiwavelength coverage from ground- and space-based facilities, extending from the ultraviolet to the infrared. The number of available bands varies by field, ranging from 18 (GOODS-North) to 43 (GOODS-South and COSMOS). Fluxes in HST bands are typically measured with aperture photometry after PSF-matching, while fluxes in ground-based and Spitzer bands are derived using template-fitting techniques \citep[e.g., \textsc{t-phot}][]{merlin15, merlin16}.

We mainly used the $H$ band from the CANDELS catalogs for our analysis. Specifically, we exploited the ASTRODEEP-GS43 catalog \footnote{\url{https://astrodeep.eu}} \citep{merlin21} for GOODS-South, which includes photometry and photometric redshifts (and spectroscopic, when available), as well as physical properties, for approximately 35000 $H$-detected sources. For the other fields, we used the official CANDELS catalogs from \citet[COSMOS]{nayyeri17}, \citet[EGS]{stefanon17}, \citet[GOODS-North]{barro19}, and \citet[UDS]{galametz13}.

In addition to the $H$ band, we used the F105W band for GOODS-South in Appendix \ref{validation_tests}. Since no official catalog exists for this band in CANDELS (where the detection was performed in $H$), we performed our own detection directly on the real image.

\subsection{Theoretical data}\label{theoretical_data}

\subsubsection{IllustrisTNG simulation}
The IllustrisTNG simulation suite builds on the original Illustris project \citep{vogelsberger14,genel14,sijacki15}, employing the moving-mesh AREPO code \citep{springel10} to solve hydrodynamic equations, allowing adaptive resolution and improved handling of shocks, gas mixing, and angular momentum conservation. 
We chose the IllustrisTNG simulation suite for this work, as it was also the basis of our previous study, where it was selected for being the state of the art and publicly available.\\
\indent TNG spans a variety of box sizes and resolutions, with its free parameters tuned to reproduce key galaxy population observables, including the cosmic star formation rate, the stellar mass function, and the stellar-to-halo mass relation at $z$=0. The TNG simulations are based on the \cite{planck16} cosmology ($\Omega_m$=0.31, $\Omega_b$=0.0486, $\Omega_{\Lambda}$=0.692, $h$=0.6774, $n_s$=0.9667, $\sigma_8$=0.8159). The TNG100-1 run (intermediate box with highest resolution) used in this work simulates a comoving volume of (110.7)$^3$ cMpc$^3$, with a gas cell mass resolution of 1.40$\times$10$^6$ $M_{\odot}$ and a dark matter (DM) particle mass of 7.5$\times$10$^6$ $M_{\odot}$.\\
\indent Subgrid models include prescriptions for gas cooling, star formation, stellar feedback, and AGN feedback.
Star formation is based on \cite{springelhernquist03}, which assumes a two-phase interstellar medium, with stars forming in dense, cold gas regions, adopting a \cite{chabrier03} initial mass function (IMF). Feedback from massive stars and supernovae injects both thermal and kinetic energy into the surrounding medium \cite{pillepich18}, driving large-scale outflows and regulating star formation. Black hole (BH) growth occurs via gas accretion and mergers, with AGN feedback modeled with two distinct modes: thermal feedback at high black hole accretion rates and kinetic feedback at low rates \citep{weinberger18}. Kinetic feedback is implemented as randomly oriented winds, with the transition between modes determined by the black hole accretion rate and mass. TNG also includes a detailed model for chemical enrichment \citep{torrey14} and magnetic fields, which evolve self-consistently within the magneto-hydrodynamical framework \cite{pakmorspringel13}. Galaxies are identified using the SUBFIND algorithm \citep{springel01,dolag09}, which traces gravitationally bound substructures within large dark matter halos.

\subsubsection{EAGLE simulation}
The EAGLE project (Evolution and Assembly of GaLaxies and their Environments) is a publicly available series of CHS, carried out with a modified version of the smoothed particle hydrodynamics (SPH) code GADGET-3 \citep{springel05}. These modifications, collectively referred to as the Anarchy SPH scheme, improve the modeling of fluid instabilities, shocks, and gas mixing compared to standard SPH techniques.

EAGLE includes simulations with different box sizes and resolutions, calibrated to match key observables of the galaxy population at $z$=0.1 (GSMF and central BH masses as a function of galaxy stellar mass). The reference cosmology for EAGLE is based on the Planck 2014 results (\cite{planck14}: $\Omega_m$=0.307,$\Omega_b$=0.04825, $\Omega_{\Lambda}$=0.693, $h$=0.6777, $n_s$=0.9611, $\sigma_8$=0.8288), and the flagship box of the suite analyzed in this work, RefL0100N1504, spans a comoving volume of 100$^3$ cMpc$^3$. It has a baryonic particle mass resolution of 1.81$\times$10$^6$ $M_{\odot}$ and a dark matter particle mass of 9.7$\times$10$^6$ $M_{\odot}$.

Star formation is modeled using the pressure-dependent law described in \cite{schayedallavecchia08}, linking the local gas density to the star formation rate, with a \cite{chabrier03} IMF. Stellar feedback follows the stochastic thermal energy injection model of \cite{dallavecchiaschaye12}, where gas is stochastically heated to prevent rapid cooling and dissipation. Black holes grow through gas accretion and mergers, and AGN feedback is implemented as a stochastic thermal process, where nearby gas particles are stochastically heated to a fixed temperature ($\Delta T_{AGN}$=10$^{8.5}$ K) without preferential directionality for outflows \citep{boothscaye09}. This approach, conceptually similar to stellar feedback, prevents rapid cooling and allows feedback energy to effectively regulate star formation and black hole growth. Like in TNG, galaxies in EAGLE are identified using the SUBFIND algorithm.

\subsubsection{Forward-modeled data}\label{FM}

Observational results are routinely compared to CHS output. In this work, we adopt a forward-modeling (FM) approach, in which synthetic observations are generated from the simulation output in order to enable a direct and self-consistent comparison with observational data.
To this end, we used our FORECAST software to create mock catalogs (IU) and images that emulate real observations by forward modeling the output of the two mentioned CHS.
In F23, we presented and publicly released\footnote{\url{www.astrodeep.eu/FORECAST}} a mock dataset that includes both the IU and mock images post-processed to mimic the CANDELS GOODS-South and CEERS fields. The IU was created using the IllustrisTNG100 simulation at its highest resolution (with the -1 suffix in the official nomenclature), covering a redshift range from $z \simeq 0$ to 20 over a field of view of 200 sq. arcmin. The dimension is comparable to that of the CANDELS GOODS-South field ($\sim$173 sq. arcmin), and corresponds to a transverse comoving size that remains smaller than the simulation box at all redshifts, while repetitions along the line of sight are mitigated through the randomization of each snapshot following \cite{blaizot05}. The mock images were generated with the procedure described in F23, essentially by projecting stellar particle fluxes onto a 2D pixel grid, followed by PSF convolution and noise injection to reproduce the specific observational conditions.\\
\indent In this work, we expand upon that initial release to increase the statistical power of our analysis. We generated four additional light cones and IUs from TNG100, and five more based on the EAGLE hydrodynamical simulation. Each light cone was generated as an independent realization by adopting different random seeds in the randomization procedure. All new datasets spanned the redshift range $z\sim$ 0 -- 7 and cover 200 sq. arcmin. each, focusing on the redshift range where the comparison with the CANDELS GOODS-South dataset is more robust.
In total, we used ten combined mock datasets (hereafter \texttt{cmd}) in this study, five from TNG100 (including the one from F23) and five from EAGLE.
For all datasets, the spectral energy distributions (SEDs) of galaxies are modeled following \cite{gutkin16}, assuming a \cite{chabrier03} IMF, according to the output of the two CHSs. All subhalos included in the IU catalogs are required to contain at least ten stellar particles, in order to exclude galaxies too close to the resolution limit. This corresponds to a minimum stellar mass of about $1.4 \times 10^{7} \,\mathrm{M_\odot}$ for TNG100 and $1.8 \times 10^{7} \, \mathrm{M_\odot}$ for EAGLE.

We post-processed all the FORECAST noiseless images to emulate the depths of the real fields (see Table \ref{depths}) by adding Gaussian noise, consistent with the reported 5$\sigma$ limiting magnitude for each field and band. 

 \begin{table}[ht!]
        \caption{5$\sigma$ limiting magnitudes of the fields.}
        \label{depths}
        \centering
        \begin{threeparttable}
        \begin{tabular}[t]{cccc}
            \hline\hline
            Survey&Field&Band&$mag_{lim}$\\
            \hline
            CANDELS&UDS&F160W&$\text{27.45}$\tablefootmark{a} \\
            CANDELS&COSMOS&F160W&$\text{27.56}$\tablefootmark{a} \\
            CANDELS&GOODS-N&F160W&$\text{28.70}$\tablefootmark{a} \\
            CANDELS&GOODS-S&F105W&$\text{28.45}$\tablefootmark{a} \\
            CANDELS&GOODS-S&F160W&$\text{28.16}$\tablefootmark{a} \\
            CANDELS&EGS&F160W&$\text{27.60}$\tablefootmark{a} \\
            CEERS&EGS&F277W&$\text{29.20}$ \tablefootmark{b} \\        
            \hline
        \end{tabular} 
        \tablefoot{
         \tablefoottext{a} {Aperture magnitude at 5$\sigma$ within a fixed radius of 0.17" \citep{guo13}.\\}
        \tablefoottext{b} {Aperture magnitude at 5$\sigma$ within a fixed radius of 0.1" \citep{finkelstein23}.}
         }
        \end{threeparttable}
    \end{table}

\subsubsection{Source detection and flux measurements}

We used \textsc{SExtractor} \citep{bertin96} v2.8.6 to detect sources in both real and forward-modeled images, adopting the Hot+Cold technique used for the CANDELS catalogs \citep{galametz13}. We considered \texttt{FLUX\_AUTO} for flux estimates unless explicitly specified. 

To ensure consistent comparisons and statistical robustness, we always compared source counts in regions of equal area. We first selected several regions of 4000x4000 pixels from the real images (within the deep zone for GN and GS); this size was chosen because larger areas are difficult to single out without crossing relevant changes in depth in GS. We checked that the source density within these different regions of the real images remains consistent across these areas. 
Then, we selected regions of the same size in the corresponding mock images, and we performed the same analysis, finding similar results. This ensures that potential discrepances between the real and simulated datasets are not influenced by the specific zones chosen for the analysis. The counts are always normalized to the total area.

\section{Results}\label{results}

\subsection{The faint-end problem}

\subsubsection{$H$ counts in CANDELS fields}\label{fields}

The deficit of faint galaxies reported in F23, where the $H$-band (F160W) number counts from forward-modeled mock images fell short of those measured in the CANDELS GOODS-South field, raises an urgent question: is it a peculiarity of that field or does it characterize all five CANDELS fields?
To address this, we used the same five TNG100 mock light cones (\texttt{cmd-TNG100}) and forward modeled them to the depth of each field, creating mock images with the corresponding noise level and $5\sigma$ limiting magnitude.

\begin{figure*}[!htbp]
    \centering
        \includegraphics[width=1.\textwidth]{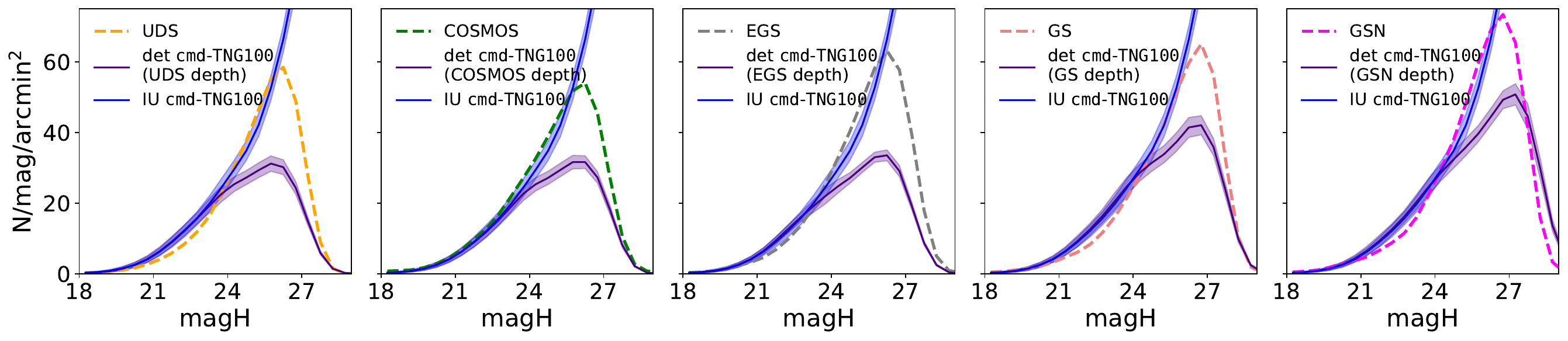}
        \caption{Number counts in the $H$ band in the CANDELS fields (from left to right: COSMOS, EGS, UDS, GS, and GSN). The mock detections (purple) and the IU counts (blue) were derived from the five \texttt{cmd-TNG100} realizations. The shaded areas indicate the 1$\sigma$ scatter across the realizations. The mock images were simulated at the corresponding survey depth (i.e., at the 5$\sigma$ limiting magnitude of the real datasets). The dashed colored lines indicate the observed counts in each field.}
        \label{all_fields}
\end{figure*}

In Fig. \ref{all_fields} we show the F160W number counts ($H$, total magnitude). The real counts, shown as single dashed lines for each field, remain generally similar in all fields. This was indeed a key feature of the survey design, to ensure homogeneity across the dataset for detection \citep{koekemoer11,merlin19}. It is immediately clear that the counts on the mock images (purple line with 1$\sigma$ shaded area) consistently deviate from real counts in the faint magnitude range; they fall below them at $H>25$ in the shallower fields (UDS, COSMOS, and EGS), and at $H>26$ in GS and GSN. The confirmation that the trend first identified in F23 for GOODS-South extends to all the fields allows us to rule out survey inhomogeneities or cosmic variance, and shows that the discrepancy is a robust feature; for this reason, in what follows, we focus the analysis on GOODS-South as a representative dataset.

\indent At this stage, the origin of the faint-end discrepancy requires further investigation. We first verified that, in the magnitude range where the discrepancy is observed, the galaxies contributing to the IU counts have stellar masses well above the resolution limit of the simulation, indicating that this regime is effectively complete in stellar mass. To rule out artifacts from the generation of the mock datasets, the pipeline used to build the light cone has been validated by verifying that the total stellar mass budget and its distribution reproduce the content of the original hydrodynamical simulation, and the final mock images have been cross-checked against an independently generated dataset \citep{snyder23}. We also checked that the shortfall is not due to biases related to the assembly of the galaxy SED by comparing counts in a bluer and a redder band: the same declining trend is observed, confirming that the discrepancy is not caused by flux redistribution across the SED. All these checks are shown in Appendix \ref{validation_tests}.

\subsubsection{$H$ counts across redshift bins}\label{zbins}
We further extend the analysis by dividing the datasets into redshift bins, utilizing both the IU and the mock images generated from two independent hydrodynamical simulations, TNG100 and EAGLE. While the light cones extend up to $z \sim 7$, we restrict the analysis to $z \leq 5$, where the statistics from CANDELS GS remain robust. We consider three redshift intervals: $z$ = 0.0 -- 2.9, $z$ = 3.0 -- 3.6, and $z$ = 3.7 -- 5.0.

\begin{figure}[!htbp]
    \centering
        \includegraphics[width=0.53\textwidth]{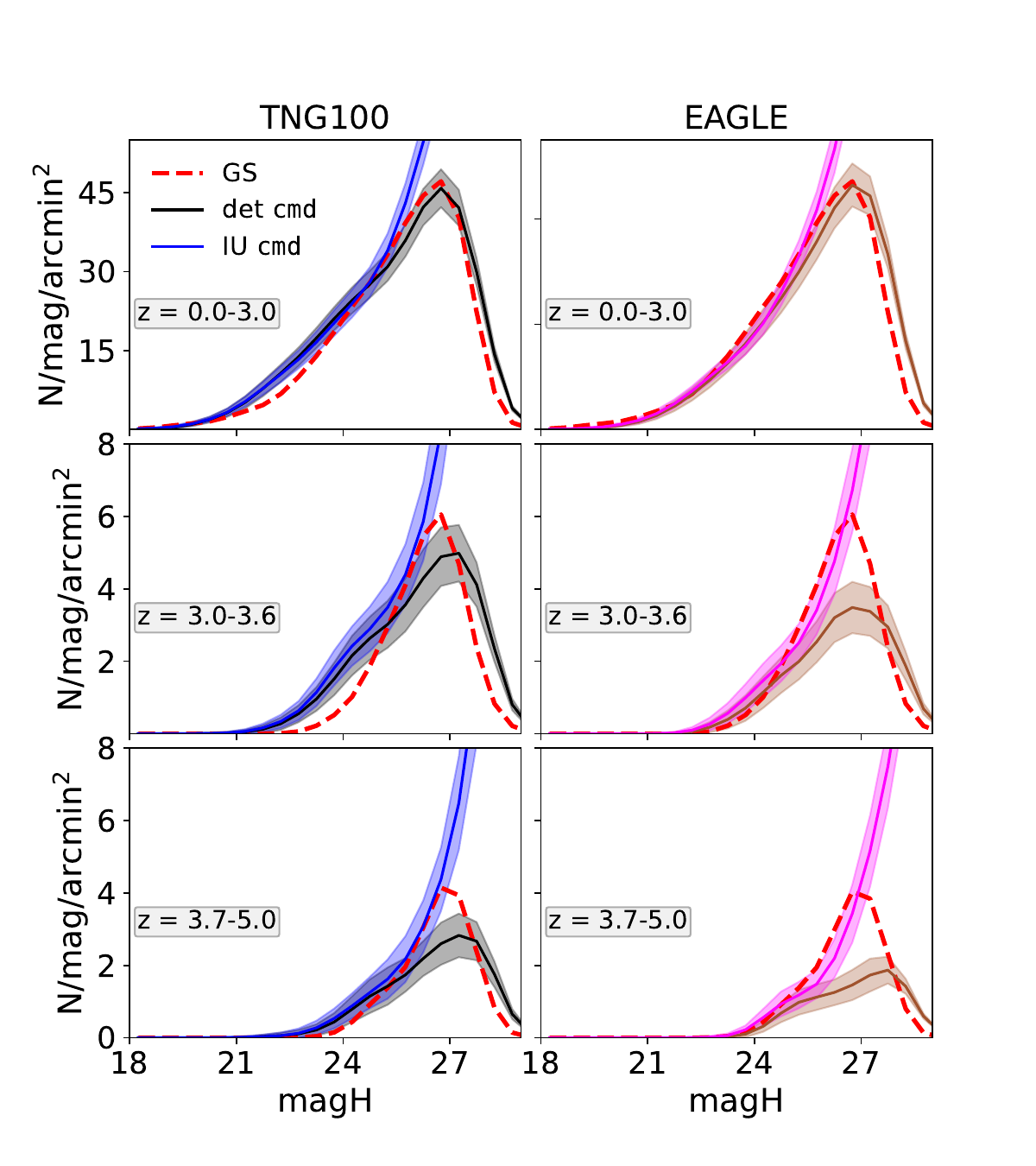}
        \caption{$H$-band number counts divided into redshift bins (from top to bottom: $z$=0.0--2.9, $z$=3.0--3.6, and $z$=3.7--5.0). The left panels show results from TNG100, right panels to EAGLE. Mock datasets, obtained from the five \texttt{cmd-} realizations, are represented by solid line with 1$\sigma$ shaded area. The IU (blue for TNG100 and magenta for EAGLE) and the sources detected on the mock image (black for TNG100 and brown for EAGLE) are compared to the CANDELS GS counts (dashed red). In GS, sources are selected with the best redshift estimate from \cite{merlin21}.}
        \label{dz}
\end{figure}

The results are shown in Fig. \ref{dz}, where the left column refers to TNG100, while the right on to EAGLE. In each panel, we compare the CANDELS GS counts (dashed red) with the IU counts (blue for TNG100 and magenta for EAGLE) and with the corresponding detections on the mock images (black for TNG100 and beige for EAGLE, with the usual 1$\sigma$ error). 
For the CANDELS dataset, we adopted the best redshift estimates from the ASTRODEEP-GS43 catalog, while for the FORECAST datasets we employed the IU redshifts for both the IU and the detections.

In the lowest bin ($z$ = 0.0 -- 2.9), TNG100 and EAGLE mock detections remain in excellent agreement with the CANDELS counts in the full magnitude range. 
At $z>3$, the behavior of the two simulations starts to diverge, and the source deficiency arises in both mock datasets, with the last bin ($z$ = 3.7 -- 5.0) being the most extreme case. 
In TNG100, the mock counts reveal an overproduction of bright sources up to $H=25$, followed by a rapid drop below CANDELS at fainter magnitudes. In EAGLE, instead, the IU counts fall below CANDELS already at $H\simeq24.8$: this deficiency must originate upstream of the image generation or source extraction, since at this regime the mock counts are complete. It therefore reflects either differences in the underlying galaxy population, as set by the underlying CHS, or in the mapping between physical properties and observed luminosities, which is determined by the adopted post-processing (e.g., dust attenuation), or a combination of the two. 
 
When comparing the two simulations, however, the differences in their behavior likely arise from differences in the galaxy populations produced by the simulations, as reflected by the normalization of the mass density and mass function in Sect. \ref{sect:md}). In TNG100, the excess of bright galaxies suggests that massive systems form too efficiently, or put in other words, that feedback at high redshift is not strong enough to suppress their luminosities at these wavelengths. Conversely, in EAGLE the deficit of sources is endemic in the full range, pointing instead to an overall underproduction of galaxies in this redshift regime.

The general trend is that both simulations underestimate the number of sources compared to CANDELS at $z>$ 3, and the shortfall appears already at intermediate magnitudes, not just at the very faint end. On top of that, EAGLE counts are systematically lower than TNG at all redshifts, so its difference with CANDELS is even stronger. The fact that the shortfall persists even where the completeness of the mocks is not an issue strengthens the conclusion that both original simulations struggle to reproduce the observed population at high redshift.

We remark that at low redshifts ($z$ = 0.5 -- 1.0), the statistics are less significant due to the limited field of view sampled by FORECAST for the specific emulation of the CANDELS fields, yet the counts are in good agreement (we point out that FORECAST allows for larger field of view configurations, which would improve the statistical significance of the results at these redshifts).

\subsection{The origin of the shortfall: Completeness and depth}

\subsubsection{Missing sources in IU}\label{sect:missing}
In F23, we proposed that the mismatch in the $H$ counts between GS and the mock GS-like image generated with FORECAST may arise either from limitations in the simulations used to generate the mock images (with the TNG100 IU being less populated than expected) or from a problem in the detection process (e.g., a significant fraction of simulated galaxies that are either blended with larger objects or too faint to be detected due to their fragmented morphologies). 
However, just considering the IU counts, one can immediately rule out the hypothesis of an issue with the detection procedure or with the depth of the images. In fact, the IU represents an upper limit on the detection counts, as it includes all sources in the light cone, regardless of their detectability: therefore, a direct comparison between the IU and the GS-corrected counts allows to assess whether the simulated population itself is sufficient to reproduce the observations.

We compute the completeness correction for the $H$ counts in both GS and one of the five mock \texttt{cmd-TNG100}-based images. We extracted stamps of realistic light profiles from the mock image, selecting galaxies with elliptical and disc-like morphologies and fixed half-light radii of $\log_{10}(R_\mathrm{hl}/\mathrm{pix})=$ 0.3, 0.6, and 0.9 to match values adopted in \citet[differently, they injected analytical Sérsic profiles]{guo13}. For each size and morphological class, we scaled the profiles to a grid of magnitudes and inject $\sim 300$ simulated sources into empty regions of the GS images (in the deep area only). We then ran \texttt{SExtractor} setting Hot parameters to optimize the detection of faint sources (see Table \ref{SEx} for the full set of parameters). 

\begin{figure}
        \centering
        \includegraphics[width=0.40\textwidth]{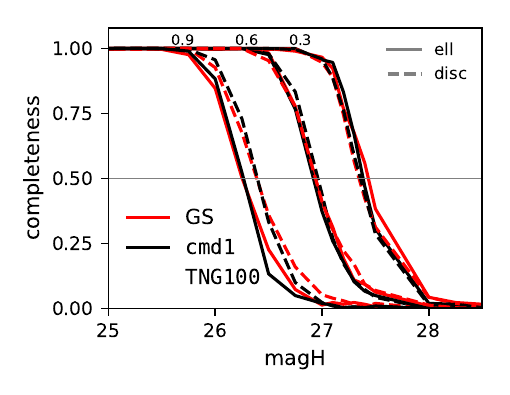}
        \caption{Completeness curves for simulated sources injected into the CANDELS GS $H$ image (red) and one of the five \texttt{cmd-TNG100} mock images (black), computed with realistic light profiles with fixed half-light radii $\log_{10}(R_\mathrm{hl}/\mathrm{pix}) = 0.3$, 0.6, and 0.9. Sources were injected in empty regions of the GS deep area and detected with \texttt{SExtractor} using optimized (hot) parameters. For $\log_{10}(R_\mathrm{hl}/\mathrm{pix})=0.6$, the 50\% completeness magnitude is consistent with that reported by \citet{guo13}.}
        \label{curve_comp}
\end{figure}

The resulting completeness curves in Fig. \ref{curve_comp} show that, as expected, sources with a smaller half-light radius are easily recovered, reaching 50\% completeness at fainter magnitudes than less compact ones. As a side note, the similarity of the completeness curves between GS and the mock image in each bin of size suggests that the noise in the mock image closely resembles the real one in terms of its impact on source detectability. 

\begin{figure}
        \centering
        \includegraphics[width=0.48\textwidth]{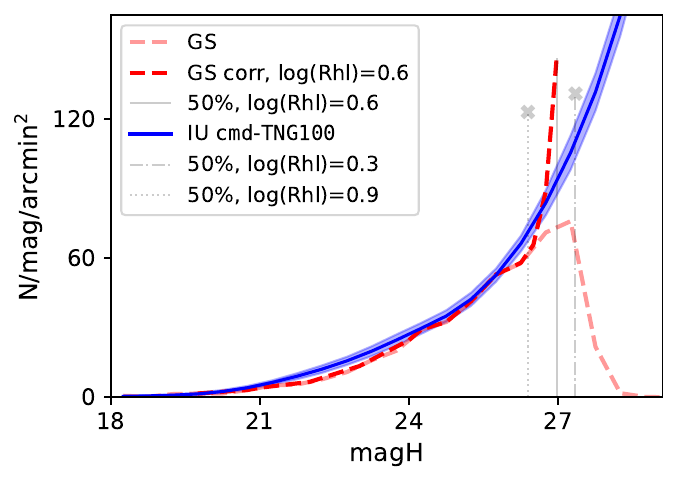}
        \caption{Comparison between observed $H$-band number counts in the CANDELS GS field (dashed bright red) and IU counts from five \texttt{cmd-TNG100} realizations (blue line with 1$\sigma$ shaded area). The dashed dark red line shows the completeness-corrected GS counts, obtained using the completeness curve at $\log_{10}(R_\mathrm{hl}/\mathrm{pix}) = 0.6$. Vertical lines indicate the 50\% completeness magnitudes derived for different source sizes ($\log_{10}(R_\mathrm{hl}/\mathrm{pix}) = 0.3$, 0.6, and 0.9).}
        \label{corrected_compl}
\end{figure}

In Fig. \ref{corrected_compl} we corrected the GS counts using the intermediate correction curve ($\log_{10}(R_\mathrm{hl}/\mathrm{pix}) = 0.6$) as the fiducial case, since the smaller size curve yields a more optimistic completeness correction, while the large size one is too conservative. Moreover, the 0.6 case is directly comparable to \citet{guo13}, with the 50\% completeness magnitude in excellent agreement with their estimate.
Using this correction, we found that the corrected GS counts significantly exceed the IU counts already at $H \gtrsim 27$. In the magnitude range where GS is complete, instead, the IU counts are consistent with the observed counts. We recall that the light-cone construction has been validated against the original simulation (Appendix \ref{validation_tests}); this ensures that the IU itself faithfully represents the simulated galaxy population and that the discrepancy is not an artifact of the light-cone generation. The result suggests that the faint-end shortfall cannot be explained by detection or post-processing effects alone. Nevertheless, completeness corrections are inevitably model-dependent, since they rely on the assumed size and light profile of the injected sources. This implies that while the discrepancy is unlikely to be an observational artifact, its quantitative extent may still depend on how observational limitations interact with the intrinsic properties of the simulated galaxies.

\subsubsection{Effect of the noise}\label{sec:deeper}

Noise may play a significant role in suppressing the mock faint population. 
To test this, we asked how much deeper the mock images would need to be in order to recover counts consistent with the observations. We found that matching the GS counts up to $H \sim 27$ would require a reduction of the noise by a factor of 10, that is, 2.5 magnitudes deeper than CANDELS at nominal depth. We refer to this configuration as the deeper dataset.

\begin{figure}[!htbp]
    \centering
        \includegraphics[width=0.48\textwidth]{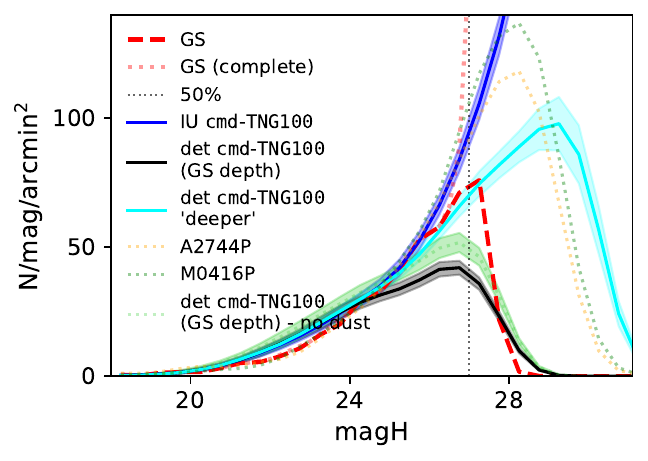}
        \caption{Comparison of $H$-band number counts between the CANDELS GS field (dashed red) and the FORECAST mock dataset based on five \texttt{cmd-TNG100} realizations. Mock detections at GS depth are shown in black with 1$\sigma$ shaded area, while the deeper mock (cyan with shaded area) is obtained by reducing the noise by a factor of 10. IU counts are shown as the solid blue line. The dashed red line indicates GS counts corrected for completeness. Despite the improved depth, a mismatch between the deeper mock and GS persists at ultra-faint magnitudes. It also fails to reproduce the observed number counts of the deeper Frontier Fields datasets A2744 Parallel (depth 29.06; yellow) and M0416 Parallel (depth 29.14; green; \citealt{merlin16_ff, castellano16}) within their respective completeness ranges. The vertical dotted line marks the 50\% completeness magnitude of GS ($\log_{10}(R_\mathrm{hl}/\mathrm{pix}) = 0.6$). All datasets are selected in the redshift range $z \simeq 0-7$.}
        \label{synthTU}
\end{figure}

Figure \ref{synthTU} shows the outcome of this experiment. Reducing the noise by a factor of 10 (cyan line) produces artificially deeper mock images that reproduce GS counts up to $H \sim 27$, corresponding to the peak of the observed distribution. However, for magnitudes fainter than this peak, a new discrepancy emerges: while the real counts begin to decline, the deeper mock counts continue to rise, producing an excess of ultra-faint detections not seen in the real dataset. 

The comparison with the Frontier Fields (FF) parallel fields A2744 and M0416 \citep{merlin16_ff}, which are deeper than GS ($\sim$29.1) but shallower than our deeper dataset, further highlights the mismatch: even with the deeper mock image, at $H>$ 26 we detect fewer sources per unit area than in the FF images. Although a different turnover is expected due to the difference in the limiting magnitudes, the mock deeper counts remain systematically lower in the range where both datasets are complete.\\
\indent This result indicates that incompleteness due to noise accounts for only part of the discrepancy. When images are simulated at higher depth, the shortfall at faint magnitudes is alleviated, at the cost of generating a surplus of ultra-faint detections that have no observational counterpart at nominal depth. In other words: either the mock images should be deeper, so that the counts at $H\simeq27$ match the observed ones - but this way a large excess of fainter sources would be detected; or, the depth of the mock images is correct, and we are back to the original problem. Ultimately, the fact that the discrepancy persists even within the completeness regime of the FF suggests that the faint-end deficit is linked to limitations in reproducing the observed faint population, rather than to noise or detection effects alone.\\
\indent Some of the missing sources are indeed present in the IU, but they are not detected because they stochastically fall below the noise level (at these magnitudes the completeness is $\approx 50\%$). However, this test shows that deeper mock images would increase the counts of very faint sources well above the observations, while at the same time failing to reproduce the counts from real deep fields.

\section{Discussion}\label{discussion}

\subsection{The nature of the missing sources}

From our analysis in Sect. \ref{sect:missing}, it appears that the faint galaxies deficit in the mock images compared to the CANDELS GS (and we checked this across all fields) is robust. At $z>$3, it occurs consistently across different filters, from F105W to F277W, showing that the discrepancy is not due to issues in the SED modeling. 
After applying completeness corrections to GS, the corrected counts exceed those of IU already at the magnitude where GS is only 50\% complete ($H\simeq$ 27, see Fig. \ref{corrected_compl}). Finally, the deeper mock experiment shows that when artificially reducing the noise to a level far below that of the real data, the counts follow GS closely, but a large excess of ultra-faint mock sources is detected. All of this suggests that the discrepancy cannot be explained by detection effects alone, and must arise from a combination of intrinsic properties of the simulated population and the way they are mapped into observable quantities. 

\subsubsection{Structural differences between CANDELS and mocks}
Since we do not have a ground-truth GS-corrected catalog to compare with the IU to identify which sources are missing, we rely on the available data. We compare structural properties (i.e., morphological and photometric tracers) of detected objects in GS and the mock images, as measured by \texttt{SExtractor}, to have hints on what kind of sources are missing in the IU. Both groups of sources are subjected to the same PSF, noise, and detection parameters. 
We focus on faint galaxies ($26<H<28$) at $z>3$, using the best redshift estimates for GS, and the IU redshift for mock detections. This selection is chosen to probe the regime where the discrepancy between mock and observed counts emerges and approaches the observational limit (see Table \ref{depths}).

\begin{figure*}[!htbp]
    \centering
        \includegraphics[width=\textwidth]{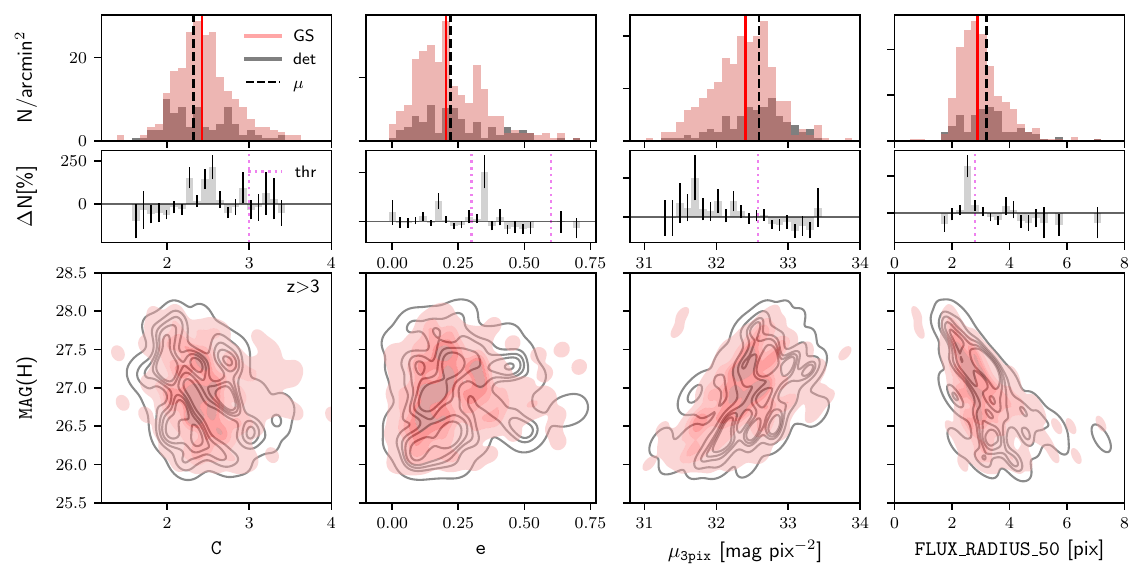}
        \caption{Characterization of the sources in GOODS-South (red) and in a mock image created from one of the \texttt{cmd-TNG100} realizations (black) at $z>3$. 
        From left to right: Concentration ($C = 5 \log_{10}(r_{80}/r_{20})$, where $r_{80}$ and $r_{20}$ are the radii containing 80\% and 20\% of the $H$ flux, respectively); ellipticity ($e = 1-b/a$); surface brightness within circular apertures of radius $R$=3 pixels ($\mu_{R} = {\rm mag}H_{R} - 2.5 \log_{10} (\pi R^2)$); and radius enclosing 50\% of the total $H$ flux (\texttt{FLUX\_RADIUS\_50}).
        From top to bottom: (i) One-dimensional distributions of the properties; (ii) fractional residuals $\Delta \text{N}_i=(\text{N}_{GS,i}-x*\text{N}_{mock,i})/(x*\text{N}_{mock,i})$ in bins $i$, with $x=\text{N}_{GS}^{tot}/\text{N}_{mock}^{tot}$; vertical magenta dotted lines indicate the thresholds defining $\Delta f$ in each parameter space. (iii) Contour plots of property against $H$-band magnitude. The comparison shows that the two datasets populate similar regions of parameter space, but with different relative abundances across them.}
        \label{fig:MS_comb}
\end{figure*}

Figure \ref{fig:MS_comb} shows how the two datasets distribute across these quantities: concentration $C$, ellipticity $e$, central surface brightness (SB) $\mu_{3pix}$, and half-light radius in $H$ $\texttt{FLUX\_RADIUS\_50}$ (or $r_{50}$). The panels in the top rows, which show the normalized distributions of these properties, seem to suggest that the deficit is broadly transversal to every property: the mock simply contains fewer faint sources than GS. The bottom panel appears to reinforce this hypothesis, showing that GS and the mock occupy essentially the same region of the parameter space as a function of the $H$ band magnitude. Read at these scales, the discrepancy can be interpreted primarily as a simple abundance offset. 

To understand whether the shortfall is purely global or also reflects differences in the relative mix of galaxy types, we rescale the mock counts to match the total number of GS sources. The scaling factor $x=\text{N}_{GS}/\text{N}_{mock}= 3.12$ is computed across the entire sample and applied uniformly. Thus, we compute the fractional residual $\Delta \text{N} [\%]=(\text{N}_{GS} - x \cdot \text{N}_{mock})/(x \cdot \text{N}_{mock})$ within each bin of a given structural property.  
If the shortfall were purely global, these residuals would scatter around zero; systematic deviations, instead, would indicate that certain galaxy types are relatively over- or under-represented in the mock image (at nominal depth). The central row panels of Fig. \ref{fig:MS_comb} show that the residuals remain positive where the light profiles are more centrally peaked with brighter $\mu_{3\rm pix}$, smaller $\texttt{FLUX\_RADIUS\_50}$ (at the PSF scale), and become negative in the diffuse/low-SB regime. The trend indicates that the faint-end shortfall is not solely a problem of fewer sources.

We make this pattern quantitative by comparing, for each property, the difference in the fraction of GS and mock galaxies that occupy a specific region $R$ of the parameter space using $\Delta f = f_{\text{GS}|R} - f_{\text{mock}|R}$ (thresholds indicated by magenta dotted lines in Fig. \ref{fig:MS_comb}). Uncertainties are simply 1$\sigma$ binomial errors, obtained by adding in quadrature the variances of the two independent fractions. 
For the thresholds that define the regions, we adopt physically motivated values: ultra-compact if $r_{50}/FWHM<1$, because at these redshifts such systems are effectively PSF-limited (stars are excluded); compact when $1<r_{50}/FWHM<1.5$, and extended when $r_{50}/FWHM>2$. 
For the concentration, we base on the classification of \cite{conselice03CAS}, with the caveat that it holds in the local Universe and has been shown to remain valid up to $z\sim3$, but not tested upward. In their work, bulge-dominated systems typically show $C\gtrsim4$, while late-type disks and irregulars have $C\lesssim3$; thus, we adopt $C>3$ as a practical indication to flag highly concentrated profiles. For ellipticity, we distinguish round ($e<0.3$) and elongated ($e>0.6$) shapes. For the central surface brightness, we adopt the median $\tilde{\mu}_{3pix}=32.59$ of compact sources ($r_{50}/FWHM<1$) as a threshold to identify objects with bright central regions. All thresholds are fixed on the mock and then applied to GS, ensuring a conservative comparison. 

With these definitions, the central surface brightness provides the most significant indication. Galaxies with bright nuclei have $\Delta f_{\mu_{3pix}} = 0.18 \pm 0.05$, which is about a $3\sigma$ excess in GS, highlighting that the difference between the datasets is pronounced between high-SB nuclei. The size tells the same story. Ultra-compact objects, with $r_{50}/FWHM<$1, have $\Delta f_{r_{50}} = 0.17 \pm 0.05$ (again, a $\sim3\sigma$ excess) in GS, meaning that the mock lacks roughly one fifth of the PSF-dominated sources. When the size cut is relaxed to $1-1.5\times FWHM$ the excess becomes much smaller, and extended systems ($r_{50}/FWHM>2$) show no significant difference in fraction. Finally, concentration and ellipticity show negligible differences between the two datasets: fractions of highly concentrated, round, or elongated profiles are nearly identical in the two datasets, so the (visual, not real) shape seems not to be a driver. Overall, the faint-end shortfall is global, but not neutral in its composition: the mock image is selectively deficient in very compact and bright cores.

\subsubsection{Persistence of the shortfall of bright compact galaxies beyond detection limits}
It is important to note that this analysis probes the mock images, and not the IU directly: it can only give hints on what the IU itself fails to reproduce. To dig further in this direction, we exploit the deeper catalog (see Sect. \ref{sec:deeper}), matched to the IU to have the redshift of these sources. We perform the same $\Delta f$ test, but limited to $r_{50}$ and $\mu_{3pix}$, since in the tested regions they are stable against the change in depth of the deeper dataset. 

For $r_{50}$, $\Delta f_{r_{50}} = 0.437 \pm 0.029$ ($\sim15\sigma$), meaning that when the observational limitations are relaxed (even though within a safe zone), the discrepancy does not shrink, rather it becomes far more emphasized. Similar results are for $\mu_{3pix}$, where $\Delta f_{\mu_{3pix}} = 0.614 \pm 0.027$. For safety, we checked the other parameters and regions: the extended and low-SB regimes show negative $\Delta f$, as expected when a deeper image recovers faint diffuse wings; ellipticity shows no significant differences; the behavior of the concentration changes since it is very sensitive to depth (as said, fainter outer wings are detected in this configuration); therefore, it is not reliable for our scopes. Finally, combining the two thresholds ($\mu_{3pix}<\tilde{\mu}_{3pix} \,\,\,\&\,\,\ r_{50}/FWHM<1 $) gives an even clearer picture: in GS 160 out of 340 sources fall in this region, whereas the deeper mock contain only 8 out of 362 ($\Delta f = 0.448 \pm 0.028$). Note that this $\Delta f$ is telling us that among the galaxies each catalog actually contains, the fraction of compact and bright nuclei systems in GS is higher; it does not mean that 45\% of all faint galaxies are missing in the mock, nor does it quantify the overall deficit in the faint counts.

An important consideration follows. The mock image does not show a generic shortage of galaxies, but specifically lacks a population of $z>3$ sources with very compact and bright nuclei. This deficit persists in the deeper mock image, where detection limits are effectively relaxed, indicating that the deficit is not a uniform offset in number counts, but reflects a selective difference in the structural properties of the faint population.

\subsection{Lost in detection}
 
The deeper experiment in Sect. \ref{sec:deeper} shows that additional galaxies emerge once the noise is suppressed. This establishes that, in addition to the intrinsic deficit discussed above, part of the discrepancy between GS and the mock images is due to sources that exist in the IU but fail to be detected at nominal depth.
To isolate this contribution, we exploited the deeper dataset, identifying galaxies that are recovered once the noise is suppressed but disappear at nominal depth. We refer to these as lost in detection, or simply lost, while those recovered in both images are labeled kept.

\begin{figure*}[htbp]
    \centering

    \begin{subfigure}{0.9\textwidth}
        \centering
        \includegraphics[width=\linewidth]{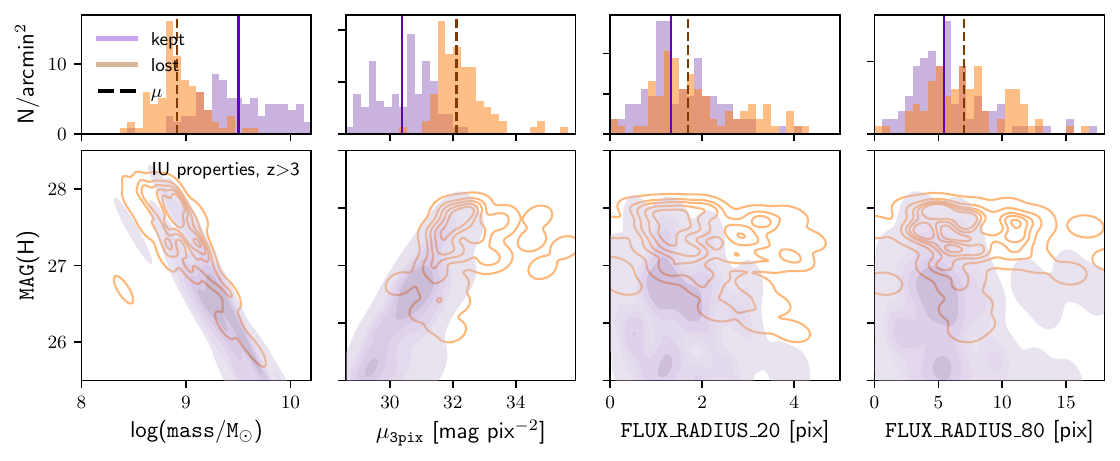}
        \label{suba}
    \end{subfigure}

    \vspace{0.5cm}

    \begin{subfigure}{0.9\textwidth}
        \centering
        \includegraphics[width=\linewidth]{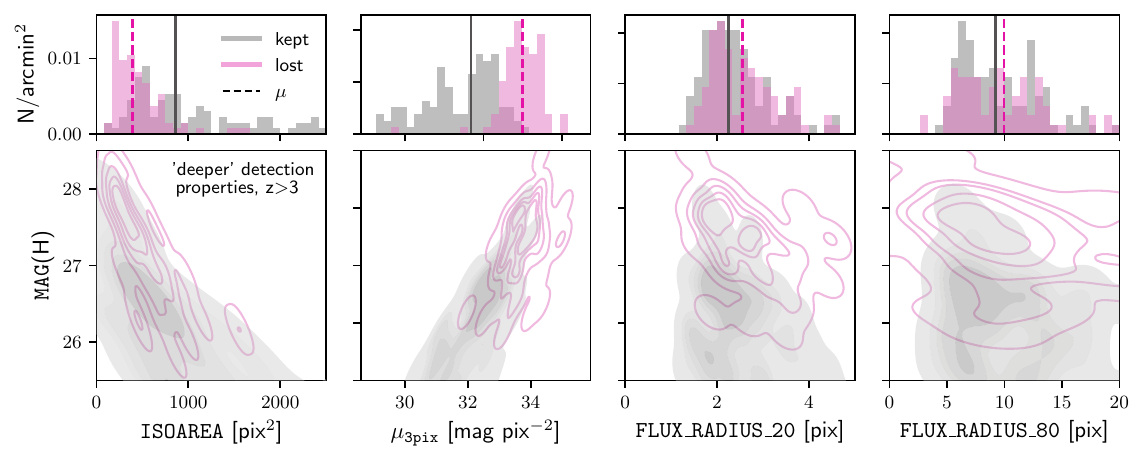}
        \label{subb}
    \end{subfigure}

    \caption{Properties of galaxies at $z>3$ detected in mock images with noise reduced by a factor of 10 relative to the nominal depth (deeper image; kept sources, violet and gray) and missed at nominal depth (lost, orange and pink). The upper panels show the intrinsic Input Universe properties of these galaxies: Stellar mass, surface brightness within $R=3$ pixels ($\mu_{3pix}$), \texttt{FLUX\_RADIUS\_20} and \texttt{FLUX\_RADIUS\_80}, i.e., the radii enclosing 20\% and 80\% of the $H$ flux. The bottom panels show the quantities measured by \texttt{SExtractor} on the deeper mock images: Isophotal area, $\mu_{3pix}$, \texttt{FLUX\_RADIUS\_20}, and \texttt{FLUX\_RADIUS\_80}. The comparison highlights the physical nature of the sources that are lost when moving from deeper to nominal depth, and how their intrinsic properties map onto the measured ones.}
    \label{fig:lost_kept}
\end{figure*}

Since the shortfall in number counts emerges only beyond $z>3$, we focused the comparison at this redshift regime (using $z_{IU}$) and restricted the analysis to the magnitude range where the discrepancy is observed, following the same selection adopted above. Both the nominal and deeper IU catalogs extend to fainter magnitudes, but these sources lie beyond the detection limit and are not considered here. Figure \ref{fig:lost_kept} shows the characterization of lost and kept galaxies with intrinsic properties (top panel) and the detection properties measured by \texttt{SExtractor} on the deeper image (bottom panel); in each figure, the top panels show the distributions of the considered properties and the bottom panels display their relation with $H$. By construction, the lost sources are present in the Input Universe; the question is what distinguishes them from the kept sample and makes them fail at nominal depth. 

Looking at IU properties, the bottom row shows that, at fixed $H$, the lost sources are less massive (and consequently, with lower metallicity). Their structural parameters depict them as fainter in their cores and, on average, more extended, having larger fractional light radii in the $H$ band (\texttt{FLUX\_RADIUS\_{\%}} or $r_{\%}$) compared to kept galaxies. 

When using the estimates from the deeper catalogs, where the S/N is high, the same properties translate into the empirical signature of the loss: lost sources show smaller isophotal areas\footnote{``An isophotal area is defined as the number of pixels with values exceeding some threshold above the background'', from the \texttt{SExtractor} \textit{User Manual}.}, fainter central surface brightness, and more extended centers and outskirts compared to kept. This is precisely the combination of parameters that are expected to fail in detection at nominal depth, especially after the PSF smears out light over more pixels.

This analysis confirms that part of the deficit is technical: the detection step, as is obvious, preferentially removes galaxies with diffuse light profiles and low surface brightness, which are indeed present in the IU and reappear in the deeper catalogs. Yet this technical issue, together with the analysis of missing sources, indicates that the discrepancy is not limited to this effect. The IU does not simply lack faint galaxies; rather, it populates the faint end with systems that are too diffuse on average: cores that are less centrally peaked (large \texttt{FLUX\_RADIUS\_20}), outskirts that are overextended (large \texttt{FLUX\_RADIUS\_80}), and correspondingly faint central surface brightness. What is underrepresented, instead, are the galaxies that GS does show at $z>3$: faint but more compact systems, with brighter bulges and outskirts that are not excessively spread out. In other words, what is missing in the mock is not simply more flux, but a specific structural class.
In summary, the IU in the faint regime is dominated by diffuse, low-SB systems that can be lost at nominal depth, but fails to reproduce the population of compact bright-core galaxies that GS contains at $z>3$, which would instead be the easiest to detect.

Several factors may contribute to the observed discrepancy. Numerical resolution imposes a first limitation: at $z>3$, the physical softening length in TNG100 and EAGLE, as well as in simulations with similar resolutions, is of the order of a few hundred parsecs, comparable to the inner (physical) scales of the galaxies probed at these redshifts. This inevitably smooths their central mass profiles and can prevent the survival of compact cores, as previously discussed by \cite{wellons15, genel18, pillepich18}. In particular, \cite{pillepich18_mf} report that in TNG100 the stellar mass profiles do not achieve numerical convergence at radii smaller than about 2 kpc. Our high-redshift galaxies typically have low stellar masses and effective radii ($r_{50, M_*}$) close to or below this scale. Therefore, they lie exactly in the regime where the inner stellar mass distribution is not numerically robust. The PSF already smooths our forward-modeled light profiles, so at these redshifts, the measured fractional radii are primarily dominated by this image smoothing. The numerical smoothing, instead, acts upstream by smoothing the stellar mass and limiting the formation of dense cores before the images are created. The few ultra-compact systems that we do find do not contradict this picture, because gravitational softening does not impose a strict lower limit on the size of individual objects but instead prevents the buildup of a statistically significant population of truly compact bulges. The observed deficit of compact galaxies at $z>3$ is therefore, at least partly, a consequence of the finite spatial resolution of the hydrodynamical model itself. 

As said, the IU galaxies on which we measured the structural parameters lie predominantly in the low-mass regime ($\tilde{log_{10}(M*/M_{\odot})} \approx 8.7$); according to the GS mass completeness reported by \cite{grazian15}, such stellar masses are above the threshold of completeness in the deep region at $z\gtrsim3$. This means that if galaxies with these masses had the structural properties of GOODS-South, they would be detected.

In the low-mass high-redshift regime considered here, stellar feedback is a natural candidate to shape the simulated galaxy structure. Large volume cosmological simulations include explicit models of stellar feedback \citep{schaye15, pillepich18}, calibrated to match galaxy statistics at low redshift; their efficiency at higher redshifts is therefore a genuine prediction of the models rather than an observationally constrained quantity. This predictive nature is a strength of hydrodynamical models, and it makes any discrepancy with the data physically meaningful. However, there are still a few systematic studies that determine whether these prescriptions predict the observationally consistent effect of stellar feedback in such early times ($z\gtrsim$3), and forward modeling could provide a direct way to test this connection.
In the FIRE high-resolution zoom simulation, \cite{elbadry16} showed that bursty stellar feedback in low-mass progenitors at early epochs can induce large fluctuations in the central potential, inducing strong radial migration of stars and causing effective radii to vary by factors of order two on timescales of a few hundred million years. Recent observations based on JWST spectroscopy confirm that such bursty star formation histories are common at these mass regimes at $z>$3 \citep{looser25}. \cite{mitchell20} found that in EAGLE the characteristic velocity of the outflows increases toward high redshifts. Their comparison with other simulations reveals that in TNG most of the gas driven out by stellar feedback remains gravitationally bound to the halo at large radii and later falls back: this recurrent outflow and inflow cycle of the gas naturally creates the kind of fluctuations in the central gravitational potential described by \cite{elbadry16}, causing the stellar component to migrate outward and the time-averaged central stellar density to reduce. If the coupling between stellar feedback and the gravitational potential is too efficient or too persistent, the net effect over several burst cycles is to dilute the central stellar densities and inflate the light profiles, producing galaxies whose stellar light is more extended and whose central surface brightness is fainter than that observed in GS.\\
\indent Dust attenuation could also play a role. In our mock light cones, dust is treated with an on-the-fly model based on the gas distribution, rather than with full radiative transfer, which is physically more robust. \cite{byun25} show that different dust implementations can substantially alter the morphology of galaxies at $1<z<3$, producing brighter bulges or, conversely, more diffuse ones depending on whether dust is linked to dust-to-metal ratios, dust-to-gas, or radiative transfer models. Although we cannot exclude that our dust treatment affects the results, the fiducial dust model adopted in this work shows reasonable agreement with the observations in the magnitude range where the GOODS-South offers a robust constraint. Further hints on the impact of dust come from the dust-free case: recomputing the IU number counts without attenuation produces a systematic increase of the counts across magnitudes, worsening the agreement with the observations where the data are complete. This shows that dust has a non-negligible impact on the counts; at the same time, the discrepancy identified in this work cannot be reduced to a simple lack of flux, but is instead connected to the structural properties of the galaxies. Overall, the role of dust is nontrivial, but its contribution cannot be completely ruled out without testing alternative dust implementations, a task that we leave for future work.\\
\indent In summary, the faint-end shortfall assessed in the mock images at  $z>3$ has a complex origin. On the observational side, PSF and noise, combined with the threshold of the detection, remove a subset of galaxies that the simulation massively produces: low surface brightness systems with diffuse light profiles with masses near the resolution limit. On top of that, the simulations show a deficit of faint galaxies with moderate concentrations and central surface brightness that in GOODS-South are detected at $z>$3. 
Bridging this gap will require advances in computational resolution, in the modeling of feedback, and in the treatment of dust within the forward-modeling pipeline. A natural next step will be to extend this analysis to a broader set of JWST observations, and to compare with the latest high-redshift simulations that incorporate advanced modeling of radiation, feedback, and dust (e.g., \citealt{kannan22,schaye23,pakmor23}).

\section{Summary and conclusions} \label{conclusions} 

We investigated the origin of the discrepancy in the number counts at the faint end of the observed near-infrared ($H$) distribution between observations (using CANDELS data, focusing on the GOODS-South field) and forward-modeled mock images generated with FORECAST, and we extended the analysis of \citet[F23]{fortuni23}. We analyzed the entire forward-modeling process from the assembly of the input light-cone catalog (i.e., the IU), which contains all sources in the simulated field of view to the final detection catalogs, obtained by running \texttt{SExtractor} on the mock images. To reduce cosmic variance and give robustness to our results, we built ten independent realizations of the light cone and mock images from two advanced CHSs, TNG100 and EAGLE. The following are our main conclusions.

\begin{itemize}
\item The forward-modeling process is robust. 
The IU catalogs accurately reproduce the intrinsic galaxy population of the original cosmological simulations in terms of stellar mass density (Fig. \ref{mdensity}) and stellar mass function (Fig. \ref{mf}), proving that the light-cone construction process preserves the underlying physical distributions across all redshifts we tested and does not introduce mass-dependent biases. 

A direct comparison with an independently generated set of TNG100 mock images \citep{snyder23} further showed that the mock detection counts in the $H$ band agree within the statistical uncertainties (Fig. \ref{p:snyder23}). 
These results assess the robustness of the forward-modeling process and underscore, in line with \cite{lovell21,leja19}, the importance of cross-validation between independent synthetic datasets to separate technical modeling issues from more fundamental limitations in the simulations themselves.

\item The faint-end discrepancy is persistent and emerges at high redshifts. In all five CANDELS fields, the mock detections fall below the observed counts at the faint end, by $H\gtrsim$26 in the deepest fields and already at $H\gtrsim25$ in the shallower ones (Fig. \ref{all_fields}). The shortfall persists for TNG100 and EAGLE (Fig. \ref{eagle}) and is therefore not tied to a specific CHS. 
When we binned the counts into redshift intervals (Fig. \ref{dz}), both simulations faithfully reproduced the GOODS-South counts up to $z\sim3$; beyond these redshifts, a deficit emerged, with a more severe effect in EAGLE. The details differ between the two simulations (as does their subgrid physics implementation), with TNG100 detecting a slight excess of bright galaxies, while EAGLE is systematically slightly lower at almost all magnitudes, even in the complete regime.
The persistence of the deficit across multiple detection bands rules out SED modeling as the main driver. Since the synthetic galaxy fluxes in the IU were derived from stellar population synthesis of individual stellar particles, biases in the resulting SEDs (e.g., systematically redder or bluer colors) might in principle affect detection in different bands. This effect is particularly critical at $z>2$, where the observed near-infrared bands probe the rest-frame UV and optical bands, making the detectability of galaxies very sensitive to stellar age, metallicity, and dust attenuation. However, the persistence of the faint-end discrepancy across three detection bands (CANDELS GS F105W, $H$, and CEERS F277W; Fig. \ref{multi-band}) indicates that the missing galaxies are not preferentially lost in a wavelength-dependent way.

\item The faint-end deficit is already present in the modeled galaxy population. The completeness test (Fig. \ref{corrected_compl}), performed by injecting true TNG100 light profiles in the mock image (Fig. \ref{curve_comp}) and finding concordance with \cite{guo13}, showed that GS-corrected counts remain well above the IU counts already at the magnitude at which it is only 50\% complete ($H\simeq27$). Because the IU contains all simulated galaxies regardless of detectability, this gap cannot be explained solely by observational incompleteness or the detection pipeline, indicating that faint galaxies are already underrepresented at the level of the modeled population.
An increase in the depth of the mock image by 2.5 magnitudes brought the mock counts in agreement with GS around the peak ($H\sim27$). At fainter magnitudes, however, the deeper dataset overproduced ultra-faint galaxies not in place in the real GS dataset, but still underestimated the counts from genuinely deeper fields such as the Frontier Fields (A2744 Parallel and M0416 Parallel).
This shows that simply increasing the survey depth is not sufficient to reconcile the simulations with the data: the shortfall of moderately faint galaxies at $z>3$ (at the proper depth) reflects a genuine limitation of the simulated galaxy population, as the additional sources recovered at higher depth do not reconcile the mock counts with the observations. 

\item The discrepancy is driven by specific structural properties. 
The faint-end shortfall is not uniform across the structural parameter space ($C$, $e$, $\mu_R$, $r_{\%}$), but is driven by a specific class of compact systems with high central surface brightness (Fig. \ref{fig:MS_comb}).
GOODS-South shows an excess of galaxies with bright central regions ($\Delta f_{\mu_{3pix}}=0.18\pm0.05$) and of PSF-dominated sizes ($r_{50}/FWHM<1$, $\Delta f_{r_{50}}=0.17\pm0.05$), whereas the mock detections are dominated by diffuse profiles with a low surface brightness. 
When observational limitations were relaxed in the deeper mock, the discrepancy became even stronger and reached $\Delta f_{(\mu_{3pix}<\tilde{\mu}_{3pix} \, \& \, r_{50}/\text{FWHM}<1)} \simeq0.45\pm0.03$ for compact and bright-core galaxies. This demonstrates that the shortfall persists beyond the detection limits.

The analysis of the structural properties of the lost (i.e., undetected) and kept sources in the detection stage showed that undetected sources have a fainter central surface brightness and larger fractional light radii than those recovered (Fig. \ref{fig:lost_kept}).

This shows a twofold origin of the faint-end shortfall.
(i) A structural component: The modeled galaxy population produces too few genuinely compact faint galaxies.
(ii) A detection component: Too many simulated galaxies are diffuse systems with a low surface brightness, which are preferentially lost at the nominal survey depth but reappear in the deeper mock.
The structural bias likely reflects the combined effect of limited spatial and mass resolution, stellar feedback that dilutes central densities, and simplified dust modeling in the forward-modeling pipeline, all of which cause the simulated faint galaxies to become systematically more extended and fainter in their cores than those observed at $z>3$.
\end{itemize}

Our results show that the faint-end shortfall of simulated number counts, compared to GOODS-South, is largely driven by a structural mismatch between the observed and modeled galaxy populations, while forward-modeling and detection effects alone are insufficient to account for the observed discrepancy. This supports the interpretation proposed in F23 that the deficit reflects the difficulty of current hydrodynamical simulations in producing a sufficient population of compact faint galaxies at $z>3$, at least in part. At the same time, uncertainties related to the adopted dust treatment might still contribute and deserve further investigation in future work.

In a broader context, our results also demonstrated the diagnostic power of forward modeling in separating observational biases from physical discrepances, turning the agreement or difference with observations into a test of the physical modeling itself. 
The next step is to apply this framework to the new generation of large-volume simulations tailored for the early Universe \citep[e.g.,][]{kannan22,rosdahl22,bird22,jones24} to assess whether their updated prescriptions for star formation and feedback (and, possibly, dust) can reproduce the abundance and structural diversity of the faint high-redshift galaxies that are now revealed with unprecedented detail by JWST and, soon, by next-generation facilities such as ELT and Roman.

\begin{acknowledgements}
The research activities in this paper were carried out with the contribution of the NextGenerationEU funds within the National Recovery and Resilience Plan (PNRR), Mission 4 - Education and Research, Component 2 - From Research to Business (M4C2), Investment Line 3.1 - Strengthening and creation of Research Infrastructures, Project IR0000034 - “STILES - Strengthening the Italian Leadership in ELT and SKA”, CUP C33C22000640006.\\

The IllustrisTNG simulations were undertaken with computational time awarded by the Gauss Centre for Supercomputing (GCS) under GCS Large-Scale Projects GCS-ILLU and GCS-DWAR on the GCS share of the supercomputer Hazel Hen at the High-Performance Computing Center Stuttgart (HLRS), as well as on the machines of the Max Planck Computing and Data Facility (MPCDF) in Garching, Germany. We acknowledge the Virgo Consortium for making their simulation data available.\\ 

The EAGLE simulations were performed using the DiRAC-2 facility at Durham, managed by the ICC, and the PRACE facility Curie based in France at TGCC, CEA, Bruy\`eres-le-Ch\^atel.

\end{acknowledgements}

\bibliographystyle{aa} 
\bibliography{bibliography}

@ARTICLE{schaye15,
       author = {{Schaye}, Joop and {Crain}, Robert A. and {Bower}, Richard G. and
         {Furlong}, Michelle and {Schaller}, Matthieu and {Theuns}, Tom and
         {Dalla Vecchia}, Claudio and {Frenk}, Carlos S. and {McCarthy}, I.~G. and
         {Helly}, John C.},
        title = "{The EAGLE project: simulating the evolution and assembly of galaxies and their environments}",
      journal = {\mnras},
         year = "2015",
        month = "Jan",
       volume = {446},
       number = {1},
        pages = {521-554},
          doi = {10.1093/mnras/stu2058},
archivePrefix = {arXiv},
       eprint = {1407.7040},
 primaryClass = {astro-ph.GA},
       adsurl = {https://ui.adsabs.harvard.edu/abs/2015MNRAS.446..521S}
}

@ARTICLE{crain23,
       author = {{Crain}, Robert A. and {van de Voort}, Freeke},
        title = "{Hydrodynamical Simulations of the Galaxy Population: Enduring Successes and Outstanding Challenges}",
      journal = {\araa},
         year = 2023,
        month = aug,
       volume = {61},
        pages = {473-515},
          doi = {10.1146/annurev-astro-041923-043618},
archivePrefix = {arXiv},
       eprint = {2309.17075},
 primaryClass = {astro-ph.GA},
       adsurl = {https://ui.adsabs.harvard.edu/abs/2023ARA&A..61..473C}
}

@ARTICLE{weinberger18,
       author = {{Weinberger}, Rainer and {Springel}, Volker and {Pakmor}, R{\"u}diger and
         {Nelson}, Dylan and {Genel}, Shy and {Pillepich}, Annalisa and
         {Vogelsberger}, Mark and {Marinacci}, Federico and {Naiman}, Jill and
         {Torrey}, Paul},
        title = "{Supermassive black holes and their feedback effects in the IllustrisTNG simulation}",
      journal = {\mnras},
         year = "2018",
        month = "Sep",
       volume = {479},
       number = {3},
        pages = {4056-4072},
          doi = {10.1093/mnras/sty1733},
archivePrefix = {arXiv},
       eprint = {1710.04659},
 primaryClass = {astro-ph.GA},
       adsurl = {https://ui.adsabs.harvard.edu/abs/2018MNRAS.479.4056W}
}

@ARTICLE{pillepich18,
       author = {{Pillepich}, Annalisa and {Springel}, Volker and {Nelson}, Dylan and
         {Genel}, Shy and {Naiman}, Jill and {Pakmor}, R{\"u}diger and
         {Hernquist}, Lars and {Torrey}, Paul and {Vogelsberger}, Mark and
         {Weinberger}, Rainer},
        title = "{Simulating galaxy formation with the IllustrisTNG model}",
      journal = {\mnras},
         year = "2018",
        month = "Jan",
       volume = {473},
       number = {3},
        pages = {4077-4106},
          doi = {10.1093/mnras/stx2656},
archivePrefix = {arXiv},
       eprint = {1703.02970},
 primaryClass = {astro-ph.GA},
       adsurl = {https://ui.adsabs.harvard.edu/abs/2018MNRAS.473.4077P}
}

@ARTICLE{laceysilk91,
       author = {{Lacey}, Cedric and {Silk}, Joseph},
        title = "{Tidally Triggered Galaxy Formation. I. Evolution of the Galaxy Luminosity Function}",
      journal = {\apj},
         year = 1991,
        month = nov,
       volume = {381},
        pages = {14},
          doi = {10.1086/170625},
       adsurl = {https://ui.adsabs.harvard.edu/abs/1991ApJ...381...14L}
}

@ARTICLE{whitefrenk91,
       author = {{White}, Simon D.~M. and {Frenk}, Carlos S.},
        title = "{Galaxy Formation through Hierarchical Clustering}",
      journal = {\apj},
         year = 1991,
        month = sep,
       volume = {379},
        pages = {52},
          doi = {10.1086/170483},
       adsurl = {https://ui.adsabs.harvard.edu/abs/1991ApJ...379...52W}
}

@ARTICLE{navarro91,
       author = {{Navarro}, Julio F. and {Benz}, Willy},
        title = "{Dynamics of Cooling Gas in Galactic Dark Halos}",
      journal = {\apj},
         year = 1991,
        month = oct,
       volume = {380},
        pages = {320},
          doi = {10.1086/170590},
       adsurl = {https://ui.adsabs.harvard.edu/abs/1991ApJ...380..320N}
}

@ARTICLE{cole94,
       author = {{Cole}, S. and {Aragon-Salamanca}, A. and {Frenk}, C.~S. and {Navarro}, J.~F. and {Zepf}, S.~E.},
        title = "{A recipe for galaxy formation.}",
      journal = {\mnras},
         year = 1994,
        month = dec,
       volume = {271},
        pages = {781-806},
          doi = {10.1093/mnras/271.4.781},
archivePrefix = {arXiv},
       eprint = {astro-ph/9402001},
 primaryClass = {astro-ph},
       adsurl = {https://ui.adsabs.harvard.edu/abs/1994MNRAS.271..781C}
}

@ARTICLE{somerville99,
   author = {{Somerville}, R.~S. and {Primack}, J.~R.},
    title = "{Semi-analytic modelling of galaxy formation: the local Universe}",
  journal = {\mnras},
   eprint = {astro-ph/9802268},
     year = 1999,
    month = dec,
   volume = 310,
    pages = {1087-1110},
      doi = {10.1046/j.1365-8711.1999.03032.x},
   adsurl = {https://ui.adsabs.harvard.edu/abs/1999MNRAS.310.1087S}
}

@ARTICLE{cole00,
       author = {{Cole}, Shaun and {Lacey}, Cedric G. and {Baugh}, Carlton M. and {Frenk}, Carlos S.},
        title = "{Hierarchical galaxy formation}",
      journal = {\mnras},
         year = 2000,
        month = nov,
       volume = {319},
       number = {1},
        pages = {168-204},
          doi = {10.1046/j.1365-8711.2000.03879.x},
archivePrefix = {arXiv},
       eprint = {astro-ph/0007281},
 primaryClass = {astro-ph},
       adsurl = {https://ui.adsabs.harvard.edu/abs/2000MNRAS.319..168C}
}

@INPROCEEDINGS{lacey01,
       author = {{Lacey}, C.},
        title = "{Semi-Analytical Models of Galaxy Formation}",
    booktitle = {The Physics of Galaxy Formation},
         year = "2001",
       editor = {{Umemura}, Masayuki and {Susa}, Hajime},
       series = {Astronomical Society of the Pacific Conference Series},
       volume = {222},
        month = "Jan",
        pages = {273},
       adsurl = {https://ui.adsabs.harvard.edu/abs/2001ASPC..222..273L}
}

@ARTICLE{lacey16,
       author = {{Lacey}, Cedric G. and {Baugh}, Carlton M. and {Frenk}, Carlos S. and {Benson}, Andrew J. and {Bower}, Richard G. and {Cole}, Shaun and {Gonzalez-Perez}, Violeta and {Helly}, John C. and {Lagos}, Claudia D.~P. and {Mitchell}, Peter D.},
        title = "{A unified multiwavelength model of galaxy formation}",
      journal = {\mnras},
         year = 2016,
        month = nov,
       volume = {462},
       number = {4},
        pages = {3854-3911},
          doi = {10.1093/mnras/stw1888},
archivePrefix = {arXiv},
       eprint = {1509.08473},
 primaryClass = {astro-ph.GA},
       adsurl = {https://ui.adsabs.harvard.edu/abs/2016MNRAS.462.3854L}
}

@ARTICLE{somerville15,
       author = {{Somerville}, Rachel S. and {Popping}, Gerg{\"o} and {Trager}, Scott C.},
        title = "{Star formation in semi-analytic galaxy formation models with multiphase gas}",
      journal = {\mnras},
         year = 2015,
        month = nov,
       volume = {453},
       number = {4},
        pages = {4337-4367},
          doi = {10.1093/mnras/stv1877},
archivePrefix = {arXiv},
       eprint = {1503.00755},
 primaryClass = {astro-ph.GA},
       adsurl = {https://ui.adsabs.harvard.edu/abs/2015MNRAS.453.4337S}
}

@ARTICLE{efstathiou85,
       author = {{Efstathiou}, G. and {Davis}, M. and {White}, S.~D.~M. and {Frenk}, C.~S.},
        title = "{Numerical techniques for large cosmological N-body simulations}",
      journal = {\apjs},
         year = 1985,
        month = feb,
       volume = {57},
        pages = {241-260},
          doi = {10.1086/191003},
       adsurl = {https://ui.adsabs.harvard.edu/abs/1985ApJS...57..241E}
}

@ARTICLE{gutkin16,
       author = {{Gutkin}, Julia and {Charlot}, St{\'e}phane and {Bruzual}, Gustavo},
        title = "{Modelling the nebular emission from primeval to present-day star-forming galaxies}",
      journal = {\mnras},
         year = 2016,
        month = oct,
       volume = {462},
       number = {2},
        pages = {1757-1774},
          doi = {10.1093/mnras/stw1716},
archivePrefix = {arXiv},
       eprint = {1607.06086},
 primaryClass = {astro-ph.GA},
       adsurl = {https://ui.adsabs.harvard.edu/abs/2016MNRAS.462.1757G}
}

@ARTICLE{chabrier03,
       author = {{Chabrier}, Gilles},
        title = "{Galactic Stellar and Substellar Initial Mass Function}",
      journal = {\pasp},
         year = 2003,
        month = jul,
       volume = {115},
       number = {809},
        pages = {763-795},
          doi = {10.1086/376392},
archivePrefix = {arXiv},
       eprint = {astro-ph/0304382},
 primaryClass = {astro-ph},
       adsurl = {https://ui.adsabs.harvard.edu/abs/2003PASP..115..763C}
}

@ARTICLE{grazian15,
       author = {{Grazian}, A. and {Fontana}, A. and {Santini}, P. and {Dunlop}, J.~S. and {Ferguson}, H.~C. and {Castellano}, M. and {Amorin}, R. and {Ashby}, M.~L.~N. and {Barro}, G. and {Behroozi}, P. and {Boutsia}, K. and {Caputi}, K.~I. and {Chary}, R.~R. and {Dekel}, A. and {Dickinson}, M.~E. and {Faber}, S.~M. and {Fazio}, G.~G. and {Finkelstein}, S.~L. and {Galametz}, A. and {Giallongo}, E. and {Giavalisco}, M. and {Grogin}, N.~A. and {Guo}, Y. and {Kocevski}, D. and {Koekemoer}, A.~M. and {Koo}, D.~C. and {Lee}, K. -S. and {Lu}, Y. and {Merlin}, E. and {Mobasher}, B. and {Nonino}, M. and {Papovich}, C. and {Paris}, D. and {Pentericci}, L. and {Reddy}, N. and {Renzini}, A. and {Salmon}, B. and {Salvato}, M. and {Sommariva}, V. and {Song}, M. and {Vanzella}, E.},
        title = "{The galaxy stellar mass function at 3.5 {\ensuremath{\leq}}z {\ensuremath{\leq}} 7.5 in the CANDELS/UDS, GOODS-South, and HUDF fields}",
      journal = {\aap},
         year = 2015,
        month = mar,
       volume = {575},
          eid = {A96},
        pages = {A96},
          doi = {10.1051/0004-6361/201424750},
archivePrefix = {arXiv},
       eprint = {1412.0532},
 primaryClass = {astro-ph.GA},
       adsurl = {https://ui.adsabs.harvard.edu/abs/2015A&A...575A..96G}
}

@ARTICLE{castellano16,
       author = {{Castellano}, M. and {Amor{\'\i}n}, R. and {Merlin}, E. and {Fontana}, A. and {McLure}, R.~J. and {M{\'a}rmol-Queralt{\'o}}, E. and {Mortlock}, A. and {Parsa}, S. and {Dunlop}, J.~S. and {Elbaz}, D. and {Balestra}, I. and {Boucaud}, A. and {Bourne}, N. and {Boutsia}, K. and {Brammer}, G. and {Bruce}, V.~A. and {Buitrago}, F. and {Capak}, P. and {Cappelluti}, N. and {Ciesla}, L. and {Comastri}, A. and {Cullen}, F. and {Derriere}, S. and {Faber}, S.~M. and {Giallongo}, E. and {Grazian}, A. and {Grillo}, C. and {Mercurio}, A. and {Micha{\l}owski}, M.~J. and {Nonino}, M. and {Paris}, D. and {Pentericci}, L. and {Pilo}, S. and {Rosati}, P. and {Santini}, P. and {Schreiber}, C. and {Shu}, X. and {Wang}, T.},
        title = "{The ASTRODEEP Frontier Fields catalogues. II. Photometric redshifts and rest frame properties in Abell-2744 and MACS-J0416}",
      journal = {\aap},
         year = 2016,
        month = may,
       volume = {590},
          eid = {A31},
        pages = {A31},
          doi = {10.1051/0004-6361/201527514},
archivePrefix = {arXiv},
       eprint = {1603.02461},
 primaryClass = {astro-ph.GA},
       adsurl = {https://ui.adsabs.harvard.edu/abs/2016A&A...590A..31C}
}

@ARTICLE{finkelstein23,
       author = {{Finkelstein}, Steven L. and {Bagley}, Micaela B. and {Ferguson}, Henry C. and {Wilkins}, Stephen M. and {Kartaltepe}, Jeyhan S. and {Papovich}, Casey and {Yung}, L.~Y. Aaron and {Haro}, Pablo Arrabal and {Behroozi}, Peter and {Dickinson}, Mark and {Kocevski}, Dale D. and {Koekemoer}, Anton M. and {Larson}, Rebecca L. and {Le Bail}, Aur{\'e}lien and {Morales}, Alexa M. and {P{\'e}rez-Gonz{\'a}lez}, Pablo G. and {Burgarella}, Denis and {Dav{\'e}}, Romeel and {Hirschmann}, Michaela and {Somerville}, Rachel S. and {Wuyts}, Stijn and {Bromm}, Volker and {Casey}, Caitlin M. and {Fontana}, Adriano and {Fujimoto}, Seiji and {Gardner}, Jonathan P. and {Giavalisco}, Mauro and {Grazian}, Andrea and {Grogin}, Norman A. and {Hathi}, Nimish P. and {Hutchison}, Taylor A. and {Jha}, Saurabh W. and {Jogee}, Shardha and {Kewley}, Lisa J. and {Kirkpatrick}, Allison and {Long}, Arianna S. and {Lotz}, Jennifer M. and {Pentericci}, Laura and {Pierel}, Justin D.~R. and {Pirzkal}, Nor and {Ravindranath}, Swara and {Ryan}, Russell E. and {Trump}, Jonathan R. and {Yang}, Guang and {Bhatawdekar}, Rachana and {Bisigello}, Laura and {Buat}, V{\'e}ronique and {Calabr{\`o}}, Antonello and {Castellano}, Marco and {Cleri}, Nikko J. and {Cooper}, M.~C. and {Croton}, Darren and {Daddi}, Emanuele and {Dekel}, Avishai and {Elbaz}, David and {Franco}, Maximilien and {Gawiser}, Eric and {Holwerda}, Benne W. and {Huertas-Company}, Marc and {Jaskot}, Anne E. and {Leung}, Gene C.~K. and {Lucas}, Ray A. and {Mobasher}, Bahram and {Pandya}, Viraj and {Tacchella}, Sandro and {Weiner}, Benjamin J. and {Zavala}, Jorge A.},
        title = "{CEERS Key Paper. I. An Early Look into the First 500 Myr of Galaxy Formation with JWST}",
      journal = {\apjl},
         year = 2023,
        month = mar,
       volume = {946},
       number = {1},
          eid = {L13},
        pages = {L13},
          doi = {10.3847/2041-8213/acade4},
archivePrefix = {arXiv},
       eprint = {2211.05792},
 primaryClass = {astro-ph.GA},
       adsurl = {https://ui.adsabs.harvard.edu/abs/2023ApJ...946L..13F}
}

@ARTICLE{finkelstein25,
       author = {{Finkelstein}, Steven L. and {Bagley}, Micaela B. and {Arrabal Haro}, Pablo and {Dickinson}, Mark and {Ferguson}, Henry C. and {Kartaltepe}, Jeyhan S. and {Kocevski}, Dale D. and {Koekemoer}, Anton M. and {Lotz}, Jennifer M. and {Papovich}, Casey and {P{\'e}rez-Gonz{\'a}lez}, Pablo G. and {Pirzkal}, Nor and {Somerville}, Rachel S. and {Trump}, Jonathan R. and {Yang}, Guang and {Yung}, L.~Y. Aaron and {Fontana}, Adriano and {Grazian}, Andrea and {Grogin}, Norman A. and {Kewley}, Lisa J. and {Kirkpatrick}, Allison and {Larson}, Rebecca L. and {Pentericci}, Laura and {Ravindranath}, Swara and {Wilkins}, Stephen M. and {Almaini}, Omar and {Amor{\'\i}n}, Ricardo O. and {Barro}, Guillermo and {Bhatawdekar}, Rachana and {Bisigello}, Laura and {Brooks}, Madisyn and {Buat}, V{\'e}ronique and {Buitrago}, Fernando and {Burgarella}, Denis and {Calabr{\`o}}, Antonello and {Castellano}, Marco and {Cheng}, Yingjie and {Cleri}, Nikko J. and {Cole}, Justin W. and {Cooper}, M.~C. and {Cooper}, Olivia R. and {Costantin}, Luca and {Cox}, Isa G. and {Croton}, Darren and {Daddi}, Emanuele and {Davis}, Kelcey and {Dekel}, Avishai and {Elbaz}, David and {Fern{\'a}ndez}, Vital and {Fujimoto}, Seiji and {Gandolfi}, Giovanni and {Gardner}, Jonathan P. and {Gawiser}, Eric and {Giavalisco}, Mauro and {G{\'o}mez-Guijarro}, Carlos and {Guo}, Yuchen and {Gupta}, Ansh R. and {Hathi}, Nimish P. and {Harish}, Santosh and {Henry}, Aur{\'e}lien and {Hirschmann}, Michaela and {Hu}, Weida and {Hutchison}, Taylor A. and {Iyer}, Kartheik G. and {Jaskot}, Anne E. and {Jha}, Saurabh W. and {Jung}, Intae and {Kassin}, Susan A. and {Kokorev}, Vasily and {Kurczynski}, Peter and {Leung}, Gene C.~K. and {Llerena}, Mario and {Long}, Arianna S. and {Lucas}, Ray A. and {Lu}, Shiying and {McGrath}, Elizabeth J. and {McIntosh}, Daniel H. and {Merlin}, Emiliano and {Mobasher}, Bahram and {Morales}, Alexa M. and {Napolitano}, Lorenzo and {Pacucci}, Fabio and {Pandya}, Viraj and {Rafelski}, Marc and {Rodighiero}, Giulia and {Rose}, Caitlin and {Santini}, Paola and {Seill{\'e}}, Lise-Marie and {Simons}, Raymond C. and {Shen}, Lu and {Straughn}, Amber N. and {Tacchella}, Sandro and {Taylor}, Anthony J. and {Vanderhoof}, Brittany N. and {Vega-Ferrero}, Jes{\'u}s and {Weiner}, Benjamin J. and {Willmer}, Christopher N.~A. and {Zhu}, Peixin and {Bell}, Eric F. and {Wuyts}, Stijn and {Holwerda}, Benne W. and {Wang}, Xin and {Wang}, Weichen and {Zavala}, Jorge A. and {CEERS Collaboration}},
        title = "{The Cosmic Evolution Early Release Science Survey (CEERS)}",
      journal = {\apjl},
         year = 2025,
        month = apr,
       volume = {983},
       number = {1},
          eid = {L4},
        pages = {L4},
          doi = {10.3847/2041-8213/adbbd3},
archivePrefix = {arXiv},
       eprint = {2501.04085},
 primaryClass = {astro-ph.GA},
       adsurl = {https://ui.adsabs.harvard.edu/abs/2025ApJ...983L...4F}
}

@ARTICLE{guo13,
       author = {{Guo}, Yicheng and {Ferguson}, Henry C. and {Giavalisco}, Mauro and {Barro}, Guillermo and {Willner}, S.~P. and {Ashby}, Matthew L.~N. and {Dahlen}, Tomas and {Donley}, Jennifer L. and {Faber}, Sandra M. and {Fontana}, Adriano and {Galametz}, Audrey and {Grazian}, Andrea and {Huang}, Kuang-Han and {Kocevski}, Dale D. and {Koekemoer}, Anton M. and {Koo}, David C. and {McGrath}, Elizabeth J. and {Peth}, Michael and {Salvato}, Mara and {Wuyts}, Stijn and {Castellano}, Marco and {Cooray}, Asantha R. and {Dickinson}, Mark E. and {Dunlop}, James S. and {Fazio}, G.~G. and {Gardner}, Jonathan P. and {Gawiser}, Eric and {Grogin}, Norman A. and {Hathi}, Nimish P. and {Hsu}, Li-Ting and {Lee}, Kyoung-Soo and {Lucas}, Ray A. and {Mobasher}, Bahram and {Nandra}, Kirpal and {Newman}, Jeffery A. and {van der Wel}, Arjen},
        title = "{CANDELS Multi-wavelength Catalogs: Source Detection and Photometry in the GOODS-South Field}",
      journal = {\apjs},
         year = 2013,
        month = aug,
       volume = {207},
       number = {2},
          eid = {24},
        pages = {24},
          doi = {10.1088/0067-0049/207/2/24},
archivePrefix = {arXiv},
       eprint = {1308.4405},
 primaryClass = {astro-ph.CO},
       adsurl = {https://ui.adsabs.harvard.edu/abs/2013ApJS..207...24G}
}

@ARTICLE{merlin21,
       author = {{Merlin}, E. and {Castellano}, M. and {Santini}, P. and {Cipolletta}, G. and {Boutsia}, K. and {Schreiber}, C. and {Buitrago}, F. and {Fontana}, A. and {Elbaz}, D. and {Dunlop}, J. and {Grazian}, A. and {McLure}, R. and {McLeod}, D. and {Nonino}, M. and {Milvang-Jensen}, B. and {Derriere}, S. and {Hathi}, N.~P. and {Pentericci}, L. and {Fortuni}, F. and {Calabr{\`o}}, A.},
        title = "{The ASTRODEEP-GS43 catalogue: New photometry and redshifts for the CANDELS GOODS-South field}",
      journal = {\aap},
         year = 2021,
        month = may,
       volume = {649},
          eid = {A22},
        pages = {A22},
          doi = {10.1051/0004-6361/202140310},
archivePrefix = {arXiv},
       eprint = {2103.09246},
 primaryClass = {astro-ph.GA},
       adsurl = {https://ui.adsabs.harvard.edu/abs/2021A&A...649A..22M}
}

@ARTICLE{merlin24,
       author = {{Merlin}, E. and {Santini}, P. and {Paris}, D. and {Castellano}, M. and {Fontana}, A. and {Treu}, T. and {Finkelstein}, S.~L. and {Dunlop}, J.~S. and {Arrabal Haro}, P. and {Bagley}, M. and {Boyett}, K. and {Calabr{\`o}}, A. and {Correnti}, M. and {Davis}, K. and {Dickinson}, M. and {Donnan}, C.~T. and {Ferguson}, H.~C. and {Fortuni}, F. and {Giavalisco}, M. and {Glazebrook}, K. and {Grazian}, A. and {Grogin}, N.~A. and {Hathi}, N. and {Hirschmann}, M. and {Kartaltepe}, J.~S. and {Kewley}, L.~J. and {Kirkpatrick}, A. and {Kocevski}, D.~D. and {Koekemoer}, A.~M. and {Leung}, G. and {Lotz}, J.~M. and {Lucas}, R.~A. and {Magee}, D.~K. and {Marchesini}, D. and {Mascia}, S. and {McLeod}, D.~J. and {McLure}, R.~J. and {Nanayakkara}, T. and {Napolitano}, L. and {Nonino}, M. and {Papovich}, C. and {Pentericci}, L. and {P{\'e}rez-Gonz{\'a}lez}, P.~G. and {Pirzkal}, N. and {Ravindranath}, S. and {Roberts-Borsani}, G. and {Somerville}, R.~S. and {Trenti}, M. and {Trump}, J.~R. and {Vulcani}, B. and {Wang}, X. and {Watson}, P.~J. and {Wilkins}, S.~M. and {Yang}, G. and {Yung}, L.~Y.~A.},
        title = "{ASTRODEEP-JWST: NIRCam-HST multi-band photometry and redshifts for half a million sources in six extragalactic deep fields}",
      journal = {\aap},
         year = 2024,
        month = nov,
       volume = {691},
          eid = {A240},
        pages = {A240},
          doi = {10.1051/0004-6361/202451409},
archivePrefix = {arXiv},
       eprint = {2409.00169},
 primaryClass = {astro-ph.GA},
       adsurl = {https://ui.adsabs.harvard.edu/abs/2024A&A...691A.240M}
}

@ARTICLE{merlin25,
       author = {{Merlin}, Emiliano and {Fortuni}, Flaminia and {Calabr{\'o}}, Antonello and {Castellano}, Marco and {Santini}, Paola and {Fontana}, Adriano and {Kimmig}, Lucas C. and {Shankar}, Francesco and {Napolitano}, Lorenzo and {Koekemoer}, Anton M. and {Lucas}, Ray A. and {Pacucci}, Fabio and {Cooper}, Michael C. and {Hirschmann}, Michaela and {P{\'e}rez-Gonz{\'a}lez}, Pablo G. and {Barro}, Guillermo and {Dickinson}, Mark and {Gandolfi}, Giovanni and {Paris}, Diego and {Grogin}, Norman A. and {Wang}, Xin},
        title = "{Witnessing downsizing in the making: quiescent and breathing galaxies at the dawn of the Universe}",
      journal = {The Open Journal of Astrophysics},
         year = 2025,
        month = nov,
       volume = {8},
        pages = {E170},
          doi = {10.33232/001c.147267},
archivePrefix = {arXiv},
       eprint = {2509.09764},
 primaryClass = {astro-ph.GA},
       adsurl = {https://ui.adsabs.harvard.edu/abs/2025OJAp....8E.170M}
}

@ARTICLE{fortuni23,
       author = {{Fortuni}, Flaminia and {Merlin}, Emiliano and {Fontana}, Adriano and {Giocoli}, Carlo and {Romelli}, Erik and {Graziani}, Luca and {Santini}, Paola and {Castellano}, Marco and {Charlot}, St{\'e}phane and {Chevallard}, Jacopo},
        title = "{FORECAST: A flexible software to forward model cosmological hydrodynamical simulations mimicking real observations}",
      journal = {\aap},
         year = 2023,
        month = sep,
       volume = {677},
          eid = {A102},
        pages = {A102},
          doi = {10.1051/0004-6361/202346725},
archivePrefix = {arXiv},
       eprint = {2305.19166},
 primaryClass = {astro-ph.IM},
       adsurl = {https://ui.adsabs.harvard.edu/abs/2023A&A...677A.102F}
}

@ARTICLE{snyder23,
       author = {{Snyder}, Gregory F. and {Pe{\~n}a}, Theodore and {Yung}, L.~Y. Aaron and {Rose}, Caitlin and {Kartaltepe}, Jeyhan and {Ferguson}, Harry},
        title = "{Mock galaxy surveys for HST and JWST from the IllustrisTNG simulations}",
      journal = {\mnras},
         year = 2023,
        month = feb,
       volume = {518},
       number = {4},
        pages = {6318-6324},
          doi = {10.1093/mnras/stac3397},
archivePrefix = {arXiv},
       eprint = {2211.09677},
 primaryClass = {astro-ph.GA},
       adsurl = {https://ui.adsabs.harvard.edu/abs/2023MNRAS.518.6318S}
}

@ARTICLE{bertin96,
       author = {{Bertin}, E. and {Arnouts}, S.},
        title = "{SExtractor: Software for source extraction.}",
      journal = {\aaps},
         year = 1996,
        month = jun,
       volume = {117},
        pages = {393-404},
          doi = {10.1051/aas:1996164},
       adsurl = {https://ui.adsabs.harvard.edu/abs/1996A&AS..117..393B}
}

@ARTICLE{nelson19a,
       author = {{Nelson}, Dylan and {Springel}, Volker and {Pillepich}, Annalisa and {Rodriguez-Gomez}, Vicente and {Torrey}, Paul and {Genel}, Shy and {Vogelsberger}, Mark and {Pakmor}, Ruediger and {Marinacci}, Federico and {Weinberger}, Rainer and {Kelley}, Luke and {Lovell}, Mark and {Diemer}, Benedikt and {Hernquist}, Lars},
        title = "{The IllustrisTNG simulations: public data release}",
      journal = {Computational Astrophysics and Cosmology},
         year = 2019,
        month = may,
       volume = {6},
       number = {1},
          eid = {2},
        pages = {2},
          doi = {10.1186/s40668-019-0028-x},
archivePrefix = {arXiv},
       eprint = {1812.05609},
 primaryClass = {astro-ph.GA},
       adsurl = {https://ui.adsabs.harvard.edu/abs/2019ComAC...6....2N}
}

@ARTICLE{pillepich18_mf,
       author = {{Pillepich}, Annalisa and {Nelson}, Dylan and {Hernquist}, Lars and {Springel}, Volker and {Pakmor}, R{\"u}diger and {Torrey}, Paul and {Weinberger}, Rainer and {Genel}, Shy and {Naiman}, Jill P. and {Marinacci}, Federico and {Vogelsberger}, Mark},
        title = "{First results from the IllustrisTNG simulations: the stellar mass content of groups and clusters of galaxies}",
      journal = {\mnras},
         year = 2018,
        month = mar,
       volume = {475},
       number = {1},
        pages = {648-675},
          doi = {10.1093/mnras/stx3112},
archivePrefix = {arXiv},
       eprint = {1707.03406},
 primaryClass = {astro-ph.GA},
       adsurl = {https://ui.adsabs.harvard.edu/abs/2018MNRAS.475..648P}
}

@ARTICLE{galametz13,
       author = {{Galametz}, Audrey and {Grazian}, Andrea and {Fontana}, Adriano and {Ferguson}, Henry C. and {Ashby}, M.~L.~N. and {Barro}, Guillermo and {Castellano}, Marco and {Dahlen}, Tomas and {Donley}, Jennifer L. and {Faber}, Sandy M. and {Grogin}, Norman and {Guo}, Yicheng and {Huang}, Kuang-Han and {Kocevski}, Dale D. and {Koekemoer}, Anton M. and {Lee}, Kyoung-Soo and {McGrath}, Elizabeth J. and {Peth}, Michael and {Willner}, S.~P. and {Almaini}, Omar and {Cooper}, Michael and {Cooray}, Asantha and {Conselice}, Christopher J. and {Dickinson}, Mark and {Dunlop}, James S. and {Fazio}, G.~G. and {Foucaud}, Sebastien and {Gardner}, Jonathan P. and {Giavalisco}, Mauro and {Hathi}, N.~P. and {Hartley}, Will G. and {Koo}, David C. and {Lai}, Kamson and {de Mello}, Duilia F. and {McLure}, Ross J. and {Lucas}, Ray A. and {Paris}, Diego and {Pentericci}, Laura and {Santini}, Paola and {Simpson}, Chris and {Sommariva}, Veronica and {Targett}, Thomas and {Weiner}, Benjamin J. and {Wuyts}, Stijn and {CANDELS Team}},
        title = "{CANDELS Multiwavelength Catalogs: Source Identification and Photometry in the CANDELS UKIDSS Ultra-deep Survey Field}",
      journal = {\apjs},
         year = 2013,
        month = jun,
       volume = {206},
       number = {2},
          eid = {10},
        pages = {10},
          doi = {10.1088/0067-0049/206/2/10},
archivePrefix = {arXiv},
       eprint = {1305.1823},
 primaryClass = {astro-ph.CO},
       adsurl = {https://ui.adsabs.harvard.edu/abs/2013ApJS..206...10G}
}

@ARTICLE{grogin11,
       author = {{Grogin}, Norman A. and {Kocevski}, Dale D. and {Faber}, S.~M. and {Ferguson}, Henry C. and {Koekemoer}, Anton M. and {Riess}, Adam G. and {Acquaviva}, Viviana and {Alexander}, David M. and {Almaini}, Omar and {Ashby}, Matthew L.~N. and {Barden}, Marco and {Bell}, Eric F. and {Bournaud}, Fr{\'e}d{\'e}ric and {Brown}, Thomas M. and {Caputi}, Karina I. and {Casertano}, Stefano and {Cassata}, Paolo and {Castellano}, Marco and {Challis}, Peter and {Chary}, Ranga-Ram and {Cheung}, Edmond and {Cirasuolo}, Michele and {Conselice}, Christopher J. and {Roshan Cooray}, Asantha and {Croton}, Darren J. and {Daddi}, Emanuele and {Dahlen}, Tomas and {Dav{\'e}}, Romeel and {de Mello}, Du{\'\i}lia F. and {Dekel}, Avishai and {Dickinson}, Mark and {Dolch}, Timothy and {Donley}, Jennifer L. and {Dunlop}, James S. and {Dutton}, Aaron A. and {Elbaz}, David and {Fazio}, Giovanni G. and {Filippenko}, Alexei V. and {Finkelstein}, Steven L. and {Fontana}, Adriano and {Gardner}, Jonathan P. and {Garnavich}, Peter M. and {Gawiser}, Eric and {Giavalisco}, Mauro and {Grazian}, Andrea and {Guo}, Yicheng and {Hathi}, Nimish P. and {H{\"a}ussler}, Boris and {Hopkins}, Philip F. and {Huang}, Jia-Sheng and {Huang}, Kuang-Han and {Jha}, Saurabh W. and {Kartaltepe}, Jeyhan S. and {Kirshner}, Robert P. and {Koo}, David C. and {Lai}, Kamson and {Lee}, Kyoung-Soo and {Li}, Weidong and {Lotz}, Jennifer M. and {Lucas}, Ray A. and {Madau}, Piero and {McCarthy}, Patrick J. and {McGrath}, Elizabeth J. and {McIntosh}, Daniel H. and {McLure}, Ross J. and {Mobasher}, Bahram and {Moustakas}, Leonidas A. and {Mozena}, Mark and {Nandra}, Kirpal and {Newman}, Jeffrey A. and {Niemi}, Sami-Matias and {Noeske}, Kai G. and {Papovich}, Casey J. and {Pentericci}, Laura and {Pope}, Alexandra and {Primack}, Joel R. and {Rajan}, Abhijith and {Ravindranath}, Swara and {Reddy}, Naveen A. and {Renzini}, Alvio and {Rix}, Hans-Walter and {Robaina}, Aday R. and {Rodney}, Steven A. and {Rosario}, David J. and {Rosati}, Piero and {Salimbeni}, Sara and {Scarlata}, Claudia and {Siana}, Brian and {Simard}, Luc and {Smidt}, Joseph and {Somerville}, Rachel S. and {Spinrad}, Hyron and {Straughn}, Amber N. and {Strolger}, Louis-Gregory and {Telford}, Olivia and {Teplitz}, Harry I. and {Trump}, Jonathan R. and {van der Wel}, Arjen and {Villforth}, Carolin and {Wechsler}, Risa H. and {Weiner}, Benjamin J. and {Wiklind}, Tommy and {Wild}, Vivienne and {Wilson}, Grant and {Wuyts}, Stijn and {Yan}, Hao-Jing and {Yun}, Min S.},
        title = "{CANDELS: The Cosmic Assembly Near-infrared Deep Extragalactic Legacy Survey}",
      journal = {\apjs},
         year = 2011,
        month = dec,
       volume = {197},
       number = {2},
          eid = {35},
        pages = {35},
          doi = {10.1088/0067-0049/197/2/35},
archivePrefix = {arXiv},
       eprint = {1105.3753},
 primaryClass = {astro-ph.CO},
       adsurl = {https://ui.adsabs.harvard.edu/abs/2011ApJS..197...35G}
}

@ARTICLE{koekemoer11,
       author = {{Koekemoer}, Anton M. and {Faber}, S.~M. and {Ferguson}, Henry C. and {Grogin}, Norman A. and {Kocevski}, Dale D. and {Koo}, David C. and {Lai}, Kamson and {Lotz}, Jennifer M. and {Lucas}, Ray A. and {McGrath}, Elizabeth J. and {Ogaz}, Sara and {Rajan}, Abhijith and {Riess}, Adam G. and {Rodney}, Steve A. and {Strolger}, Louis and {Casertano}, Stefano and {Castellano}, Marco and {Dahlen}, Tomas and {Dickinson}, Mark and {Dolch}, Timothy and {Fontana}, Adriano and {Giavalisco}, Mauro and {Grazian}, Andrea and {Guo}, Yicheng and {Hathi}, Nimish P. and {Huang}, Kuang-Han and {van der Wel}, Arjen and {Yan}, Hao-Jing and {Acquaviva}, Viviana and {Alexander}, David M. and {Almaini}, Omar and {Ashby}, Matthew L.~N. and {Barden}, Marco and {Bell}, Eric F. and {Bournaud}, Fr{\'e}d{\'e}ric and {Brown}, Thomas M. and {Caputi}, Karina I. and {Cassata}, Paolo and {Challis}, Peter J. and {Chary}, Ranga-Ram and {Cheung}, Edmond and {Cirasuolo}, Michele and {Conselice}, Christopher J. and {Roshan Cooray}, Asantha and {Croton}, Darren J. and {Daddi}, Emanuele and {Dav{\'e}}, Romeel and {de Mello}, Duilia F. and {de Ravel}, Loic and {Dekel}, Avishai and {Donley}, Jennifer L. and {Dunlop}, James S. and {Dutton}, Aaron A. and {Elbaz}, David and {Fazio}, Giovanni G. and {Filippenko}, Alexei V. and {Finkelstein}, Steven L. and {Frazer}, Chris and {Gardner}, Jonathan P. and {Garnavich}, Peter M. and {Gawiser}, Eric and {Gruetzbauch}, Ruth and {Hartley}, Will G. and {H{\"a}ussler}, Boris and {Herrington}, Jessica and {Hopkins}, Philip F. and {Huang}, Jia-Sheng and {Jha}, Saurabh W. and {Johnson}, Andrew and {Kartaltepe}, Jeyhan S. and {Khostovan}, Ali A. and {Kirshner}, Robert P. and {Lani}, Caterina and {Lee}, Kyoung-Soo and {Li}, Weidong and {Madau}, Piero and {McCarthy}, Patrick J. and {McIntosh}, Daniel H. and {McLure}, Ross J. and {McPartland}, Conor and {Mobasher}, Bahram and {Moreira}, Heidi and {Mortlock}, Alice and {Moustakas}, Leonidas A. and {Mozena}, Mark and {Nandra}, Kirpal and {Newman}, Jeffrey A. and {Nielsen}, Jennifer L. and {Niemi}, Sami and {Noeske}, Kai G. and {Papovich}, Casey J. and {Pentericci}, Laura and {Pope}, Alexandra and {Primack}, Joel R. and {Ravindranath}, Swara and {Reddy}, Naveen A. and {Renzini}, Alvio and {Rix}, Hans-Walter and {Robaina}, Aday R. and {Rosario}, David J. and {Rosati}, Piero and {Salimbeni}, Sara and {Scarlata}, Claudia and {Siana}, Brian and {Simard}, Luc and {Smidt}, Joseph and {Snyder}, Diana and {Somerville}, Rachel S. and {Spinrad}, Hyron and {Straughn}, Amber N. and {Telford}, Olivia and {Teplitz}, Harry I. and {Trump}, Jonathan R. and {Vargas}, Carlos and {Villforth}, Carolin and {Wagner}, Cory R. and {Wandro}, Pat and {Wechsler}, Risa H. and {Weiner}, Benjamin J. and {Wiklind}, Tommy and {Wild}, Vivienne and {Wilson}, Grant and {Wuyts}, Stijn and {Yun}, Min S.},
        title = "{CANDELS: The Cosmic Assembly Near-infrared Deep Extragalactic Legacy Survey{\textemdash}The Hubble Space Telescope Observations, Imaging Data Products, and Mosaics}",
      journal = {\apjs},
         year = 2011,
        month = dec,
       volume = {197},
       number = {2},
          eid = {36},
        pages = {36},
          doi = {10.1088/0067-0049/197/2/36},
archivePrefix = {arXiv},
       eprint = {1105.3754},
 primaryClass = {astro-ph.CO},
       adsurl = {https://ui.adsabs.harvard.edu/abs/2011ApJS..197...36K}
}

@ARTICLE{laigle19,
       author = {{Laigle}, C. and {Davidzon}, I. and {Ilbert}, O. and {Devriendt}, J. and {Kashino}, D. and {Pichon}, C. and {Capak}, P. and {Arnouts}, S. and {de la Torre}, S. and {Dubois}, Y. and {Gozaliasl}, G. and {Le Borgne}, D. and {Lilly}, S. and {McCracken}, H.~J. and {Salvato}, M. and {Slyz}, A.},
        title = "{Horizon-AGN virtual observatory - 1. SED-fitting performance and forecasts for future imaging surveys}",
      journal = {\mnras},
         year = 2019,
        month = jul,
       volume = {486},
       number = {4},
        pages = {5104-5123},
          doi = {10.1093/mnras/stz1054},
archivePrefix = {arXiv},
       eprint = {1903.10934},
 primaryClass = {astro-ph.GA},
       adsurl = {https://ui.adsabs.harvard.edu/abs/2019MNRAS.486.5104L}
}

@ARTICLE{drakos22,
       author = {{Drakos}, Nicole E. and {Villasenor}, Bruno and {Robertson}, Brant E. and {Hausen}, Ryan and {Dickinson}, Mark E. and {Ferguson}, Henry C. and {Furlanetto}, Steven R. and {Greene}, Jenny E. and {Madau}, Piero and {Shapley}, Alice E. and {Stark}, Daniel P. and {Wechsler}, Risa H.},
        title = "{Deep Realistic Extragalactic Model (DREaM) Galaxy Catalogs: Predictions for a Roman Ultra-deep Field}",
      journal = {\apj},
         year = 2022,
        month = feb,
       volume = {926},
       number = {2},
          eid = {194},
        pages = {194},
          doi = {10.3847/1538-4357/ac46fb},
archivePrefix = {arXiv},
       eprint = {2110.10703},
 primaryClass = {astro-ph.GA},
       adsurl = {https://ui.adsabs.harvard.edu/abs/2022ApJ...926..194D}
}

@ARTICLE{marshall22,
       author = {{Marshall}, Madeline A. and {Watts}, Katelyn and {Wilkins}, Stephen and {Di Matteo}, Tiziana and {Kuusisto}, Jussi K. and {Roper}, William J. and {Vijayan}, Aswin P. and {Ni}, Yueying and {Feng}, Yu and {Croft}, Rupert A.~C.},
        title = "{The BLUETIDES mock image catalogue: simulated observations of high-redshift galaxies and predictions for JWST imaging surveys}",
      journal = {\mnras},
         year = 2022,
        month = oct,
       volume = {516},
       number = {1},
        pages = {1047-1061},
          doi = {10.1093/mnras/stac2111},
archivePrefix = {arXiv},
       eprint = {2206.08941},
 primaryClass = {astro-ph.GA},
       adsurl = {https://ui.adsabs.harvard.edu/abs/2022MNRAS.516.1047M}
}

@ARTICLE{lovell21,
       author = {{Lovell}, Christopher C. and {Geach}, James E. and {Dav{\'e}}, Romeel and {Narayanan}, Desika and {Li}, Qi},
        title = "{Reproducing submillimetre galaxy number counts with cosmological hydrodynamic simulations}",
      journal = {\mnras},
         year = 2021,
        month = mar,
       volume = {502},
       number = {1},
        pages = {772-793},
          doi = {10.1093/mnras/staa4043},
archivePrefix = {arXiv},
       eprint = {2006.15156},
 primaryClass = {astro-ph.GA},
       adsurl = {https://ui.adsabs.harvard.edu/abs/2021MNRAS.502..772L}
}

@ARTICLE{ishikawa24,
       author = {{Ishikawa}, Shogo and {Okumura}, Teppei and {Nishimichi}, Takahiro},
        title = "{MOCK OBSERVATORY: Two thousand light-cone mock catalogues of luminous red galaxies from the Hyper Suprime-Cam Survey for the cosmological large-scale analysis}",
      journal = {\mnras},
         year = 2024,
        month = apr,
       volume = {529},
       number = {2},
        pages = {1839-1851},
          doi = {10.1093/mnras/stae648},
archivePrefix = {arXiv},
       eprint = {2308.03871},
 primaryClass = {astro-ph.CO},
       adsurl = {https://ui.adsabs.harvard.edu/abs/2024MNRAS.529.1839I}
}

@ARTICLE{aihara18,
       author = {{Aihara}, Hiroaki and {Arimoto}, Nobuo and {Armstrong}, Robert and {Arnouts}, St{\'e}phane and {Bahcall}, Neta A. and {Bickerton}, Steven and {Bosch}, James and {Bundy}, Kevin and {Capak}, Peter L. and {Chan}, James H.~H. and {Chiba}, Masashi and {Coupon}, Jean and {Egami}, Eiichi and {Enoki}, Motohiro and {Finet}, Francois and {Fujimori}, Hiroki and {Fujimoto}, Seiji and {Furusawa}, Hisanori and {Furusawa}, Junko and {Goto}, Tomotsugu and {Goulding}, Andy and {Greco}, Johnny P. and {Greene}, Jenny E. and {Gunn}, James E. and {Hamana}, Takashi and {Harikane}, Yuichi and {Hashimoto}, Yasuhiro and {Hattori}, Takashi and {Hayashi}, Masao and {Hayashi}, Yusuke and {He{\l}miniak}, Krzysztof G. and {Higuchi}, Ryo and {Hikage}, Chiaki and {Ho}, Paul T.~P. and {Hsieh}, Bau-Ching and {Huang}, Kuiyun and {Huang}, Song and {Ikeda}, Hiroyuki and {Imanishi}, Masatoshi and {Inoue}, Akio K. and {Iwasawa}, Kazushi and {Iwata}, Ikuru and {Jaelani}, Anton T. and {Jian}, Hung-Yu and {Kamata}, Yukiko and {Karoji}, Hiroshi and {Kashikawa}, Nobunari and {Katayama}, Nobuhiko and {Kawanomoto}, Satoshi and {Kayo}, Issha and {Koda}, Jin and {Koike}, Michitaro and {Kojima}, Takashi and {Komiyama}, Yutaka and {Konno}, Akira and {Koshida}, Shintaro and {Koyama}, Yusei and {Kusakabe}, Haruka and {Leauthaud}, Alexie and {Lee}, Chien-Hsiu and {Lin}, Lihwai and {Lin}, Yen-Ting and {Lupton}, Robert H. and {Mandelbaum}, Rachel and {Matsuoka}, Yoshiki and {Medezinski}, Elinor and {Mineo}, Sogo and {Miyama}, Shoken and {Miyatake}, Hironao and {Miyazaki}, Satoshi and {Momose}, Rieko and {More}, Anupreeta and {More}, Surhud and {Moritani}, Yuki and {Moriya}, Takashi J. and {Morokuma}, Tomoki and {Mukae}, Shiro and {Murata}, Ryoma and {Murayama}, Hitoshi and {Nagao}, Tohru and {Nakata}, Fumiaki and {Niida}, Mana and {Niikura}, Hiroko and {Nishizawa}, Atsushi J. and {Obuchi}, Yoshiyuki and {Oguri}, Masamune and {Oishi}, Yukie and {Okabe}, Nobuhiro and {Okamoto}, Sakurako and {Okura}, Yuki and {Ono}, Yoshiaki and {Onodera}, Masato and {Onoue}, Masafusa and {Osato}, Ken and {Ouchi}, Masami and {Price}, Paul A. and {Pyo}, Tae-Soo and {Sako}, Masao and {Sawicki}, Marcin and {Shibuya}, Takatoshi and {Shimasaku}, Kazuhiro and {Shimono}, Atsushi and {Shirasaki}, Masato and {Silverman}, John D. and {Simet}, Melanie and {Speagle}, Joshua and {Spergel}, David N. and {Strauss}, Michael A. and {Sugahara}, Yuma and {Sugiyama}, Naoshi and {Suto}, Yasushi and {Suyu}, Sherry H. and {Suzuki}, Nao and {Tait}, Philip J. and {Takada}, Masahiro and {Takata}, Tadafumi and {Tamura}, Naoyuki and {Tanaka}, Manobu M. and {Tanaka}, Masaomi and {Tanaka}, Masayuki and {Tanaka}, Yoko and {Terai}, Tsuyoshi and {Terashima}, Yuichi and {Toba}, Yoshiki and {Tominaga}, Nozomu and {Toshikawa}, Jun and {Turner}, Edwin L. and {Uchida}, Tomohisa and {Uchiyama}, Hisakazu and {Umetsu}, Keiichi and {Uraguchi}, Fumihiro and {Urata}, Yuji and {Usuda}, Tomonori and {Utsumi}, Yousuke and {Wang}, Shiang-Yu and {Wang}, Wei-Hao and {Wong}, Kenneth C. and {Yabe}, Kiyoto and {Yamada}, Yoshihiko and {Yamanoi}, Hitomi and {Yasuda}, Naoki and {Yeh}, Sherry and {Yonehara}, Atsunori and {Yuma}, Suraphong},
        title = "{The Hyper Suprime-Cam SSP Survey: Overview and survey design}",
      journal = {\pasj},
         year = 2018,
        month = jan,
       volume = {70},
          eid = {S4},
        pages = {S4},
          doi = {10.1093/pasj/psx066},
archivePrefix = {arXiv},
       eprint = {1704.05858},
 primaryClass = {astro-ph.IM},
       adsurl = {https://ui.adsabs.harvard.edu/abs/2018PASJ...70S...4A}
}

@ARTICLE{cochrane23,
       author = {{Cochrane}, R.~K. and {Angl{\'e}s-Alc{\'a}zar}, D. and {Mercedes-Feliz}, J. and {Hayward}, C.~C. and {Faucher-Gigu{\`e}re}, C. -A. and {Wellons}, S. and {Terrazas}, B.~A. and {Wetzel}, A. and {Hopkins}, P.~F. and {Moreno}, J. and {Su}, K. -Y. and {Somerville}, R.~S.},
        title = "{The impact of AGN-driven winds on physical and observable galaxy sizes}",
      journal = {\mnras},
         year = 2023,
        month = aug,
       volume = {523},
       number = {2},
        pages = {2409-2421},
          doi = {10.1093/mnras/stad1528},
archivePrefix = {arXiv},
       eprint = {2303.12858},
 primaryClass = {astro-ph.GA},
       adsurl = {https://ui.adsabs.harvard.edu/abs/2023MNRAS.523.2409C}
}

@ARTICLE{cochrane24,
       author = {{Cochrane}, R.~K. and {Angl{\'e}s-Alc{\'a}zar}, D. and {Cullen}, F. and {Hayward}, C.~C.},
        title = "{Disappearing Galaxies: The Orientation Dependence of JWST-bright, HST-dark, Star-forming Galaxy Selection}",
      journal = {\apj},
         year = 2024,
        month = jan,
       volume = {961},
       number = {1},
          eid = {37},
        pages = {37},
          doi = {10.3847/1538-4357/ad02f8},
archivePrefix = {arXiv},
       eprint = {2310.08829},
 primaryClass = {astro-ph.GA},
       adsurl = {https://ui.adsabs.harvard.edu/abs/2024ApJ...961...37C}
}

@ARTICLE{nanni24,
       author = {{Nanni}, Lorenza and {Neumann}, Justus and {Thomas}, Daniel and {Maraston}, Claudia and {Trayford}, James and {Lovell}, Christopher C. and {Law}, David R. and {Yan}, Renbin and {Chen}, Yanping},
        title = "{iMaNGA: mock MaNGA galaxies based on IllustrisTNG and MaStar SSPs. - III. Stellar metallicity drivers in MaNGA and TNG50}",
      journal = {\mnras},
         year = 2024,
        month = jan,
       volume = {527},
       number = {3},
        pages = {6419-6438},
          doi = {10.1093/mnras/stad3599},
archivePrefix = {arXiv},
       eprint = {2309.14257},
 primaryClass = {astro-ph.GA},
       adsurl = {https://ui.adsabs.harvard.edu/abs/2024MNRAS.527.6419N}
}

@ARTICLE{okegunn83,
       author = {{Oke}, J.~B. and {Gunn}, J.~E.},
        title = "{Secondary standard stars for absolute spectrophotometry.}",
      journal = {\apj},
         year = 1983,
        month = mar,
       volume = {266},
        pages = {713-717},
          doi = {10.1086/160817},
       adsurl = {https://ui.adsabs.harvard.edu/abs/1983ApJ...266..713O}
}

@ARTICLE{furlong15,
       author = {{Furlong}, M. and {Bower}, R.~G. and {Theuns}, T. and {Schaye}, J. and {Crain}, R.~A. and {Schaller}, M. and {Dalla Vecchia}, C. and {Frenk}, C.~S. and {McCarthy}, I.~G. and {Helly}, J. and {Jenkins}, A. and {Rosas-Guevara}, Y.~M.},
        title = "{Evolution of galaxy stellar masses and star formation rates in the EAGLE simulations}",
      journal = {\mnras},
         year = 2015,
        month = jul,
       volume = {450},
       number = {4},
        pages = {4486-4504},
          doi = {10.1093/mnras/stv852},
archivePrefix = {arXiv},
       eprint = {1410.3485},
 primaryClass = {astro-ph.GA},
       adsurl = {https://ui.adsabs.harvard.edu/abs/2015MNRAS.450.4486F}
}

@ARTICLE{eagle17,
       author = {{The EAGLE team}},
        title = "{The EAGLE simulations of galaxy formation: Public release of particle data}",
      journal = {arXiv e-prints},
         year = 2017,
        month = jun,
          eid = {arXiv:1706.09899},
        pages = {arXiv:1706.09899},
          doi = {10.48550/arXiv.1706.09899},
archivePrefix = {arXiv},
       eprint = {1706.09899},
 primaryClass = {astro-ph.GA},
       adsurl = {https://ui.adsabs.harvard.edu/abs/2017arXiv170609899T}
}

@ARTICLE{dubois14,
       author = {{Dubois}, Y. and {Pichon}, C. and {Welker}, C. and {Le Borgne}, D. and {Devriendt}, J. and {Laigle}, C. and {Codis}, S. and {Pogosyan}, D. and {Arnouts}, S. and {Benabed}, K. and {Bertin}, E. and {Blaizot}, J. and {Bouchet}, F. and {Cardoso}, J. -F. and {Colombi}, S. and {de Lapparent}, V. and {Desjacques}, V. and {Gavazzi}, R. and {Kassin}, S. and {Kimm}, T. and {McCracken}, H. and {Milliard}, B. and {Peirani}, S. and {Prunet}, S. and {Rouberol}, S. and {Silk}, J. and {Slyz}, A. and {Sousbie}, T. and {Teyssier}, R. and {Tresse}, L. and {Treyer}, M. and {Vibert}, D. and {Volonteri}, M.},
        title = "{Dancing in the dark: galactic properties trace spin swings along the cosmic web}",
      journal = {\mnras},
         year = 2014,
        month = oct,
       volume = {444},
       number = {2},
        pages = {1453-1468},
          doi = {10.1093/mnras/stu1227},
archivePrefix = {arXiv},
       eprint = {1402.1165},
 primaryClass = {astro-ph.CO},
       adsurl = {https://ui.adsabs.harvard.edu/abs/2014MNRAS.444.1453D}
}

@ARTICLE{dubois21,
       author = {{Dubois}, Yohan and {Beckmann}, Ricarda and {Bournaud}, Fr{\'e}d{\'e}ric and {Choi}, Hoseung and {Devriendt}, Julien and {Jackson}, Ryan and {Kaviraj}, Sugata and {Kimm}, Taysun and {Kraljic}, Katarina and {Laigle}, Clotilde and {Martin}, Garreth and {Park}, Min-Jung and {Peirani}, S{\'e}bastien and {Pichon}, Christophe and {Volonteri}, Marta and {Yi}, Sukyoung K.},
        title = "{Introducing the NEWHORIZON simulation: Galaxy properties with resolved internal dynamics across cosmic time}",
      journal = {\aap},
         year = 2021,
        month = jul,
       volume = {651},
          eid = {A109},
        pages = {A109},
          doi = {10.1051/0004-6361/202039429},
archivePrefix = {arXiv},
       eprint = {2009.10578},
 primaryClass = {astro-ph.GA},
       adsurl = {https://ui.adsabs.harvard.edu/abs/2021A&A...651A.109D}
}

@ARTICLE{dave19,
       author = {{Dav{\'e}}, Romeel and {Angl{\'e}s-Alc{\'a}zar}, Daniel and {Narayanan}, Desika and {Li}, Qi and {Rafieferantsoa}, Mika H. and {Appleby}, Sarah},
        title = "{SIMBA: Cosmological simulations with black hole growth and feedback}",
      journal = {\mnras},
         year = 2019,
        month = jun,
       volume = {486},
       number = {2},
        pages = {2827-2849},
          doi = {10.1093/mnras/stz937},
archivePrefix = {arXiv},
       eprint = {1901.10203},
 primaryClass = {astro-ph.GA},
       adsurl = {https://ui.adsabs.harvard.edu/abs/2019MNRAS.486.2827D}
}

@ARTICLE{schaye23,
       author = {{Schaye}, Joop and {Kugel}, Roi and {Schaller}, Matthieu and {Helly}, John C. and {Braspenning}, Joey and {Elbers}, Willem and {McCarthy}, Ian G. and {van Daalen}, Marcel P. and {Vandenbroucke}, Bert and {Frenk}, Carlos S. and {Kwan}, Juliana and {Salcido}, Jaime and {Bah{\'e}}, Yannick M. and {Borrow}, Josh and {Chaikin}, Evgenii and {Hahn}, Oliver and {Hu{\v{s}}ko}, Filip and {Jenkins}, Adrian and {Lacey}, Cedric G. and {Nobels}, Folkert S.~J.},
        title = "{The FLAMINGO project: cosmological hydrodynamical simulations for large-scale structure and galaxy cluster surveys}",
      journal = {\mnras},
         year = 2023,
        month = dec,
       volume = {526},
       number = {4},
        pages = {4978-5020},
          doi = {10.1093/mnras/stad2419},
archivePrefix = {arXiv},
       eprint = {2306.04024},
 primaryClass = {astro-ph.CO},
       adsurl = {https://ui.adsabs.harvard.edu/abs/2023MNRAS.526.4978S}
}

@ARTICLE{pakmor23,
       author = {{Pakmor}, R{\"u}diger and {Springel}, Volker and {Coles}, Jonathan P. and {Guillet}, Thomas and {Pfrommer}, Christoph and {Bose}, Sownak and {Barrera}, Monica and {Delgado}, Ana Maria and {Ferlito}, Fulvio and {Frenk}, Carlos and {Hadzhiyska}, Boryana and {Hern{\'a}ndez-Aguayo}, C{\'e}sar and {Hernquist}, Lars and {Kannan}, Rahul and {White}, Simon D.~M.},
        title = "{The MillenniumTNG Project: the hydrodynamical full physics simulation and a first look at its galaxy clusters}",
      journal = {\mnras},
         year = 2023,
        month = sep,
       volume = {524},
       number = {2},
        pages = {2539-2555},
          doi = {10.1093/mnras/stac3620},
archivePrefix = {arXiv},
       eprint = {2210.10060},
 primaryClass = {astro-ph.CO},
       adsurl = {https://ui.adsabs.harvard.edu/abs/2023MNRAS.524.2539P}
}

@ARTICLE{kannan22,
       author = {{Kannan}, R. and {Garaldi}, E. and {Smith}, A. and {Pakmor}, R. and {Springel}, V. and {Vogelsberger}, M. and {Hernquist}, L.},
        title = "{Introducing the THESAN project: radiation-magnetohydrodynamic simulations of the epoch of reionization}",
      journal = {\mnras},
         year = 2022,
        month = apr,
       volume = {511},
       number = {3},
        pages = {4005-4030},
          doi = {10.1093/mnras/stab3710},
archivePrefix = {arXiv},
       eprint = {2110.00584},
 primaryClass = {astro-ph.GA},
       adsurl = {https://ui.adsabs.harvard.edu/abs/2022MNRAS.511.4005K}
}

@ARTICLE{bird22,
       author = {{Bird}, Simeon and {Ni}, Yueying and {Di Matteo}, Tiziana and {Croft}, Rupert and {Feng}, Yu and {Chen}, Nianyi},
        title = "{The ASTRID simulation: galaxy formation and reionization}",
      journal = {\mnras},
         year = 2022,
        month = may,
       volume = {512},
       number = {3},
        pages = {3703-3716},
          doi = {10.1093/mnras/stac648},
archivePrefix = {arXiv},
       eprint = {2111.01160},
 primaryClass = {astro-ph.GA},
       adsurl = {https://ui.adsabs.harvard.edu/abs/2022MNRAS.512.3703B}
}

@ARTICLE{planck14,
       author = {{Planck Collaboration} and {Ade}, P.~A.~R. and {Aghanim}, N. and {Alves}, M.~I.~R. and {Armitage-Caplan}, C. and {Arnaud}, M. and {Ashdown}, M. and {Atrio-Barandela}, F. and {Aumont}, J. and {Aussel}, H. and {Baccigalupi}, C. and {Banday}, A.~J. and {Barreiro}, R.~B. and {Barrena}, R. and {Bartelmann}, M. and {Bartlett}, J.~G. and {Bartolo}, N. and {Basak}, S. and {Battaner}, E. and {Battye}, R. and {Benabed}, K. and {Beno{\^\i}t}, A. and {Benoit-L{\'e}vy}, A. and {Bernard}, J. -P. and {Bersanelli}, M. and {Bertincourt}, B. and {Bethermin}, M. and {Bielewicz}, P. and {Bikmaev}, I. and {Blanchard}, A. and {Bobin}, J. and {Bock}, J.~J. and {B{\"o}hringer}, H. and {Bonaldi}, A. and {Bonavera}, L. and {Bond}, J.~R. and {Borrill}, J. and {Bouchet}, F.~R. and {Boulanger}, F. and {Bourdin}, H. and {Bowyer}, J.~W. and {Bridges}, M. and {Brown}, M.~L. and {Bucher}, M. and {Burenin}, R. and {Burigana}, C. and {Butler}, R.~C. and {Calabrese}, E. and {Cappellini}, B. and {Cardoso}, J. -F. and {Carr}, R. and {Carvalho}, P. and {Casale}, M. and {Castex}, G. and {Catalano}, A. and {Challinor}, A. and {Chamballu}, A. and {Chary}, R. -R. and {Chen}, X. and {Chiang}, H.~C. and {Chiang}, L. -Y. and {Chon}, G. and {Christensen}, P.~R. and {Churazov}, E. and {Church}, S. and {Clemens}, M. and {Clements}, D.~L. and {Colombi}, S. and {Colombo}, L.~P.~L. and {Combet}, C. and {Comis}, B. and {Couchot}, F. and {Coulais}, A. and {Crill}, B.~P. and {Cruz}, M. and {Curto}, A. and {Cuttaia}, F. and {Da Silva}, A. and {Dahle}, H. and {Danese}, L. and {Davies}, R.~D. and {Davis}, R.~J. and {de Bernardis}, P. and {de Rosa}, A. and {de Zotti}, G. and {D{\'e}chelette}, T. and {Delabrouille}, J. and {Delouis}, J. -M. and {D{\'e}mocl{\`e}s}, J. and {D{\'e}sert}, F. -X. and {Dick}, J. and {Dickinson}, C. and {Diego}, J.~M. and {Dolag}, K. and {Dole}, H. and {Donzelli}, S. and {Dor{\'e}}, O. and {Douspis}, M. and {Ducout}, A. and {Dunkley}, J. and {Dupac}, X. and {Efstathiou}, G. and {Elsner}, F. and {En{\ss}lin}, T.~A. and {Eriksen}, H.~K. and {Fabre}, O. and {Falgarone}, E. and {Falvella}, M.~C. and {Fantaye}, Y. and {Fergusson}, J. and {Filliard}, C. and {Finelli}, F. and {Flores-Cacho}, I. and {Foley}, S. and {Forni}, O. and {Fosalba}, P. and {Frailis}, M. and {Fraisse}, A.~A. and {Franceschi}, E. and {Freschi}, M. and {Fromenteau}, S. and {Frommert}, M. and {Gaier}, T.~C. and {Galeotta}, S. and {Gallegos}, J. and {Galli}, S. and {Gandolfo}, B. and {Ganga}, K. and {Gauthier}, C. and {G{\'e}nova-Santos}, R.~T. and {Ghosh}, T. and {Giard}, M. and {Giardino}, G. and {Gilfanov}, M. and {Girard}, D. and {Giraud-H{\'e}raud}, Y. and {Gjerl{\o}w}, E. and {Gonz{\'a}lez-Nuevo}, J. and {G{\'o}rski}, K.~M. and {Gratton}, S. and {Gregorio}, A. and {Gruppuso}, A. and {Gudmundsson}, J.~E. and {Haissinski}, J. and {Hamann}, J. and {Hansen}, F.~K. and {Hansen}, M. and {Hanson}, D. and {Harrison}, D.~L. and {Heavens}, A. and {Helou}, G. and {Hempel}, A. and {Henrot-Versill{\'e}}, S. and {Hern{\'a}ndez-Monteagudo}, C. and {Herranz}, D. and {Hildebrandt}, S.~R. and {Hivon}, E. and {Ho}, S. and {Hobson}, M. and {Holmes}, W.~A. and {Hornstrup}, A. and {Hou}, Z. and {Hovest}, W. and {Huey}, G. and {Huffenberger}, K.~M. and {Hurier}, G. and {Ili{\'c}}, S. and {Jaffe}, A.~H. and {Jaffe}, T.~R. and {Jasche}, J. and {Jewell}, J. and {Jones}, W.~C. and {Juvela}, M. and {Kalberla}, P. and {Kangaslahti}, P. and {Keih{\"a}nen}, E. and {Kerp}, J. and {Keskitalo}, R. and {Khamitov}, I. and {Kiiveri}, K. and {Kim}, J. and {Kisner}, T.~S. and {Kneissl}, R. and {Knoche}, J. and {Knox}, L. and {Kunz}, M. and {Kurki-Suonio}, H. and {Lacasa}, F. and {Lagache}, G. and {L{\"a}hteenm{\"a}ki}, A. and {Lamarre}, J. -M. and {Langer}, M. and {Lasenby}, A. and {Lattanzi}, M. and {Laureijs}, R.~J. and {Lavabre}, A. and {Lawrence}, C.~R. and {Le Jeune}, M. and {Leach}, S. and {Leahy}, J.~P. and {Leonardi}, R. and {Le{\'o}n-Tavares}, J. and {Leroy}, C. and {Lesgourgues}, J. and {Lewis}, A. and {Li}, C. and {Liddle}, A. and {Liguori}, M. and {Lilje}, P.~B. and {Linden-V{\o}rnle}, M. and {Lindholm}, V. and {L{\'o}pez-Caniego}, M. and {Lowe}, S. and {Lubin}, P.~M. and {Mac{\'\i}as-P{\'e}rez}, J.~F. and {MacTavish}, C.~J. and {Maffei}, B. and {Maggio}, G. and {Maino}, D. and {Mandolesi}, N. and {Mangilli}, A. and {Marcos-Caballero}, A. and {Marinucci}, D. and {Maris}, M. and {Marleau}, F. and {Marshall}, D.~J. and {Martin}, P.~G. and {Mart{\'\i}nez-Gonz{\'a}lez}, E. and {Masi}, S. and {Massardi}, M. and {Matarrese}, S. and {Matsumura}, T. and {Matthai}, F. and {Maurin}, L. and {Mazzotta}, P. and {McDonald}, A. and {McEwen}, J.~D. and {McGehee}, P. and {Mei}, S. and {Meinhold}, P.~R. and {Melchiorri}, A. and {Melin}, J. -B. and {Mendes}, L. and {Menegoni}, E. and {Mennella}, A. and {Migliaccio}, M. and {Mikkelsen}, K. and {Millea}, M. and {Miniscalco}, R. and {Mitra}, S. and {Miville-Desch{\^e}nes}, M. -A. and {Molinari}, D. and {Moneti}, A. and {Montier}, L. and {Morgante}, G. and {Morisset}, N. and {Mortlock}, D. and {Moss}, A. and {Munshi}, D. and {Murphy}, J.~A. and {Naselsky}, P. and {Nati}, F. and {Natoli}, P. and {Negrello}, M. and {Nesvadba}, N.~P.~H. and {Netterfield}, C.~B. and {N{\o}rgaard-Nielsen}, H.~U. and {North}, C. and {Noviello}, F. and {Novikov}, D. and {Novikov}, I. and {O'Dwyer}, I.~J. and {Orieux}, F. and {Osborne}, S. and {O'Sullivan}, C. and {Oxborrow}, C.~A. and {Paci}, F. and {Pagano}, L. and {Pajot}, F. and {Paladini}, R. and {Pandolfi}, S. and {Paoletti}, D. and {Partridge}, B. and {Pasian}, F. and {Patanchon}, G. and {Paykari}, P. and {Pearson}, D. and {Pearson}, T.~J. and {Peel}, M. and {Peiris}, H.~V. and {Perdereau}, O. and {Perotto}, L. and {Perrotta}, F. and {Pettorino}, V. and {Piacentini}, F. and {Piat}, M. and {Pierpaoli}, E. and {Pietrobon}, D. and {Plaszczynski}, S. and {Platania}, P. and {Pogosyan}, D. and {Pointecouteau}, E. and {Polenta}, G. and {Ponthieu}, N. and {Popa}, L. and {Poutanen}, T. and {Pratt}, G.~W. and {Pr{\'e}zeau}, G. and {Prunet}, S. and {Puget}, J. -L. and {Pullen}, A.~R. and {Rachen}, J.~P. and {Racine}, B. and {Rahlin}, A. and {R{\"a}th}, C. and {Reach}, W.~T. and {Rebolo}, R. and {Reinecke}, M. and {Remazeilles}, M. and {Renault}, C. and {Renzi}, A. and {Riazuelo}, A. and {Ricciardi}, S. and {Riller}, T. and {Ringeval}, C. and {Ristorcelli}, I. and {Robbers}, G. and {Rocha}, G. and {Roman}, M. and {Rosset}, C. and {Rossetti}, M. and {Roudier}, G. and {Rowan-Robinson}, M. and {Rubi{\~n}o-Mart{\'\i}n}, J.~A. and {Ruiz-Granados}, B. and {Rusholme}, B. and {Salerno}, E. and {Sandri}, M. and {Sanselme}, L. and {Santos}, D. and {Savelainen}, M. and {Savini}, G. and {Schaefer}, B.~M. and {Schiavon}, F. and {Scott}, D. and {Seiffert}, M.~D. and {Serra}, P. and {Shellard}, E.~P.~S. and {Smith}, K. and {Smoot}, G.~F. and {Souradeep}, T. and {Spencer}, L.~D. and {Starck}, J. -L. and {Stolyarov}, V. and {Stompor}, R. and {Sudiwala}, R. and {Sunyaev}, R. and {Sureau}, F. and {Sutter}, P. and {Sutton}, D. and {Suur-Uski}, A. -S. and {Sygnet}, J. -F. and {Tauber}, J.~A. and {Tavagnacco}, D. and {Taylor}, D. and {Terenzi}, L. and {Texier}, D. and {Toffolatti}, L. and {Tomasi}, M. and {Torre}, J. -P. and {Tristram}, M. and {Tucci}, M. and {Tuovinen}, J. and {T{\"u}rler}, M. and {Tuttlebee}, M. and {Umana}, G. and {Valenziano}, L. and {Valiviita}, J. and {Van Tent}, B. and {Varis}, J. and {Vibert}, L. and {Viel}, M. and {Vielva}, P. and {Villa}, F. and {Vittorio}, N. and {Wade}, L.~A. and {Wandelt}, B.~D. and {Watson}, C. and {Watson}, R. and {Wehus}, I.~K. and {Welikala}, N. and {Weller}, J. and {White}, M. and {White}, S.~D.~M. and {Wilkinson}, A. and {Winkel}, B. and {Xia}, J. -Q. and {Yvon}, D. and {Zacchei}, A. and {Zibin}, J.~P. and {Zonca}, A.},
        title = "{Planck 2013 results. I. Overview of products and scientific results}",
      journal = {\aap},
         year = 2014,
        month = nov,
       volume = {571},
          eid = {A1},
        pages = {A1},
          doi = {10.1051/0004-6361/201321529},
archivePrefix = {arXiv},
       eprint = {1303.5062},
 primaryClass = {astro-ph.CO},
       adsurl = {https://ui.adsabs.harvard.edu/abs/2014A&A...571A...1P}
}

@ARTICLE{springel01,
       author = {{Springel}, Volker and {White}, Simon D.~M. and {Tormen}, Giuseppe and {Kauffmann}, Guinevere},
        title = "{Populating a cluster of galaxies - I. Results at z=0}",
      journal = {\mnras},
         year = 2001,
        month = dec,
       volume = {328},
       number = {3},
        pages = {726-750},
          doi = {10.1046/j.1365-8711.2001.04912.x},
archivePrefix = {arXiv},
       eprint = {astro-ph/0012055},
 primaryClass = {astro-ph},
       adsurl = {https://ui.adsabs.harvard.edu/abs/2001MNRAS.328..726S}
}

@ARTICLE{dolag09,
       author = {{Dolag}, K. and {Borgani}, S. and {Murante}, G. and {Springel}, V.},
        title = "{Substructures in hydrodynamical cluster simulations}",
      journal = {\mnras},
         year = 2009,
        month = oct,
       volume = {399},
       number = {2},
        pages = {497-514},
          doi = {10.1111/j.1365-2966.2009.15034.x},
archivePrefix = {arXiv},
       eprint = {0808.3401},
 primaryClass = {astro-ph},
       adsurl = {https://ui.adsabs.harvard.edu/abs/2009MNRAS.399..497D}
}

@ARTICLE{springel05,
       author = {{Springel}, Volker},
        title = "{The cosmological simulation code GADGET-2}",
      journal = {\mnras},
         year = 2005,
        month = dec,
       volume = {364},
       number = {4},
        pages = {1105-1134},
          doi = {10.1111/j.1365-2966.2005.09655.x},
archivePrefix = {arXiv},
       eprint = {astro-ph/0505010},
 primaryClass = {astro-ph},
       adsurl = {https://ui.adsabs.harvard.edu/abs/2005MNRAS.364.1105S}
}

@ARTICLE{springel10,
       author = {{Springel}, Volker},
        title = "{E pur si muove: Galilean-invariant cosmological hydrodynamical simulations on a moving mesh}",
      journal = {\mnras},
         year = 2010,
        month = jan,
       volume = {401},
       number = {2},
        pages = {791-851},
          doi = {10.1111/j.1365-2966.2009.15715.x},
archivePrefix = {arXiv},
       eprint = {0901.4107},
 primaryClass = {astro-ph.CO},
       adsurl = {https://ui.adsabs.harvard.edu/abs/2010MNRAS.401..791S}
}

@ARTICLE{planck16,
       author = {{Planck Collaboration} and {Adam}, R. and {Ade}, P.~A.~R. and {Aghanim}, N. and {Akrami}, Y. and {Alves}, M.~I.~R. and {Arg{\"u}eso}, F. and {Arnaud}, M. and {Arroja}, F. and {Ashdown}, M. and {Aumont}, J. and {Baccigalupi}, C. and {Ballardini}, M. and {Banday}, A.~J. and {Barreiro}, R.~B. and {Bartlett}, J.~G. and {Bartolo}, N. and {Basak}, S. and {Battaglia}, P. and {Battaner}, E. and {Battye}, R. and {Benabed}, K. and {Beno{\^\i}t}, A. and {Benoit-L{\'e}vy}, A. and {Bernard}, J. -P. and {Bersanelli}, M. and {Bertincourt}, B. and {Bielewicz}, P. and {Bikmaev}, I. and {Bock}, J.~J. and {B{\"o}hringer}, H. and {Bonaldi}, A. and {Bonavera}, L. and {Bond}, J.~R. and {Borrill}, J. and {Bouchet}, F.~R. and {Boulanger}, F. and {Bucher}, M. and {Burenin}, R. and {Burigana}, C. and {Butler}, R.~C. and {Calabrese}, E. and {Cardoso}, J. -F. and {Carvalho}, P. and {Casaponsa}, B. and {Castex}, G. and {Catalano}, A. and {Challinor}, A. and {Chamballu}, A. and {Chary}, R. -R. and {Chiang}, H.~C. and {Chluba}, J. and {Chon}, G. and {Christensen}, P.~R. and {Church}, S. and {Clemens}, M. and {Clements}, D.~L. and {Colombi}, S. and {Colombo}, L.~P.~L. and {Combet}, C. and {Comis}, B. and {Contreras}, D. and {Couchot}, F. and {Coulais}, A. and {Crill}, B.~P. and {Cruz}, M. and {Curto}, A. and {Cuttaia}, F. and {Danese}, L. and {Davies}, R.~D. and {Davis}, R.~J. and {de Bernardis}, P. and {de Rosa}, A. and {de Zotti}, G. and {Delabrouille}, J. and {Delouis}, J. -M. and {D{\'e}sert}, F. -X. and {Di Valentino}, E. and {Dickinson}, C. and {Diego}, J.~M. and {Dolag}, K. and {Dole}, H. and {Donzelli}, S. and {Dor{\'e}}, O. and {Douspis}, M. and {Ducout}, A. and {Dunkley}, J. and {Dupac}, X. and {Efstathiou}, G. and {Eisenhardt}, P.~R.~M. and {Elsner}, F. and {En{\ss}lin}, T.~A. and {Eriksen}, H.~K. and {Falgarone}, E. and {Fantaye}, Y. and {Farhang}, M. and {Feeney}, S. and {Fergusson}, J. and {Fernandez-Cobos}, R. and {Feroz}, F. and {Finelli}, F. and {Florido}, E. and {Forni}, O. and {Frailis}, M. and {Fraisse}, A.~A. and {Franceschet}, C. and {Franceschi}, E. and {Frejsel}, A. and {Frolov}, A. and {Galeotta}, S. and {Galli}, S. and {Ganga}, K. and {Gauthier}, C. and {G{\'e}nova-Santos}, R.~T. and {Gerbino}, M. and {Ghosh}, T. and {Giard}, M. and {Giraud-H{\'e}raud}, Y. and {Giusarma}, E. and {Gjerl{\o}w}, E. and {Gonz{\'a}lez-Nuevo}, J. and {G{\'o}rski}, K.~M. and {Grainge}, K.~J.~B. and {Gratton}, S. and {Gregorio}, A. and {Gruppuso}, A. and {Gudmundsson}, J.~E. and {Hamann}, J. and {Handley}, W. and {Hansen}, F.~K. and {Hanson}, D. and {Harrison}, D.~L. and {Heavens}, A. and {Helou}, G. and {Henrot-Versill{\'e}}, S. and {Hern{\'a}ndez-Monteagudo}, C. and {Herranz}, D. and {Hildebrandt}, S.~R. and {Hivon}, E. and {Hobson}, M. and {Holmes}, W.~A. and {Hornstrup}, A. and {Hovest}, W. and {Huang}, Z. and {Huffenberger}, K.~M. and {Hurier}, G. and {Ili{\'c}}, S. and {Jaffe}, A.~H. and {Jaffe}, T.~R. and {Jin}, T. and {Jones}, W.~C. and {Juvela}, M. and {Karakci}, A. and {Keih{\"a}nen}, E. and {Keskitalo}, R. and {Khamitov}, I. and {Kiiveri}, K. and {Kim}, J. and {Kisner}, T.~S. and {Kneissl}, R. and {Knoche}, J. and {Knox}, L. and {Krachmalnicoff}, N. and {Kunz}, M. and {Kurki-Suonio}, H. and {Lacasa}, F. and {Lagache}, G. and {L{\"a}hteenm{\"a}ki}, A. and {Lamarre}, J. -M. and {Langer}, M. and {Lasenby}, A. and {Lattanzi}, M. and {Lawrence}, C.~R. and {Le Jeune}, M. and {Leahy}, J.~P. and {Lellouch}, E. and {Leonardi}, R. and {Le{\'o}n-Tavares}, J. and {Lesgourgues}, J. and {Levrier}, F. and {Lewis}, A. and {Liguori}, M. and {Lilje}, P.~B. and {Lilley}, M. and {Linden-V{\o}rnle}, M. and {Lindholm}, V. and {Liu}, H. and {L{\'o}pez-Caniego}, M. and {Lubin}, P.~M. and {Ma}, Y. -Z. and {Mac{\'\i}as-P{\'e}rez}, J.~F. and {Maggio}, G. and {Maino}, D. and {Mak}, D.~S.~Y. and {Mandolesi}, N. and {Mangilli}, A. and {Marchini}, A. and {Marcos-Caballero}, A. and {Marinucci}, D. and {Maris}, M. and {Marshall}, D.~J. and {Martin}, P.~G. and {Martinelli}, M. and {Mart{\'\i}nez-Gonz{\'a}lez}, E. and {Masi}, S. and {Matarrese}, S. and {Mazzotta}, P. and {McEwen}, J.~D. and {McGehee}, P. and {Mei}, S. and {Meinhold}, P.~R. and {Melchiorri}, A. and {Melin}, J. -B. and {Mendes}, L. and {Mennella}, A. and {Migliaccio}, M. and {Mikkelsen}, K. and {Millea}, M. and {Mitra}, S. and {Miville-Desch{\^e}nes}, M. -A. and {Molinari}, D. and {Moneti}, A. and {Montier}, L. and {Moreno}, R. and {Morgante}, G. and {Mortlock}, D. and {Moss}, A. and {Mottet}, S. and {M{\"u}nchmeyer}, M. and {Munshi}, D. and {Murphy}, J.~A. and {Narimani}, A. and {Naselsky}, P. and {Nastasi}, A. and {Nati}, F. and {Natoli}, P. and {Negrello}, M. and {Netterfield}, C.~B. and {N{\o}rgaard-Nielsen}, H.~U. and {Noviello}, F. and {Novikov}, D. and {Novikov}, I. and {Olamaie}, M. and {Oppermann}, N. and {Orlando}, E. and {Oxborrow}, C.~A. and {Paci}, F. and {Pagano}, L. and {Pajot}, F. and {Paladini}, R. and {Pandolfi}, S. and {Paoletti}, D. and {Partridge}, B. and {Pasian}, F. and {Patanchon}, G. and {Pearson}, T.~J. and {Peel}, M. and {Peiris}, H.~V. and {Pelkonen}, V. -M. and {Perdereau}, O. and {Perotto}, L. and {Perrott}, Y.~C. and {Perrotta}, F. and {Pettorino}, V. and {Piacentini}, F. and {Piat}, M. and {Pierpaoli}, E. and {Pietrobon}, D. and {Plaszczynski}, S. and {Pogosyan}, D. and {Pointecouteau}, E. and {Polenta}, G. and {Popa}, L. and {Pratt}, G.~W. and {Pr{\'e}zeau}, G. and {Prunet}, S. and {Puget}, J. -L. and {Rachen}, J.~P. and {Racine}, B. and {Reach}, W.~T. and {Rebolo}, R. and {Reinecke}, M. and {Remazeilles}, M. and {Renault}, C. and {Renzi}, A. and {Ristorcelli}, I. and {Rocha}, G. and {Roman}, M. and {Romelli}, E. and {Rosset}, C. and {Rossetti}, M. and {Rotti}, A. and {Roudier}, G. and {Rouill{\'e} d'Orfeuil}, B. and {Rowan-Robinson}, M. and {Rubi{\~n}o-Mart{\'\i}n}, J.~A. and {Ruiz-Granados}, B. and {Rumsey}, C. and {Rusholme}, B. and {Said}, N. and {Salvatelli}, V. and {Salvati}, L. and {Sandri}, M. and {Sanghera}, H.~S. and {Santos}, D. and {Saunders}, R.~D.~E. and {Sauv{\'e}}, A. and {Savelainen}, M. and {Savini}, G. and {Schaefer}, B.~M. and {Schammel}, M.~P. and {Scott}, D. and {Seiffert}, M.~D. and {Serra}, P. and {Shellard}, E.~P.~S. and {Shimwell}, T.~W. and {Shiraishi}, M. and {Smith}, K. and {Souradeep}, T. and {Spencer}, L.~D. and {Spinelli}, M. and {Stanford}, S.~A. and {Stern}, D. and {Stolyarov}, V. and {Stompor}, R. and {Strong}, A.~W. and {Sudiwala}, R. and {Sunyaev}, R. and {Sutter}, P. and {Sutton}, D. and {Suur-Uski}, A. -S. and {Sygnet}, J. -F. and {Tauber}, J.~A. and {Tavagnacco}, D. and {Terenzi}, L. and {Texier}, D. and {Toffolatti}, L. and {Tomasi}, M. and {Tornikoski}, M. and {Tramonte}, D. and {Tristram}, M. and {Troja}, A. and {Trombetti}, T. and {Tucci}, M. and {Tuovinen}, J. and {T{\"u}rler}, M. and {Umana}, G. and {Valenziano}, L. and {Valiviita}, J. and {Van Tent}, F. and {Vassallo}, T. and {Vibert}, L. and {Vidal}, M. and {Viel}, M. and {Vielva}, P. and {Villa}, F. and {Wade}, L.~A. and {Walter}, B. and {Wandelt}, B.~D. and {Watson}, R. and {Wehus}, I.~K. and {Welikala}, N. and {Weller}, J. and {White}, M. and {White}, S.~D.~M. and {Wilkinson}, A. and {Yvon}, D. and {Zacchei}, A. and {Zibin}, J.~P. and {Zonca}, A.},
        title = "{Planck 2015 results. I. Overview of products and scientific results}",
      journal = {\aap},
         year = 2016,
        month = sep,
       volume = {594},
          eid = {A1},
        pages = {A1},
          doi = {10.1051/0004-6361/201527101},
archivePrefix = {arXiv},
       eprint = {1502.01582},
 primaryClass = {astro-ph.CO},
       adsurl = {https://ui.adsabs.harvard.edu/abs/2016A&A...594A...1P}
}

@ARTICLE{schayedallavecchia08,
       author = {{Schaye}, Joop and {Dalla Vecchia}, Claudio},
        title = "{On the relation between the Schmidt and Kennicutt-Schmidt star formation laws and its implications for numerical simulations}",
      journal = {\mnras},
         year = 2008,
        month = jan,
       volume = {383},
       number = {3},
        pages = {1210-1222},
          doi = {10.1111/j.1365-2966.2007.12639.x},
archivePrefix = {arXiv},
       eprint = {0709.0292},
 primaryClass = {astro-ph},
       adsurl = {https://ui.adsabs.harvard.edu/abs/2008MNRAS.383.1210S}
}

@ARTICLE{dallavecchiaschaye12,
       author = {{Dalla Vecchia}, Claudio and {Schaye}, Joop},
        title = "{Simulating galactic outflows with thermal supernova feedback}",
      journal = {\mnras},
         year = 2012,
        month = oct,
       volume = {426},
       number = {1},
        pages = {140-158},
          doi = {10.1111/j.1365-2966.2012.21704.x},
archivePrefix = {arXiv},
       eprint = {1203.5667},
 primaryClass = {astro-ph.GA},
       adsurl = {https://ui.adsabs.harvard.edu/abs/2012MNRAS.426..140D}
}

@ARTICLE{boothscaye09,
       author = {{Booth}, C.~M. and {Schaye}, Joop},
        title = "{Cosmological simulations of the growth of supermassive black holes and feedback from active galactic nuclei: method and tests}",
      journal = {\mnras},
         year = 2009,
        month = sep,
       volume = {398},
       number = {1},
        pages = {53-74},
          doi = {10.1111/j.1365-2966.2009.15043.x},
archivePrefix = {arXiv},
       eprint = {0904.2572},
 primaryClass = {astro-ph.CO},
       adsurl = {https://ui.adsabs.harvard.edu/abs/2009MNRAS.398...53B}
}

@ARTICLE{springelhernquist03,
       author = {{Springel}, Volker and {Hernquist}, Lars},
        title = "{Cosmological smoothed particle hydrodynamics simulations: a hybrid multiphase model for star formation}",
      journal = {\mnras},
         year = 2003,
        month = feb,
       volume = {339},
       number = {2},
        pages = {289-311},
          doi = {10.1046/j.1365-8711.2003.06206.x},
archivePrefix = {arXiv},
       eprint = {astro-ph/0206393},
 primaryClass = {astro-ph},
       adsurl = {https://ui.adsabs.harvard.edu/abs/2003MNRAS.339..289S}
}

@ARTICLE{pakmorspringel13,
       author = {{Pakmor}, R{\"u}diger and {Springel}, Volker},
        title = "{Simulations of magnetic fields in isolated disc galaxies}",
      journal = {\mnras},
         year = 2013,
        month = jun,
       volume = {432},
       number = {1},
        pages = {176-193},
          doi = {10.1093/mnras/stt428},
archivePrefix = {arXiv},
       eprint = {1212.1452},
 primaryClass = {astro-ph.CO},
       adsurl = {https://ui.adsabs.harvard.edu/abs/2013MNRAS.432..176P}
}

@ARTICLE{torrey14,
       author = {{Torrey}, Paul and {Vogelsberger}, Mark and {Genel}, Shy and {Sijacki}, Debora and {Springel}, Volker and {Hernquist}, Lars},
        title = "{A model for cosmological simulations of galaxy formation physics: multi-epoch validation}",
      journal = {\mnras},
         year = 2014,
        month = mar,
       volume = {438},
       number = {3},
        pages = {1985-2004},
          doi = {10.1093/mnras/stt2295},
archivePrefix = {arXiv},
       eprint = {1305.4931},
 primaryClass = {astro-ph.CO},
       adsurl = {https://ui.adsabs.harvard.edu/abs/2014MNRAS.438.1985T}
}

@ARTICLE{santini21,
       author = {{Santini}, P. and {Castellano}, M. and {Merlin}, E. and {Fontana}, A. and {Fortuni}, F. and {Kodra}, D. and {Magnelli}, B. and {Menci}, N. and {Calabr{\`o}}, A. and {Lovell}, C.~C. and {Pentericci}, L. and {Testa}, V. and {Wilkins}, S.~M.},
        title = "{The emergence of passive galaxies in the early Universe}",
      journal = {\aap},
         year = 2021,
        month = aug,
       volume = {652},
          eid = {A30},
        pages = {A30},
          doi = {10.1051/0004-6361/202039738},
archivePrefix = {arXiv},
       eprint = {2011.10584},
 primaryClass = {astro-ph.GA},
       adsurl = {https://ui.adsabs.harvard.edu/abs/2021A&A...652A..30S}
}

@ARTICLE{merlin19,
       author = {{Merlin}, E. and {Fortuni}, F. and {Torelli}, M. and {Santini}, P. and {Castellano}, M. and {Fontana}, A. and {Grazian}, A. and {Pentericci}, L. and {Pilo}, S. and {Schmidt}, K.~B.},
        title = "{Red and dead CANDELS: massive passive galaxies at the dawn of the Universe}",
      journal = {\mnras},
         year = 2019,
        month = dec,
       volume = {490},
       number = {3},
        pages = {3309-3328},
          doi = {10.1093/mnras/stz2615},
archivePrefix = {arXiv},
       eprint = {1909.07996},
 primaryClass = {astro-ph.GA},
       adsurl = {https://ui.adsabs.harvard.edu/abs/2019MNRAS.490.3309M}
}

@ARTICLE{genel14,
       author = {{Genel}, Shy and {Vogelsberger}, Mark and {Springel}, Volker and {Sijacki}, Debora and {Nelson}, Dylan and {Snyder}, Greg and {Rodriguez-Gomez}, Vicente and {Torrey}, Paul and {Hernquist}, Lars},
        title = "{Introducing the Illustris project: the evolution of galaxy populations across cosmic time}",
      journal = {\mnras},
         year = 2014,
        month = nov,
       volume = {445},
       number = {1},
        pages = {175-200},
          doi = {10.1093/mnras/stu1654},
archivePrefix = {arXiv},
       eprint = {1405.3749},
 primaryClass = {astro-ph.CO},
       adsurl = {https://ui.adsabs.harvard.edu/abs/2014MNRAS.445..175G}
}

@ARTICLE{vogelsberger14,
       author = {{Vogelsberger}, Mark and {Genel}, Shy and {Springel}, Volker and {Torrey}, Paul and {Sijacki}, Debora and {Xu}, Dandan and {Snyder}, Greg and {Nelson}, Dylan and {Hernquist}, Lars},
        title = "{Introducing the Illustris Project: simulating the coevolution of dark and visible matter in the Universe}",
      journal = {\mnras},
         year = 2014,
        month = oct,
       volume = {444},
       number = {2},
        pages = {1518-1547},
          doi = {10.1093/mnras/stu1536},
archivePrefix = {arXiv},
       eprint = {1405.2921},
 primaryClass = {astro-ph.CO},
       adsurl = {https://ui.adsabs.harvard.edu/abs/2014MNRAS.444.1518V}
}

@ARTICLE{sijacki15,
       author = {{Sijacki}, Debora and {Vogelsberger}, Mark and {Genel}, Shy and {Springel}, Volker and {Torrey}, Paul and {Snyder}, Gregory F. and {Nelson}, Dylan and {Hernquist}, Lars},
        title = "{The Illustris simulation: the evolving population of black holes across cosmic time}",
      journal = {\mnras},
         year = 2015,
        month = sep,
       volume = {452},
       number = {1},
        pages = {575-596},
          doi = {10.1093/mnras/stv1340},
archivePrefix = {arXiv},
       eprint = {1408.6842},
 primaryClass = {astro-ph.GA},
       adsurl = {https://ui.adsabs.harvard.edu/abs/2015MNRAS.452..575S}
}

@ARTICLE{wright24,
       author = {{Wright}, Ruby J. and {Somerville}, Rachel S. and {Lagos}, Claudia del P. and {Schaller}, Matthieu and {Dav{\'e}}, Romeel and {Angl{\'e}s-Alc{\'a}zar}, Daniel and {Genel}, Shy},
        title = "{The baryon cycle in modern cosmological hydrodynamical simulations}",
      journal = {\mnras},
         year = 2024,
        month = aug,
       volume = {532},
       number = {3},
        pages = {3417-3440},
          doi = {10.1093/mnras/stae1688},
archivePrefix = {arXiv},
       eprint = {2402.08408},
 primaryClass = {astro-ph.GA},
       adsurl = {https://ui.adsabs.harvard.edu/abs/2024MNRAS.532.3417W}
}

@ARTICLE{evrard88,
       author = {{Evrard}, August E.},
        title = "{Beyond N-body: 3D cosmological gas dynamics.}",
      journal = {\mnras},
         year = 1988,
        month = dec,
       volume = {235},
        pages = {911-934},
          doi = {10.1093/mnras/235.3.911},
       adsurl = {https://ui.adsabs.harvard.edu/abs/1988MNRAS.235..911E}
}

@ARTICLE{hernquistkatz89,
       author = {{Hernquist}, Lars and {Katz}, Neal},
        title = "{TREESPH: A Unification of SPH with the Hierarchical Tree Method}",
      journal = {\apjs},
         year = 1989,
        month = jun,
       volume = {70},
        pages = {419},
          doi = {10.1086/191344},
       adsurl = {https://ui.adsabs.harvard.edu/abs/1989ApJS...70..419H}
}

@ARTICLE{springelhernquist02,
       author = {{Springel}, Volker and {Hernquist}, Lars},
        title = "{Cosmological smoothed particle hydrodynamics simulations: the entropy equation}",
      journal = {\mnras},
         year = 2002,
        month = jul,
       volume = {333},
       number = {3},
        pages = {649-664},
          doi = {10.1046/j.1365-8711.2002.05445.x},
archivePrefix = {arXiv},
       eprint = {astro-ph/0111016},
 primaryClass = {astro-ph},
       adsurl = {https://ui.adsabs.harvard.edu/abs/2002MNRAS.333..649S}
}

@ARTICLE{dave20,
       author = {{Dav{\'e}}, Romeel and {Crain}, Robert A. and {Stevens}, Adam R.~H. and {Narayanan}, Desika and {Saintonge}, Amelie and {Catinella}, Barbara and {Cortese}, Luca},
        title = "{Galaxy cold gas contents in modern cosmological hydrodynamic simulations}",
      journal = {\mnras},
         year = 2020,
        month = sep,
       volume = {497},
       number = {1},
        pages = {146-166},
          doi = {10.1093/mnras/staa1894},
archivePrefix = {arXiv},
       eprint = {2002.07226},
 primaryClass = {astro-ph.GA},
       adsurl = {https://ui.adsabs.harvard.edu/abs/2020MNRAS.497..146D}
}

@ARTICLE{weaver23,
       author = {{Weaver}, J.~R. and {Davidzon}, I. and {Toft}, S. and {Ilbert}, O. and {McCracken}, H.~J. and {Gould}, K.~M.~L. and {Jespersen}, C.~K. and {Steinhardt}, C. and {Lagos}, C.~D.~P. and {Capak}, P.~L. and {Casey}, C.~M. and {Chartab}, N. and {Faisst}, A.~L. and {Hayward}, C.~C. and {Kartaltepe}, J.~S. and {Kauffmann}, O.~B. and {Koekemoer}, A.~M. and {Kokorev}, V. and {Laigle}, C. and {Liu}, D. and {Long}, A. and {Magdis}, G.~E. and {McPartland}, C.~J.~R. and {Milvang-Jensen}, B. and {Mobasher}, B. and {Moneti}, A. and {Peng}, Y. and {Sanders}, D.~B. and {Shuntov}, M. and {Sneppen}, A. and {Valentino}, F. and {Zalesky}, L. and {Zamorani}, G.},
        title = "{COSMOS2020: The galaxy stellar mass function. The assembly and star formation cessation of galaxies at 0.2< z {\ensuremath{\leq}} 7.5}",
      journal = {\aap},
         year = 2023,
        month = sep,
       volume = {677},
          eid = {A184},
        pages = {A184},
          doi = {10.1051/0004-6361/202245581},
archivePrefix = {arXiv},
       eprint = {2212.02512},
 primaryClass = {astro-ph.GA},
       adsurl = {https://ui.adsabs.harvard.edu/abs/2023A&A...677A.184W}
}

@ARTICLE{snyder15,
       author = {{Snyder}, Gregory F. and {Torrey}, Paul and {Lotz}, Jennifer M. and {Genel}, Shy and {McBride}, Cameron K. and {Vogelsberger}, Mark and {Pillepich}, Annalisa and {Nelson}, Dylan and {Sales}, Laura V. and {Sijacki}, Debora and {Hernquist}, Lars and {Springel}, Volker},
        title = "{Galaxy morphology and star formation in the Illustris Simulation at z = 0}",
      journal = {\mnras},
         year = 2015,
        month = dec,
       volume = {454},
       number = {2},
        pages = {1886-1908},
          doi = {10.1093/mnras/stv2078},
archivePrefix = {arXiv},
       eprint = {1502.07747},
 primaryClass = {astro-ph.GA},
       adsurl = {https://ui.adsabs.harvard.edu/abs/2015MNRAS.454.1886S}
}

@ARTICLE{nayyeri17,
       author = {{Nayyeri}, H. and {Hemmati}, S. and {Mobasher}, B. and {Ferguson}, H.~C. and {Cooray}, A. and {Barro}, G. and {Faber}, S.~M. and {Dickinson}, M. and {Koekemoer}, A.~M. and {Peth}, M. and {Salvato}, M. and {Ashby}, M.~L.~N. and {Darvish}, B. and {Donley}, J. and {Durbin}, M. and {Finkelstein}, S. and {Fontana}, A. and {Grogin}, N.~A. and {Gruetzbauch}, R. and {Huang}, K. and {Khostovan}, A.~A. and {Kocevski}, D. and {Kodra}, D. and {Lee}, B. and {Newman}, J. and {Pacifici}, C. and {Pforr}, J. and {Stefanon}, M. and {Wiklind}, T. and {Willner}, S.~P. and {Wuyts}, S. and {Castellano}, M. and {Conselice}, C. and {Dolch}, T. and {Dunlop}, J.~S. and {Galametz}, A. and {Hathi}, N.~P. and {Lucas}, R.~A. and {Yan}, H.},
        title = "{CANDELS Multi-wavelength Catalogs: Source Identification and Photometry in the CANDELS COSMOS Survey Field}",
      journal = {\apjs},
         year = 2017,
        month = jan,
       volume = {228},
       number = {1},
          eid = {7},
        pages = {7},
          doi = {10.3847/1538-4365/228/1/7},
archivePrefix = {arXiv},
       eprint = {1612.07364},
 primaryClass = {astro-ph.GA},
       adsurl = {https://ui.adsabs.harvard.edu/abs/2017ApJS..228....7N}
}

@ARTICLE{stefanon17,
       author = {{Stefanon}, Mauro and {Yan}, Haojing and {Mobasher}, Bahram and {Barro}, Guillermo and {Donley}, Jennifer L. and {Fontana}, Adriano and {Hemmati}, Shoubaneh and {Koekemoer}, Anton M. and {Lee}, BoMee and {Lee}, Seong-Kook and {Nayyeri}, Hooshang and {Peth}, Michael and {Pforr}, Janine and {Salvato}, Mara and {Wiklind}, Tommy and {Wuyts}, Stijn and {Ashby}, Matthew L.~N. and {Castellano}, Marco and {Conselice}, Christopher J. and {Cooper}, Michael C. and {Cooray}, Asantha R. and {Dolch}, Timothy and {Ferguson}, Henry and {Galametz}, Audrey and {Giavalisco}, Mauro and {Guo}, Yicheng and {Willner}, Steven P. and {Dickinson}, Mark E. and {Faber}, Sandra M. and {Fazio}, Giovanni G. and {Gardner}, Jonathan P. and {Gawiser}, Eric and {Grazian}, Andrea and {Grogin}, Norman A. and {Kocevski}, Dale and {Koo}, David C. and {Lee}, Kyoung-Soo and {Lucas}, Ray A. and {McGrath}, Elizabeth J. and {Nandra}, Kirpal and {Newman}, Jeffrey A. and {van der Wel}, Arjen},
        title = "{CANDELS Multi-wavelength Catalogs: Source Identification and Photometry in the CANDELS Extended Groth Strip}",
      journal = {\apjs},
         year = 2017,
        month = apr,
       volume = {229},
       number = {2},
          eid = {32},
        pages = {32},
          doi = {10.3847/1538-4365/aa66cb},
archivePrefix = {arXiv},
       eprint = {1703.05768},
 primaryClass = {astro-ph.GA},
       adsurl = {https://ui.adsabs.harvard.edu/abs/2017ApJS..229...32S}
}

@ARTICLE{barro19,
       author = {{Barro}, Guillermo and {P{\'e}rez-Gonz{\'a}lez}, Pablo G. and {Cava}, Antonio and {Brammer}, Gabriel and {Pandya}, Viraj and {Eliche Moral}, Carmen and {Esquej}, Pilar and {Dom{\'\i}nguez-S{\'a}nchez}, Helena and {Alcalde Pampliega}, Belen and {Guo}, Yicheng and {Koekemoer}, Anton M. and {Trump}, Jonathan R. and {Ashby}, Matthew L.~N. and {Cardiel}, Nicolas and {Castellano}, Marco and {Conselice}, Christopher J. and {Dickinson}, Mark E. and {Dolch}, Timothy and {Donley}, Jennifer L. and {Espino Briones}, N{\'e}stor and {Faber}, Sandra M. and {Fazio}, Giovanni G. and {Ferguson}, Henry and {Finkelstein}, Steve and {Fontana}, Adriano and {Galametz}, Audrey and {Gardner}, Jonathan P. and {Gawiser}, Eric and {Giavalisco}, Mauro and {Grazian}, Andrea and {Grogin}, Norman A. and {Hathi}, Nimish P. and {Hemmati}, Shoubaneh and {Hern{\'a}n-Caballero}, Antonio and {Kocevski}, Dale and {Koo}, David C. and {Kodra}, Dritan and {Lee}, Kyoung-Soo and {Lin}, Lihwai and {Lucas}, Ray A. and {Mobasher}, Bahram and {McGrath}, Elizabeth J. and {Nandra}, Kirpal and {Nayyeri}, Hooshang and {Newman}, Jeffrey A. and {Pforr}, Janine and {Peth}, Michael and {Rafelski}, Marc and {Rodr{\'\i}guez-Munoz}, Lucia and {Salvato}, Mara and {Stefanon}, Mauro and {van der Wel}, Arjen and {Willner}, Steven P. and {Wiklind}, Tommy and {Wuyts}, Stijn},
        title = "{The CANDELS/SHARDS Multiwavelength Catalog in GOODS-N: Photometry, Photometric Redshifts, Stellar Masses, Emission-line Fluxes, and Star Formation Rates}",
      journal = {\apjs},
         year = 2019,
        month = aug,
       volume = {243},
       number = {2},
          eid = {22},
        pages = {22},
          doi = {10.3847/1538-4365/ab23f2},
archivePrefix = {arXiv},
       eprint = {1908.00569},
 primaryClass = {astro-ph.GA},
       adsurl = {https://ui.adsabs.harvard.edu/abs/2019ApJS..243...22B}
}

@ARTICLE{merlin15,
       author = {{Merlin}, E. and {Fontana}, A. and {Ferguson}, H.~C. and {Dunlop}, J.~S. and {Elbaz}, D. and {Bourne}, N. and {Bruce}, V.~A. and {Buitrago}, F. and {Castellano}, M. and {Schreiber}, C. and {Grazian}, A. and {McLure}, R.~J. and {Okumura}, K. and {Shu}, X. and {Wang}, T. and {Amor{\'\i}n}, R. and {Boutsia}, K. and {Cappelluti}, N. and {Comastri}, A. and {Derriere}, S. and {Faber}, S.~M. and {Santini}, P.},
        title = "{T-PHOT: A new code for PSF-matched, prior-based, multiwavelength extragalactic deconfusion photometry}",
      journal = {\aap},
         year = 2015,
        month = oct,
       volume = {582},
          eid = {A15},
        pages = {A15},
          doi = {10.1051/0004-6361/201526471},
archivePrefix = {arXiv},
       eprint = {1505.02516},
 primaryClass = {astro-ph.IM},
       adsurl = {https://ui.adsabs.harvard.edu/abs/2015A&A...582A..15M}
}

@ARTICLE{merlin16,
       author = {{Merlin}, E. and {Bourne}, N. and {Castellano}, M. and {Ferguson}, H.~C. and {Wang}, T. and {Derriere}, S. and {Dunlop}, J.~S. and {Elbaz}, D. and {Fontana}, A.},
        title = "{T-PHOT version 2.0: Improved algorithms for background subtraction, local convolution, kernel registration, and new options}",
      journal = {\aap},
         year = 2016,
        month = nov,
       volume = {595},
          eid = {A97},
        pages = {A97},
          doi = {10.1051/0004-6361/201628751},
archivePrefix = {arXiv},
       eprint = {1609.00146},
 primaryClass = {astro-ph.IM},
       adsurl = {https://ui.adsabs.harvard.edu/abs/2016A&A...595A..97M}
}

@ARTICLE{merlin16_ff,
       author = {{Merlin}, E. and {Amor{\'\i}n}, R. and {Castellano}, M. and {Fontana}, A. and {Buitrago}, F. and {Dunlop}, J.~S. and {Elbaz}, D. and {Boucaud}, A. and {Bourne}, N. and {Boutsia}, K. and {Brammer}, G. and {Bruce}, V.~A. and {Capak}, P. and {Cappelluti}, N. and {Ciesla}, L. and {Comastri}, A. and {Cullen}, F. and {Derriere}, S. and {Faber}, S.~M. and {Ferguson}, H.~C. and {Giallongo}, E. and {Grazian}, A. and {Lotz}, J. and {Micha{\l}owski}, M.~J. and {Paris}, D. and {Pentericci}, L. and {Pilo}, S. and {Santini}, P. and {Schreiber}, C. and {Shu}, X. and {Wang}, T.},
        title = "{The ASTRODEEP Frontier Fields catalogues. I. Multiwavelength photometry of Abell-2744 and MACS-J0416}",
      journal = {\aap},
         year = 2016,
        month = may,
       volume = {590},
          eid = {A30},
        pages = {A30},
          doi = {10.1051/0004-6361/201527513},
archivePrefix = {arXiv},
       eprint = {1603.02460},
 primaryClass = {astro-ph.GA},
       adsurl = {https://ui.adsabs.harvard.edu/abs/2016A&A...590A..30M}
}

@ARTICLE{merlin22,
       author = {{Merlin}, Emiliano and {Bonchi}, Andrea and {Paris}, Diego and {Belfiori}, Davide and {Fontana}, Adriano and {Castellano}, Marco and {Nonino}, Mario and {Polenta}, Gianluca and {Santini}, Paola and {Yang}, Lilan and {Glazebrook}, Karl and {Treu}, Tommaso and {Roberts-Borsani}, Guido and {Trenti}, Michele and {Birrer}, Simon and {Brammer}, Gabriel and {Grillo}, Claudio and {Calabr{\`o}}, Antonello and {Marchesini}, Danilo and {Mason}, Charlotte and {Mercurio}, Amata and {Morishita}, Takahiro and {Strait}, Victoria and {Boyett}, Kristan and {Leethochawalit}, Nicha and {Nanayakkara}, Themiya and {Vulcani}, Benedetta and {Bradac}, Marusa and {Wang}, Xin},
        title = "{Early Results from GLASS-JWST. II. NIRCam Extragalactic Imaging and Photometric Catalog}",
      journal = {\apjl},
         year = 2022,
        month = oct,
       volume = {938},
       number = {2},
          eid = {L14},
        pages = {L14},
          doi = {10.3847/2041-8213/ac8f93},
archivePrefix = {arXiv},
       eprint = {2207.11701},
 primaryClass = {astro-ph.GA},
       adsurl = {https://ui.adsabs.harvard.edu/abs/2022ApJ...938L..14M}
}

@ARTICLE{paris23,
       author = {{Paris}, Diego and {Merlin}, Emiliano and {Fontana}, Adriano and {Bonchi}, Andrea and {Brammer}, Gabriel and {Correnti}, Matteo and {Treu}, Tommaso and {Boyett}, Kristan and {Calabr{\`o}}, Antonello and {Castellano}, Marco and {Chen}, Wenlei and {Yang}, Lilan and {Glazebrook}, Karl and {Kelly}, Patrick and {Koekemoer}, Anton M. and {Leethochawalit}, Nicha and {Mascia}, Sara and {Mason}, Charlotte and {Morishita}, Takahiro and {Nonino}, Mario and {Pentericci}, Laura and {Polenta}, Gianluca and {Roberts-Borsani}, Guido and {Santini}, Paola and {Trenti}, Michele and {Vanzella}, Eros and {Vulcani}, Benedetta and {Windhorst}, Rogier A. and {Nanayakkara}, Themiya and {Wang}, Xin},
        title = "{The GLASS-JWST Early Release Science Program. II. Stage I Release of NIRCam Imaging and Catalogs in the Abell 2744 Region}",
      journal = {\apj},
         year = 2023,
        month = jul,
       volume = {952},
       number = {1},
          eid = {20},
        pages = {20},
          doi = {10.3847/1538-4357/acda8a},
archivePrefix = {arXiv},
       eprint = {2301.02179},
 primaryClass = {astro-ph.GA},
       adsurl = {https://ui.adsabs.harvard.edu/abs/2023ApJ...952...20P}
}

@ARTICLE{fontana00,
       author = {{Fontana}, Adriano and {D'Odorico}, Sandro and {Poli}, Francesco and {Giallongo}, Emanuele and {Arnouts}, Stephane and {Cristiani}, Stefano and {Moorwood}, Alan and {Saracco}, Paolo},
        title = "{Photometric Redshifts and Selection of High-Redshift Galaxies in the NTT and Hubble Deep Fields}",
      journal = {\aj},
         year = 2000,
        month = nov,
       volume = {120},
       number = {5},
        pages = {2206-2219},
          doi = {10.1086/316803},
archivePrefix = {arXiv},
       eprint = {astro-ph/0009158},
 primaryClass = {astro-ph},
       adsurl = {https://ui.adsabs.harvard.edu/abs/2000AJ....120.2206F}
}

@ARTICLE{brammer08,
       author = {{Brammer}, Gabriel B. and {van Dokkum}, Pieter G. and {Coppi}, Paolo},
        title = "{EAZY: A Fast, Public Photometric Redshift Code}",
      journal = {\apj},
         year = 2008,
        month = oct,
       volume = {686},
       number = {2},
        pages = {1503-1513},
          doi = {10.1086/591786},
archivePrefix = {arXiv},
       eprint = {0807.1533},
 primaryClass = {astro-ph},
       adsurl = {https://ui.adsabs.harvard.edu/abs/2008ApJ...686.1503B}
}

@article{wu21,
    doi = {10.3847/1538-3881/ac20d6},
    url = {https://dx.doi.org/10.3847/1538-3881/ac20d6},
    year = {2021},
    month = {oct},
    publisher = {The American Astronomical Society},
    volume = {162},
    number = {5},
    pages = {201},
    author = {Wu, Po-Feng and Nelson, Dylan and van der Wel, Arjen and Pillepich, Annalisa and Zibetti, Stefano and Bezanson, Rachel and DEugenio, Francesco and Gallazzi, Anna and Pacifici, Camilla and Straatman, Caroline M. S. and Barišić, Ivana and Bell, Eric F. and Maseda, Michael V. and Muzzin, Adam and Sobral, David and Whitaker, Katherine E.},
    title = {Toward Precise Galaxy Evolution: A Comparison between Spectral Indices of z ∼1 Galaxies in the IllustrisTNG Simulation and the LEGA-C Survey},
    journal = {The Astronomical Journal}
}

@ARTICLE{kitzbichler07,
       author = {{Kitzbichler}, M.~G. and {White}, S.~D.~M.},
        title = "{The high-redshift galaxy population in hierarchical galaxy formation models}",
      journal = {\mnras},
         year = 2007,
        month = mar,
       volume = {376},
       number = {1},
        pages = {2-12},
          doi = {10.1111/j.1365-2966.2007.11458.x},
archivePrefix = {arXiv},
       eprint = {astro-ph/0609636},
 primaryClass = {astro-ph},
       adsurl = {https://ui.adsabs.harvard.edu/abs/2007MNRAS.376....2K}
}

@ARTICLE{blaizot05,
       author = {{Blaizot}, J{\'e}r{\'e}my and {Wadadekar}, Yogesh and {Guiderdoni}, Bruno and {Colombi}, St{\'e}phane T. and {Bertin}, Emmanuel and {Bouchet}, Fran{\c{c}}ois R. and {Devriendt}, Julien E.~G. and {Hatton}, Steve},
        title = "{MoMaF: the Mock Map Facility}",
      journal = {\mnras},
         year = 2005,
        month = jun,
       volume = {360},
       number = {1},
        pages = {159-175},
          doi = {10.1111/j.1365-2966.2005.09019.x},
archivePrefix = {arXiv},
       eprint = {astro-ph/0309305},
 primaryClass = {astro-ph},
       adsurl = {https://ui.adsabs.harvard.edu/abs/2005MNRAS.360..159B}
}

@ARTICLE{leja19,
       author = {{Leja}, Joel and {Carnall}, Adam C. and {Johnson}, Benjamin D. and {Conroy}, Charlie and {Speagle}, Joshua S.},
        title = "{How to Measure Galaxy Star Formation Histories. II. Nonparametric Models}",
      journal = {\apj},
         year = 2019,
        month = may,
       volume = {876},
       number = {1},
          eid = {3},
        pages = {3},
          doi = {10.3847/1538-4357/ab133c},
archivePrefix = {arXiv},
       eprint = {1811.03637},
 primaryClass = {astro-ph.GA},
       adsurl = {https://ui.adsabs.harvard.edu/abs/2019ApJ...876....3L}
}

@ARTICLE{suresh26,
       author = {{Suresh}, Arjun and {Blanton}, Michael R. and {Rennehan}, Douglas},
        title = "{AGN Feedback Models and AGN Demographics. I. Radio-mode AGN in EAGLE, SIMBA, and TNG100 Are Inconsistent with Observations}",
      journal = {\apj},
         year = 2026,
        month = jan,
       volume = {997},
       number = {1},
          eid = {105},
        pages = {105},
          doi = {10.3847/1538-4357/ae2608},
archivePrefix = {arXiv},
       eprint = {2508.04907},
 primaryClass = {astro-ph.GA},
       adsurl = {https://ui.adsabs.harvard.edu/abs/2026ApJ...997..105S}
}

@ARTICLE{lovell25,
       author = {{Lovell}, Christopher C. and {Roper}, William J. and {Vijayan}, Aswin P. and {Wilkins}, Stephen M. and {Newman}, Sophie and {Seeyave}, Louise},
        title = "{Synthesizer: a Software Package for Synthetic Astronomical Observables}",
      journal = {The Open Journal of Astrophysics},
         year = 2025,
        month = oct,
       volume = {8},
          eid = {152},
        pages = {152},
          doi = {10.33232/001c.145766},
archivePrefix = {arXiv},
       eprint = {2508.03888},
 primaryClass = {astro-ph.IM},
       adsurl = {https://ui.adsabs.harvard.edu/abs/2025OJAp....8E.152L}
}

@ARTICLE{santini22,
       author = {{Santini}, Paola and {Castellano}, Marco and {Fontana}, Adriano and {Fortuni}, Flaminia and {Menci}, Nicola and {Merlin}, Emiliano and {Pagul}, Amanda and {Testa}, Vincenzo and {Calabr{\`o}}, Antonello and {Paris}, Diego and {Pentericci}, Laura},
        title = "{The Stellar Mass Function in CANDELS and Frontier Fields: The Buildup of Low-mass Passive Galaxies since z   3}",
      journal = {\apj},
         year = 2022,
        month = dec,
       volume = {940},
       number = {2},
          eid = {135},
        pages = {135},
          doi = {10.3847/1538-4357/ac9a48},
archivePrefix = {arXiv},
       eprint = {2209.11250},
 primaryClass = {astro-ph.GA},
       adsurl = {https://ui.adsabs.harvard.edu/abs/2022ApJ...940..135S}
}

@ARTICLE{wellons15,
       author = {{Wellons}, Sarah and {Torrey}, Paul and {Ma}, Chung-Pei and {Rodriguez-Gomez}, Vicente and {Vogelsberger}, Mark and {Kriek}, Mariska and {van Dokkum}, Pieter and {Nelson}, Erica and {Genel}, Shy and {Pillepich}, Annalisa and {Springel}, Volker and {Sijacki}, Debora and {Snyder}, Gregory and {Nelson}, Dylan and {Sales}, Laura and {Hernquist}, Lars},
        title = "{The formation of massive, compact galaxies at z = 2 in the Illustris simulation}",
      journal = {\mnras},
         year = 2015,
        month = may,
       volume = {449},
       number = {1},
        pages = {361-372},
          doi = {10.1093/mnras/stv303},
archivePrefix = {arXiv},
       eprint = {1411.0667},
 primaryClass = {astro-ph.GA},
       adsurl = {https://ui.adsabs.harvard.edu/abs/2015MNRAS.449..361W}
}

@ARTICLE{genel18,
       author = {{Genel}, Shy and {Nelson}, Dylan and {Pillepich}, Annalisa and {Springel}, Volker and {Pakmor}, R{\"u}diger and {Weinberger}, Rainer and {Hernquist}, Lars and {Naiman}, Jill and {Vogelsberger}, Mark and {Marinacci}, Federico and {Torrey}, Paul},
        title = "{The size evolution of star-forming and quenched galaxies in the IllustrisTNG simulation}",
      journal = {\mnras},
         year = 2018,
        month = mar,
       volume = {474},
       number = {3},
        pages = {3976-3996},
          doi = {10.1093/mnras/stx3078},
archivePrefix = {arXiv},
       eprint = {1707.05327},
 primaryClass = {astro-ph.GA},
       adsurl = {https://ui.adsabs.harvard.edu/abs/2018MNRAS.474.3976G}
}

@ARTICLE{conselice03CAS,
       author = {{Conselice}, Christopher J.},
        title = "{The Relationship between Stellar Light Distributions of Galaxies and Their Formation Histories}",
      journal = {\apjs},
         year = 2003,
        month = jul,
       volume = {147},
       number = {1},
        pages = {1-28},
          doi = {10.1086/375001},
archivePrefix = {arXiv},
       eprint = {astro-ph/0303065},
 primaryClass = {astro-ph},
       adsurl = {https://ui.adsabs.harvard.edu/abs/2003ApJS..147....1C}
}

@ARTICLE{lachance25,
       author = {{LaChance}, Patrick and {Croft}, Rupert and {Ni}, Yueying and {Chen}, Nianyi and {Matteo}, Tiziana Di and {Bird}, Simeon},
        title = "{The evolution of galaxy morphology from redshift z=6 to 3: Mock JWST observations of galaxies in the ASTRID simulation}",
      journal = {The Open Journal of Astrophysics},
         year = 2025,
        month = feb,
       volume = {8},
          eid = {20},
        pages = {20},
          doi = {10.33232/001c.129991},
archivePrefix = {arXiv},
       eprint = {2401.16608},
 primaryClass = {astro-ph.GA},
       adsurl = {https://ui.adsabs.harvard.edu/abs/2025OJAp....8E..20L}
}

\begin{appendix}

\section{Validation tests}\label{validation_tests}

To ensure that our analysis is not affected by artifacts of the forward-modeling procedure, we carried out a set of validation tests. These checks verify that the construction of the mock light cones and the subsequent image generation do not introduce artificial biases, and that the simulated datasets consistently reproduce the expected properties of the original (parent) hydrodynamical simulation.

\subsection{Mass budget}

We checked whether our results are consistent with those published by the TNG and EAGLE teams. To this end, we considered the mass density \citep{furlong15} and mass functions \citep[MF,][]{pillepich18_mf, schaye15}; to our knowledge, no published result directly discusses the number counts. 

As outlined in F23, FORECAST builds the light cone by stacking partitions cut out from consecutive simulation boxes (the terms ``snapshot'' and ``box'' are used interchangeably). The partitions are sub-volumes of the full box, tailored to match the field of view and the targeted redshift interval. To avoid repeated structures within the cone, each partition undergoes random transformations such as rotations, axis inversions, and shifts of the central coordinates (see F23 for further details). 
For both the mass density and mass function, the volume of each FORECAST partition was computed as the area of the field of view multiplied by the comoving $z$-depth of the partition. 
When multiple partitions (typically up to two) are required to cover the z-depth of a single CHS snapshot, we combined the mass and volume across partitions to ensure consistency.

\FloatBarrier

\subsubsection{Stellar mass density}\label{sect:md}

\begin{figure}[!htbp]
    \centering
        \includegraphics[width=0.45\textwidth]{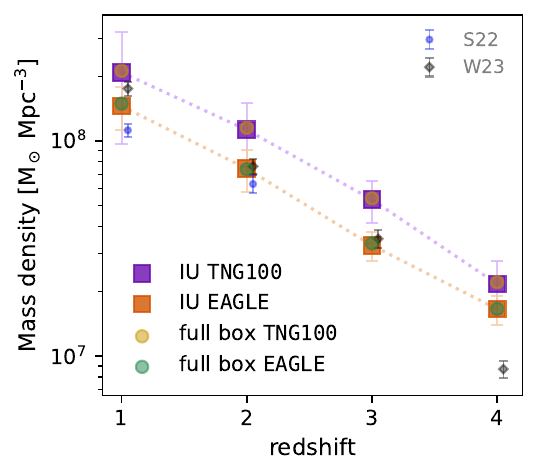}
        \caption{Stellar mass density comparison across redshifts for TNG100 and EAGLE simulations (considering only virialized objects, i.e., subhalos). Full snapshot values are shown as golden dots (TNG100) and green dots (EAGLE). Average mass densities from N=200 FORECAST realizations are represented by purple squares (TNG100) and orange squares (EAGLE), with 1$\sigma$ error bars. 
        TNG100 consistently shows stellar mass densities $1.4-1.6$ times higher than EAGLE across all redshifts, with a similar offset observed for dark matter, reflecting intrinsic differences between the simulations. Points with error bars are observations from \cite{santini22} and \cite{weaver23}.}
        \label{mdensity}
\end{figure}

We analyzed four snapshots at $z=1.0,\, 2.0,\, 3.0,\, 4.0$, and the corresponding partitions of the FORECAST light cone (i.e., we compared CHS snapshot boxes with the TNG100 and EAGLE IUs). At each redshift, we generated N=200 realizations of the partition by randomizing the initial conditions (e.g., center, axes, and orientation of the box), ensuring that the regions sampled from the snapshot are statistically independent.

For FORECAST IUs, the stellar mass density of each realization was computed as the sum of the masses of the star particles belonging to all subhalos within the partition, divided by the comoving volume of the partition.
For both simulation snapshot boxes, the stellar mass density was computed using the total stellar mass of all subhalos within the snapshot --- \texttt{SubhaloMassType(4)} for TNG100 \citep{nelson19a} and \texttt{Subhalo.MassType\_Star} for EAGLE \citep{eagle17} --- divided by their respective comoving volumes, $V_{\text{TNG}}$ = (110.7)$^3$ cMpc$^3$ and $V_{\text{EAGLE}}$ = (100)$^3$ cMpc$^3$.

In Fig. \ref{mdensity} we compare the stellar mass density of the full TNG100 snapshot (golden dots) and the EAGLE snapshot (green dots) with the average mass density of 200 realizations of the corresponding FORECAST partition (purple squares for TNG100 and orange squares for EAGLE, with 1$\sigma$ error bars) at different redshifts. 

The comparison shows that the FORECAST mass densities are consistent with the mass densities of the full simulation boxes for both TNG100 ($\times$0.98 on average) and EAGLE ($\times$0.99 on average), across all four redshifts.
At higher redshifts ($z$=3,4), the agreement between the full snapshot and the FORECAST mass densities is particularly strong, with minimal scatter between realizations. At $z$=1 and $z$=2, while the average mass densities remain consistent, the scatter slightly increases, reflecting a larger variance in the sampled regions of the simulation boxes.
Interestingly, we find that the stellar mass density of the subhalos in TNG100 is consistently $\times 1.4-1.6$ higher than that of EAGLE across all redshifts. 
To assess whether this offset could be driven by the different mass resolution of the two simulations, we repeated the computation of the stellar mass density adopting a common stellar-mass threshold for both. As expected, the absolute normalization of the stellar mass density changes with the adopted threshold, while the relative offset between TNG100 and EAGLE is preserved. This suggests that the offset is not due to TNG100 including a larger number of low-mass galaxies than EAGLE as a consequence of its higher mass resolution.
We also investigated the mass density of dark matter halos and found a similar offset, with TNG100 exceeding EAGLE by a factor of $1.25-1.55$ at all epochs. Since we are only considering the mass within collapsed halos, we interpret these offsets as arising from intrinsic differences in the distribution of baryonic and dark matter components in the two simulations. This interpretation is supported by the fact that, analyzing the total dark matter mass budget (that is, not only the virialized halos), the ratio between TNG100 and EAGLE mass density is 1 across all redshifts.
This can be understood in terms of the different implementations of star formation and feedback: two CHS based on the same cosmological framework (except for small differences in the parameters, see Sect. \ref{theoretical_data}) can yield markedly different halo assembly histories.

\begin{figure*}[b]
    \centering
        \includegraphics[width=1\textwidth]{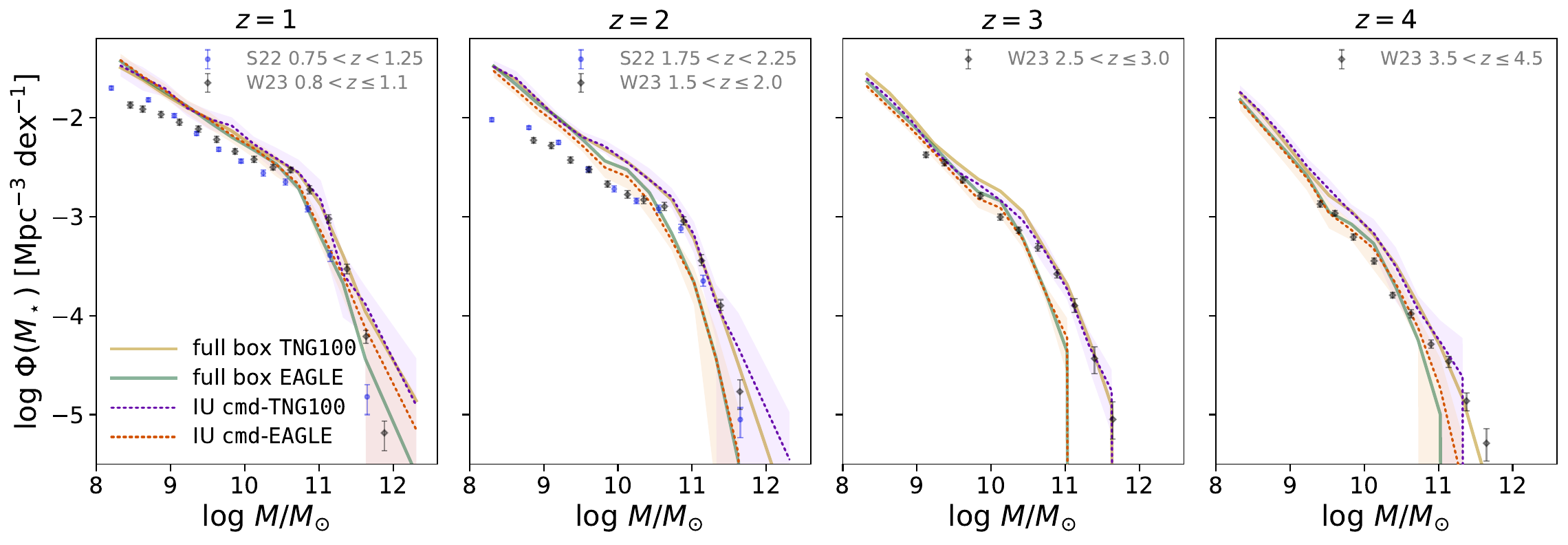}
        \caption{Stellar mass functions (MF) across four redshift bins ($z$ = 1, 2, 3, 4, from left to right). The MFs from the full snapshots are shown as solid lines, black for TNG100 and green for EAGLE. The average MFs from N=5 FORECAST realizations of the IU, with shaded area representing the 1$\sigma$ scatter, are represented as dashed lines, purple for TNG100 and orange for EAGLE. For reference, the observational data from \cite{santini22} and \cite{weaver23} are also shown. The mass functions derived from the FORECAST match those from the full snapshots across the entire mass and redshift range.}
        \label{mf}
\end{figure*}

The mass density of all stellar particles (i.e., not only those in subhalos) has a similar trend, with TNG being $\times$1.42-1.86 denser than EAGLE, at $z\leq3$; this indicates that the difference in stellar mass within subhalos is not due to a differing fraction of unbound stellar mass, but rather reflects a genuinely higher overall stellar mass budget in TNG100. Still, the fraction of stellar mass in EAGLE remains close to 1 across all redshifts; this is consistent, since most stellar particles are expected to form and stay within virialized structures. However, while for EAGLE this is true at all redshifts, TNG100 shows a larger fraction of nonvirialized stellar mass at $z=3,\,4$, with the total stellar mass exceeding the virialized one by a factor of $1.15-1.42$, implying that $15-42\%$ of the stellar mass at these epochs lies in filaments, intracluster light, and in general outside collapsed subhalos. A similar behavior is observed at lower redshifts in massive halos, where a significant fraction of the stellar content resides beyond 30-100 kpc from the center, contributing to the diffuse intracluster light \citep[see][]{pillepich18_mf}. 
Our results suggest that such extended stellar distributions are already in place at early cosmic times, likely reflecting recurring star formation activity in lower-density regions that have not collapsed into subhalos yet. These differences may arise from a combination of factors. Subgrid recipes for star formation and feedback certainly are a major cause of this \citep{wright24}. We point out that these discrepances highlight the importance of comparing the outputs of cosmological simulations not only at $z=0$, but across the whole Hubble time, paying special attention to early epochs which are now widely studied with the advent of new surveys. 

\FloatBarrier

\subsubsection{Stellar mass function}

We then compared the MF obtained with FORECAST (again, only considering the ground-truth values, i.e., the IU) with those directly extrapolated from the CHS snapshots (we verified that the latter are in perfect agreement with those published by the CHS teams), considering four redshifts: $z$=1, 2, 3, and 4.

To compute the MF from the FORECAST IUs, we used the ten mock datasets introduced in this work (\texttt{cmd}), five realizations based on TNG100 and five on EAGLE. For each dataset, we read the stellar mass of subhalos from the corresponding IU catalogs, where galaxy stellar mass is computed as the sum of stellar particle masses grouped by subhalo ID.

For the full-box reference MFs, we used the stellar masses defined in the previous section for the mass density. All masses were binned into 14 bins, each spanning 0.3 dex, with the last bin covering 0.8 dex.

In Fig. \ref{mf} we show the galaxy MF at the four considered redshifts, comparing the results from the full snapshots (black lines for TNG100 and green lines for EAGLE) and the FORECAST partitions (lines with diamonds: purple for TNG100, orange for EAGLE, with 1$\sigma$ shaded areas).

For both TNG100 and EAGLE, the FORECAST MFs are in excellent agreement with the full snapshot values across all redshifts, matching well within the 1$\sigma$ error throughout the full mass range. A slight offset is observed at $z$=3 for TNG100, where the FORECAST MF is lower by 0.04 dex in the low-mass bins, by 0.07 dex in the intermediate-mass bins, and by $\sim$ 0.04 dex in the highest-mass bins. Similarly, for EAGLE, minor deviations are visible at the high-mass end at some redshifts, though they remain within the statistical uncertainties.

Although the light-cone geometry can affect spatial clustering and cosmic variance estimates \citep{blaizot05}, it typically does not bias integrated properties when built carefully \citep{kitzbichler07}. In our case, the stacking and slicing of the snapshots made by FORECAST are sufficiently sharp to allow seamless transitions near the boundaries of the snapshots and to preserve the mass budget in the IU.

Overall, these results confirm the robustness of FORECAST in reproducing the intrinsic stellar mass distributions of both TNG100 and EAGLE. The consistency across redshifts and mass ranges demonstrates the effectiveness of the forward-modeling process adopted by FORECAST in preserving the intrinsic properties of the underlying simulations.

\FloatBarrier

\subsection{Comparison with a mock dataset in literature}  
As a check to rule out the possibility of major issues in our implementation, we compared the $H$-band counts from our FORECAST mock image with those of the synthetic data products of \citet[S23]{snyder23}.

Although the technique adopted to create the S23 light cone is essentially equivalent to ours, they produced the mock images using a slightly different method. Briefly, S23 generated the light cone (the position of the objects within the field of view) from the public TNG catalogs, then computed stellar fluxes of the subhalos in specific filters (including $H$) and created individual images of those galaxies through the public TNG application programming interface, arranging the produced individual cut-outs into a mock image according to the positions of the galaxies specified on the light-cone catalog (the procedure is described in detail in their work). FORECAST, instead, generates light cones by stacking consecutive simulation snapshots along the line of sight and assigning redshifts and positions to individual stellar particles. Stellar fluxes are computed at the particle level, assigning SEDs to each SSP based on age and metallicity, integrating through the desired filters, and projecting the resulting fluxes directly onto the image plane. This pixel-based construction provides a more direct mapping between the simulated physical properties and the resulting light distribution, preserving the spatial information of the stellar emission at the pixel level.

For this analysis, we used the two images based on TNG100 in the $H$ band released by S23 (\texttt{tng100-7-6-yxz\_137sqarcmin} and \texttt{tng100-7-6-xyz\_137sqarcmin}). These mock images cover a total area of 137 sq. arcmin. and span the redshift range $z=0.1-8.8$. We post-processed them with our pipeline to add PSF effects and noise to emulate the depth of GS; then we performed source detection with \textsc{SExtractor}. Since the S23 light-cone catalogs do not provide ground-truth $m_{AB}$ for galaxies in their corresponding images, we created an Input Universe catalog for S23 by detecting sources and measuring fluxes on their original image (noiseless and PSF-less). To guarantee a fair comparison with the S23 data, we adapted our FORECAST mock image to match the same redshift interval.

\begin{figure}[!htbp]
    \centering
        \includegraphics[width=0.48\textwidth]{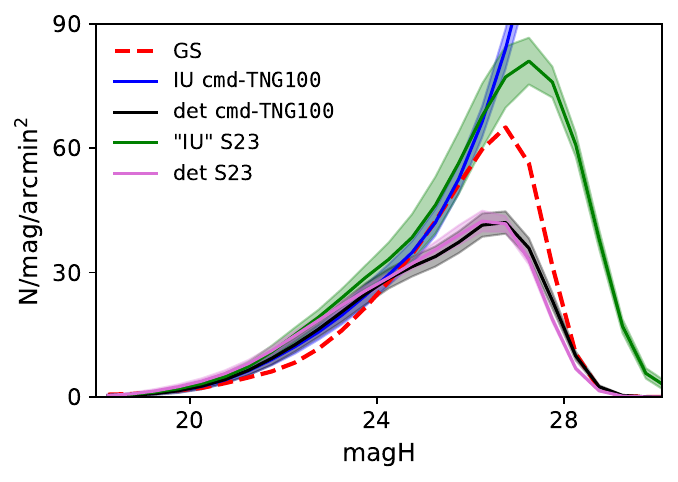}
        \caption{Comparison of the $H$-band counts from \texttt{cmd-TNG100} realizations generated with \texttt{FORECAST} and from two realizations of \citet[S23]{snyder23}, all based on the TNG100 simulation. For both datasets, IU and detection curves show the average over multiple realizations, with shaded areas indicating the 1$\sigma$ scatter. The IU from \texttt{FORECAST} (blue) is built analytically from the noiseless catalogs, whereas the IU from S23 (green) was obtained by running the detection on noiseless images. Detection counts for both datasets (black for \texttt{FORECAST}, magenta for S23) are measured on the corresponding mock images. The two datasets are in excellent agreement across the entire magnitude range, with the IU curves diverging only at the faint end, where the S23 curve turns over because of the detection procedure.}
        \label{p:snyder23}
\end{figure}

Figure \ref{p:snyder23} compares the $H$ counts derived from the IU and the detections on the mock images in both \texttt{cmd-TNG100} and S23. The detections are in excellent agreement across the entire magnitude range, confirming the consistency of the two implementations. As said, the construction of the light cone and the rendering of galaxy SEDs on the pixel grid (i.e., the creation of the noiseless, PSF-less mock images) follow different pipelines, while the final steps of the process (i.e., noise addition, PSF convolution, and source detection) are performed in the same way. The excellent agreement therefore indicates that the early stages of the forward-modeling process are consistent between the two approaches. This comparison does not directly test the impact of PSF and noise modeling, and thus does not exclude their possible contribution to the discrepancy with observations. However, \cite{lachance25} has shown that, at the level of galaxy populations, the impact of the post-processing step remains relatively minor. The IU curves also show very good agreement up to the peak of S23. At fainter magnitudes, the S23 IU counts drop due to detection on their noiseless image, which excludes sources below the threshold and results in the faint-end turnover.

Despite the differences in the methods used to create the mock images, the source counts, particularly those from detected sources, are remarkably consistent between our and S23 datasets, suggesting the robustness of the results obtained with FORECAST and ruling out the risk of a relevant bias in our analysis.

\FloatBarrier

\subsection{Testing FORECAST on the EAGLE simulation}

\begin{figure}[!htbp]
    \centering
        \includegraphics[width=0.48\textwidth]{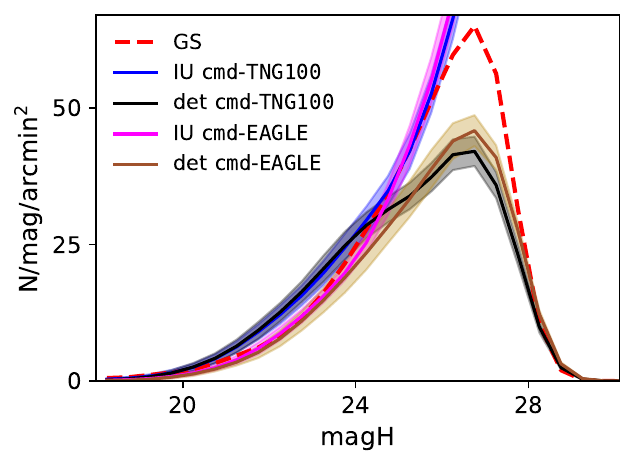}
        \caption{Comparison of $H$ counts for the CANDELS GS obtained from \texttt{cmd-TNG100} and \texttt{cmd-EAGLE} realizations. For both simulations, IU curves (blue and magenta) are built analytically from the noiseless catalogs, and detection curves (black and brown) are measured on the corresponding mock images. All curves show the average over five realizations, with shaded areas indicating the 1$\sigma$ scatter. Both simulations reproduce the overall shape of the observed GS counts (dashed red), with light differences in normalization reflecting their different intrinsic number densities and a mild excess of bright sources in TNG100.}
        \label{eagle}
\end{figure}

We asked whether the result depended on the original hydrodynamical simulation used in the forward-modeling process. In Fig. \ref{eagle} we compare the counts of the detections on the mock $H$-band images created with FORECAST using EAGLE and TNG100 for the GS field. At faint magnitudes the counts for the EAGLE IU (magenta line with 1$\sigma$ shaded area due to the five realizations) are consistent with those from the TNG100 IU (blue line), indicating that both predict a similar abundance of faint sources in the simulated field of view. Moreover, the detection counts in the EAGLE mock image (brown line) closely match those in the TNG100 mock image (black line), reinforcing their consistency under the depth limitations of the GS field.

At brighter magnitudes, marginal differences emerge between the two simulations: both the IU and the detections from TNG100 consistently exceed the corresponding counts from EAGLE. However, this is not directly relevant in our analysis: despite the differences in the bright regime, the consistent underproduction of faint sources across both mock images produced with two different simulations hints at a common, fundamental problem.

\FloatBarrier

\subsection{$H$ counts across photometric bands}

\begin{figure*}[b]
    \centering
        \includegraphics[width=1.\textwidth]{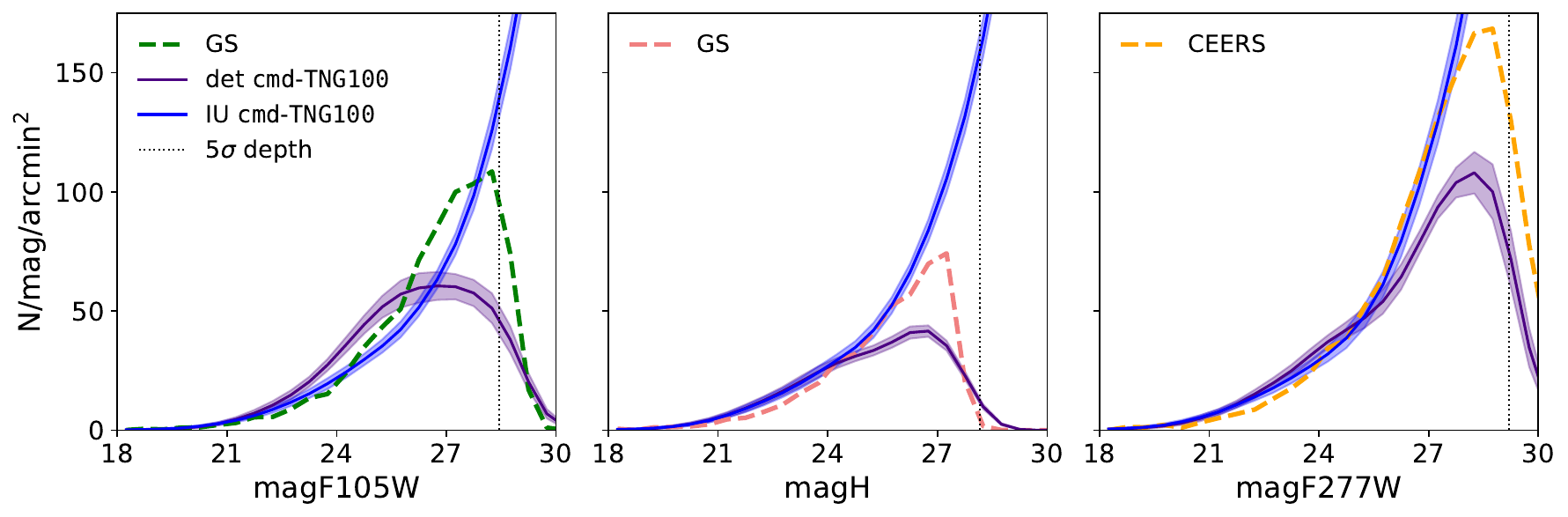}
        \caption{Galaxy counts across three bands: F105W (left) and $H$ (center) from CANDELS GOODS-South, and F277W (right) from CEERS. Observed counts are shown as green (F105W), red ($H$ or F160W), and orange (F277W) dashed lines, based on GOODS-South (\citealt{merlin21}; this work for F105W) and CEERS \citep{merlin24} catalogs. Mock counts from \texttt{cmd-TNG100} realizations are shown as blue lines for the IU and purple lines for the detections, all with 1$\sigma$ dispersion across the realizations. Vertical dotted lines mark the 5$\sigma$ depth limits reported in Table \ref{depths}. All mock datasets show a deficit at the faint end, independent of the observed region of the SED.}
        \label{multi-band}
\end{figure*}

\begin{table*}[b]
\caption{\texttt{SExtractor} Cold and Hot mode parameters adopted for the detection.}
\label{SEx}
\centering
\begin{tabular}{lcccccc}
\hline\hline
\texttt{Parameter} & \multicolumn{2}{c}{F105W} & \multicolumn{2}{c}{H} & \multicolumn{2}{c}{F277W} \\
                   & \texttt{Cold} & \texttt{Hot} & \texttt{Cold} & \texttt{Hot} & \texttt{Cold} & \texttt{Hot} \\
\hline
\texttt{DETECT\_MINAREA}      & 9.81 & 4.9 & 10.0 & 5.0 & 10.0 & 5.0 \\
\texttt{DETECT\_THRESH}       & 1.35 & 0.9 & 1.0 & 0.68 & 1.0 & 0.68 \\
\texttt{ANALYSIS\_THRESH}     & 3.0 & 0.9 & 3.0 & 0.68 & 3.0 & 0.68 \\
\texttt{DEBLEND\_NTHRESH}     & 16 & 64 & 16 & 64 & 16 & 64 \\
\texttt{DEBLEND\_MINCOUNT}    & 0.0001 & 0.001 & 0.0001 & 0.001 & 0.0001 & 0.001 \\
\texttt{BACK\_SIZE}           & 256 & 128 & 256 & 128 & 256 & 128 \\
\texttt{BACK\_FILTERSIZE}     & 9 & 5 & 9 & 5 & 9 & 5 \\
\texttt{BACKPHOTO\_TYPE}      & \texttt{local} & \texttt{local} & \texttt{local} & \texttt{local} & \texttt{local} & \texttt{local} \\
\texttt{BACKPHOTO\_THICK}     & 100 & 48 & 100 & 48 & 100 & 48 \\
\texttt{MEMORY\_OBJSTACK}     & 4000 & 4000 & 4000 & 4000 & 4000 & 4000 \\
\texttt{MEMORY\_PIXSTACK}     & 400\,000 & 400\,000 & 400\,000 & 100\,000 & 400\,000 & 100\,000 \\
\texttt{MEMORY\_BUFSIZE}      & 5000 & 5000 & 5000 & 5000 & 5000 & 5000 \\
\hline
\end{tabular}
\tablefoot{Parameters adopted for the CANDELS GOODS-South F105W and F160W ($H$) bands, and for the CEERS F277W band. A Gaussian filter (\textit{FWHM} = 4.0 pixels) was used in the Hot mode, and a top-hat PSF with a diameter of 5.0 pixels in the Cold mode.}
\end{table*}

We finally checked whether the number count issue can be due to differences in the spectral energy distribution of simulated galaxies with respect to their real counterparts: significantly bluer or redder SEDs would skew the count distribution across bands. To this end, we analyzed the counts in F105W from GS and F277W from CEERS (Cosmic Evolution Early Release Science).

The CEERS survey extends the legacy of CANDELS by exploiting the unprecedented sensitivity and resolution of JWST. CEERS (PI: S. Finkelstein; \citealt{finkelstein23,finkelstein25}) observed $\sim$94.6 sq. arcmin overlapping with the EGS field (ten pointings), using JWST’s NIRCam instrument to image seven photometric bands. For this validation test, we used the F277W band, exploiting the CEERS photometric catalog by \cite{merlin24}, which provides fluxes in 16 photometric bands combining archival HST data and JWST observations. Source detection was performed with \texttt{SExtractor} on stacks of the F356W and F444W mosaics, while fluxes were measured using aperture photometry on PSF-matched images (see also \citealt{merlin22,paris23}). Photometric redshifts were estimated using \texttt{zphot} \citep{fontana00} and \texttt{EAzY} \citep{brammer08} SED-fitting software.

We post-processed the noiseless mock images of each band to match the 5$\sigma$ depth of the corresponding observations (see Table \ref{depths} in Sect.\ref{FM}). Detection was performed independently on the F105W band for both real and mock images, since it is not used as a detection band in the official CANDELS catalogs. For the $H$ and F277W bands, instead, we relied on the published GOODS-South and CEERS catalogs introduced above, adopting consistently tuned \textsc{SExtractor} parameters (Table \ref{SEx}).
Figure \ref{multi-band} compares the number counts across the three bands, using the \texttt{cmd-TNG100} dataset. As usual, mock counts are shown for both the IU (blue lines) and the detections (purple lines), while the observed counts from the CANDELS and CEERS surveys are shown as colored lines (green for F105W, red for $H$, and orange for F277W).

Clearly, the problem is not in the SED: the counts of the mock detections in F105W and F277W show the same declining trend of F160W at the faint end, systematically underestimating the observed counts for magnitudes fainter than $\sim 26$. 

If a difference in the SED was the cause, we would expect to see an excess of faint sources in at least one of the adjacent bands, either bluer or redder. Instead, the lack of faint sources appears to be consistent across all bands.

From this result, we must conclude that the faint-end deficit is not caused by a problem introduced by the processing of the CHS output, but reflects a true limitation of the simulations in producing a sufficient number of faint galaxies at high redshift.

A caveat is that only two additional filters were tested, and not all spectral features (e.g., nebular lines) are fully constrained by this exercise. Moreover, the depths of the ancillary data differ, especially between CANDELS and CEERS, which complicates a perfectly homogeneous comparison. Nonetheless, across the broad F105W$-$F277W wavelength range, the faint-end shortfall emerges as a consistent and robust feature.
\end{appendix}

\end{document}